# GWTC-4.0: Updating the Gravitational-wave Transient Catalog with Observations from the First Part of the Fourth LIGO–Virgo–KAGRA Observing Run

A. G. Abac[1], I. Abouelfettouh[2], F. Acernese[3,4], K. Ackley[5], C. Adamcewicz[6], S. Adhicary[7], D. Adhikari[8,9], N. Adhikari[10], R. X. Adhikari[11], V. K. Adkins[12], S. Afroz[13], A. Agapito[14], D. Agarwal[15], M. Agathos[16], N. Aggarwal[17], S. Aggarwal[18], O. D. Aguiar[19], I.-L. Ahrend[20], L. Aiello[21,22], A. Ain[23], P. Ajith[24], S. Akcay[25], T. Akutsu[26,27], S. Albanesi[28,29], W. Ali[30,31], S. Al-Kershi[8,9], C. Alléné[32], A. Allocca[4,33], S. Al-Shammari[34], P. A. Altin[35], S. Alvarez-Lopez[36], W. Amar[32], O. Amarasinghe[34], A. Amato[37,38], F. Amicucci[39,40], C. Amra[41], A. Ananyeva[11], S. B. Anderson[11], W. G. Anderson[11], M. Andia[42], M. Ando[43], M. Andrés-Carcasona[44], T. Andrić[8,9,45,46], J. Anglin[47], S. Ansoldi[48,49], J. M. Antelis[50], S. Antier[42], M. Aoumi[51], E. Z. Appavuravther[52,53], S. Appert[11], S. K. Apple[54], K. Arai[11], A. Araya[43], M. C. Araya[11], M. Arca Sedda[45,46], J. S. Areeda[55], N. Aritomi[2], F. Armato[30,31], S. Armstrong[56], N. Arnaud[57], M. Arogeti[58], S. M. Aronson[12], K. G. Arun[59], G. Ashton[60], Y. Aso[26,61], L. Asprea[29], M. Assiduo[62,63], S. Assis de Souza Melo[64], S. M. Aston[65], P. Astone[39], F. Attadio[39,40], F. Aubin[66], K. AultONeal[67], G. Avallone[68], E. A. Avila[50], S. Babak[20], C. Badger[69], S. Bae[70], S. Bagnasco[29], L. Baiotti[71], R. Bajpai[72], T. Baka[38,73], A. M. Baker[6], K. A. Baker[74], T. Baker[75], G. Baldi[76,77], N. Baldicchi[52,78], M. Ball[79], G. Ballardin[64], S. W. Ballmer[80], S. Banagiri[6], B. Banerjee[45], D. Bankar[81], T. M. Baptiste[12], P. Baral[10], M. Baratti[82,83], J. C. Barayoga[11], B. C. Barish[11], D. Barker[2], N. Barman[81], P. Barneo[84,85,86], F. Barone[4,87], B. Barr[88], L. Barsotti[36], M. Barsuglia[20], D. Barta[89], A. M. Bartoletti[90], M. A. Barton[88], I. Bartos[47], A. Basalaev[8,9], R. Bassiri[91], A. Basti[82,83], M. Bawaj[52,78], P. Baxi[92], J. C. Bayley[88], A. C. Baylor[10], P. A. Baynard, II[58], M. Bazzan[93,94], V. M. Bedakihale[95], F. Beirnaert[96], M. Bejger[97], D. Belardinelli[22], A. S. Bell[88], D. S. Bellie[98], L. Bellizzi[82,83], W. Benoit[18], I. Bentara[57], J. D. Bentley[99], M. Ben Yaala[56], S. Bera[100,101], F. Bergamin[34], B. K. Berger[91], S. Bernuzzi[28], M. Beroiz[11], C. P. L. Berry[88], D. Bersanetti[30], T. Bertheas[102], A. Bertolini[37,38], J. Betzwieser[65], D. Beveridge[74], G. Bevilacqua[103], N. Bevins[104], R. Bhandare[105], S. A. Bhat[82], R. Bhatt[11], D. Bhattacharjee[106,107], S. Bhattacharyya[108], S. Bhaumik[47], V. Biancalana[103], A. Bianchi[38,109], I. A. Bilenko[110], G. Billingsley[11], A. Binetti[111], S. Bini[11,76,77], C. Binu[112], S. Biot[113], O. Birnholtz[114], S. Biscoveanu[98], A. Bisht[9], M. Bitossi[64,82], M.-A. Bizouard[115], S. Blaber[116], J. K. Blackburn[11], L. A. Blagg[79], C. D. Blair[65,74], D. G. Blair[74], N. Bode[8,9], N. Boettner[99], G. Boileau[115], M. Boldrini[39], G. N. Bolingbroke[117], A. Bolliand[41,118], L. D. Bonavena[47], R. Bondarescu[84], F. Bondu[119], E. Bonilla[91], M. S. Bonilla[55], A. Bonino[120], R. Bonnand[32,118], A. Borchers[8,9], V. Boschi[82], S. Bose[121], V. Bossilkov[65], Y. Bothra[38,109], A. Boudon[57], L. Bourg[58], M. Boyle[122], A. Bozzi[64], C. Bradaschia[82], P. R. Brady[10], A. Branch[65], M. Branchesi[45,46], I. Braun[106], T. Briant[123], A. Brillet[115], M. Brinkmann[8,9], P. Brockill[10], E. Brockmueller[8,9], A. F. Brooks[11], B. C. Brown[47], D. D. Brown[117], M. L. Brozzetti[52,78], S. Brunett[11], G. Bruno[15], R. Bruntz[124], J. Bryant[120], Y. Bu[125], F. Bucci[63], J. Buchanan[124], O. Bulashenko[84,85], T. Bulik[126], H. J. Bulten[38], A. Buonanno[1,127], K. Burtnyk[2], R. Buscicchio[128,129], D. Buskulic[32], C. Buy[102], R. L. Byer[91], G. S. Cabourn Davies[75], R. Cabrita[15], V. Cáceres-Barbosa[7], L. Cadonati[58], G. Cagnoli[130], C. Cahillane[80], A. Calafat[100], J. Calderón Bustillo[131], T. A. Callister[132], E. Calloni[4,33], S. R. Callos[79], M. Canepa[30,31], G. Caneva Santoro[44], K. C. Cannon[43], H. Cao[36], L. A. Capistran[133], E. Capocasa[20], E. Capote[2,11], G. Capurri[82,83], G. Carapella[68,134], F. Carbognani[64], M. Carlassara[8,9], J. B. Carlin[125], T. K. Carlson[135], M. F. Carney[106], M. Carpinelli[64,128], G. Carrillo[79], J. J. Carter[8,9], G. Carullo[120,136], A. Casallas-Lagos[137], J. Casanueva Diaz[64], C. Casentini[22,138], S. Y. Castro-Lucas[139], S. Caudill[135], M. Cavaglià[107], R. Cavalieri[64], A. Ceja[55], G. Cella[82], P. Cerdá-Durán[140,141], E. Cesarini[22], N. Chabbra[35], W. Chaibi[115], A. Chakraborty[13], P. Chakraborty[8,9], S. Chakraborty[105], S. Chalathadka Subrahmanya[99], J. C. L. Chan[142], M. Chan[116], K. Chang[143], S. Chao[143,144], P. Charlton[145], E. Chassande-Mottin[20], C. Chatterjee[146], Debarati Chatterjee[81], Deep Chatterjee[36], M. Chaturvedi[105], S. Chaty[20], K. Chatziioannou[11], A. Chen[147], A. H.-Y. Chen[148], D. Chen[149], H. Chen[144], H. Y. Chen[150], S. Chen[146], Yanbei Chen[151], Yitian Chen[122], H. P. Cheng[152], P. Chessa[52,78], H. T. Cheung[92], S. Y. Cheung[6], F. Chiadini[134,153], G. Chiarini[8,9,94], A. Chiba[154], A. Chincarini[30], M. L. Chiofalo[82,83], A. Chiummo[4,64], C. Chou[148], S. Choudhary[74], N. Christensen[115,155], S. S. Y. Chua[35], G. Ciani[76,77], P. Ciecielag[97], M. Cieślar[126], M. Cifaldi[22], B. Cirok[156], F. Clara[2], J. A. Clark[11,58], T. A. Clarke[6], P. Clearwater[157], S. Clesse[113], F. Cleva[115,118], S. M. Clyne[158], E. Coccia[44,45,46], E. Codazzo[159,160], P.-F. Cohadon[123], S. Colace[31], E. Colangeli[75], M. Colleoni[100], C. G. Collette[114], J. Collins[65], S. Colloms[88], A. Colombo[129,161], C. M. Compton[2], G. Connolly[79], L. Conti[94], T. R. Corbitt[12], I. Cordero-Carrión[162], S. Corezzi[52,78], N. J. Cornish[163], I. Coronado[164], A. Corsi[165], R. Cottingham[65], M. W. Coughlin[18], A. Couineaux[39], P. Couvares[11,58], D. M. Coward[74], R. Coyne[158], A. Cozzumbo[45], J. D. E. Creighton[10], T. D. Creighton[166], P. Cremonese[100], S. Crook[65], R. Crouch[2], J. Csizmazia[2], J. R. Cudell[167], T. J. Cullen[11], A. Cumming[88], E. Cuoco[168,169], M. Cusinato[140], L. V. Da Conceição[170], T. Dal Canton[42], S. Dal Pra[171], G. Dálya[102], B. D'Angelo[30], S. Danilishin[37,38], S. D'Antonio[39], K. Danzmann[8,9], K. E. Darroch[124], L. P. Dartez[65], R. Das[108], A. Dasgupta[95], V. Dattilo[64], A. Daumas[20], N. Davari[172,173], I. Dave[105], A. Davenport[139], M. Davier[42], T. F. Davies[74], D. Davis[11], L. Davis[74], M. C. Davis[18],

P. Davis[174,175], E. J. Daw[176], M. Dax[1], J. De Bolle[96], M. Deenadayalan[81], J. Degallaix[177], U. Deka[178], M. De Laurentis[4,33], F. De Lillo[23], S. Della Torre[129], W. Del Pozzo[82,83], A. Demagny[32], F. De Marco[39,40], G. Demasi[63,179], F. De Matteis[21,22], N. Demos[36], T. Dent[180], A. Depasse[15], N. DePergola[104], R. De Pietri[181,182], R. De Rosa[4,33], C. De Rossi[64], M. Desai[36], R. DeSalvo[183], A. DeSimone[184], R. De Simone[134,153], A. Dhani[1], R. Diab[47], M. C. Díaz[166], M. Di Cesare[4,33], G. Dideron[185], T. Dietrich[1], L. Di Fiore[4], C. Di Fronzo[74], M. Di Giovanni[39,40], T. Di Girolamo[4,33], D. Diksha[37,38], J. Ding[20,186], S. Di Pace[39,40], I. Di Palma[39,40], D. Di Piero[49,187], F. Di Renzo[57], Divyajyoti[34], A. Dmitriev[120], J. P. Docherty[88], Z. Doctor[98], N. Doerksen[170], E. Dohmen[2], A. Doke[135], A. Domiciano De Souza[188], L. D'Onofrio[39], F. Donovan[36], K. L. Dooley[34], T. Dooney[73], S. Doravari[81], O. Dorosh[189], W. J. D. Doyle[124], M. Drago[39,40], J. C. Driggers[2], L. Dunn[125], U. Dupletsa[45], P.-A. Duverne[20], D. D'Urso[159,172], P. Dutta Roy[47], H. Duval[190], S. E. Dwyer[2], C. Eassa[2], M. Ebersold[32,191], T. Eckhardt[99], G. Eddolls[80], A. Effler[65], J. Eichholz[35], H. Einsle[115], M. Eisenmann[26], M. Emma[60], K. Endo[154], R. Enficiaud[1], L. Errico[4,33], R. Espinosa[166], M. Esposito[4,33], R. C. Essick[192], H. Estellés[1], T. Etzel[11], M. Evans[36], T. Evstafyeva[185], B. E. Ewing[7], J. M. Ezquiaga[142], F. Fabrizi[62,63], V. Fafone[21,22], S. Fairhurst[34], A. M. Farah[132], B. Farr[79], W. M. Farr[193,194], G. Favaro[93], M. Favata[195], M. Fays[167], M. Fazio[56], J. Feicht[11], M. M. Fejer[91], R. Felicetti[49,187], E. Fenyvesi[89,196], J. Fernandes[197], T. Fernandes[140,198], D. Fernando[112], S. Ferraiuolo[39,40,199], T. A. Ferreira[12], F. Fidecaro[82,83], P. Figura[97], A. Fiori[82,83], I. Fiori[64], M. Fishbach[192], R. P. Fisher[124], R. Fittipaldi[134,200], V. Fiumara[134,201], R. Flaminio[32], S. M. Fleischer[202], L. S. Fleming[203], E. Floden[18], H. Fong[116], J. A. Font[140,141], F. Fontinele-Nunes[18], C. Foo[1], B. Fornal[204], K. Franceschetti[181], F. Frappez[32], S. Frasca[39,40], F. Frasconi[82], J. P. Freed[67], Z. Frei[205], A. Freise[38,109], O. Freitas[140,198], R. Frey[79], W. Frischhertz[65], P. Fritschel[36], V. V. Frolov[65], G. G. Fronzé[29], M. Fuentes-Garcia[11], S. Fujii[206], T. Fujimori[207], P. Fulda[47], M. Fyffe[65], B. Gadre[73], J. R. Gair[1], S. Galaudage[188], V. Galdi[208,209], R. Gamba[7], A. Gamboa[1], S. Gamoji[183], D. Ganapathy[210], A. Ganguly[81], B. Garaventa[30], J. García-Bellido[211], C. García-Quirós[191], J. W. Gardner[35], K. A. Gardner[116], S. Garg[43], J. Gargiulo[64], X. Garrido[42], A. Garron[100], F. Garufi[4,33], P. A. Garver[91], C. Gasbarra[21,22], B. Gateley[2], F. Gautier[212], V. Gayathri[10], T. Gayer[80], G. Gemme[30], A. Gennai[82], V. Gennari[102], J. George[105], R. George[150], O. Gerberding[99], L. Gergely[156], Archisman Ghosh[96], Sayantan Ghosh[197], Shaon Ghosh[195], Shrobana Ghosh[8,9], Suprovo Ghosh[213], Tathagata Ghosh[81], J. A. Giaime[12,65], K. D. Giardina[65], D. R. Gibson[203], C. Gier[56], S. Gkaitatzis[82,83], J. Glanzer[11], F. Glotin[42], J. Godfrey[79], R. V. Godley[8,9], P. Godwin[11], A. S. Goettel[34], E. Goetz[116], J. Golomb[11], S. Gomez Lopez[39,40], B. Goncharov[45], G. González[12], P. Goodarzi[214], S. Goode[6], A. W. Goodwin-Jones[15], M. Gosselin[64], R. Gouaty[32], D. W. Gould[35], K. Govorkova[36], A. Grado[52,78], V. Graham[88], A. E. Granados[18], M. Granata[177], V. Granata[134,215], S. Gras[36], P. Grassia[11], J. Graves[58], C. Gray[2], R. Gray[88], G. Greco[52], A. C. Green[38,109], L. Green[216], S. M. Green[75], S. R. Green[217], C. Greenberg[135], A. M. Gretarsson[67], H. K. Griffin[18], D. Griffith[11], H. L. Griggs[58], G. Grignani[52,78], C. Grimaud[32], H. Grote[34], S. Grunewald[1], D. Guerra[140], D. Guetta[218], G. M. Guidi[62,63], A. R. Guimaraes[12], H. K. Gulati[95], F. Gulminelli[174,175], H. Guo[147], W. Guo[74], Y. Guo[37,38], Anuradha Gupta[219], I. Gupta[7], N. C. Gupta[95], S. K. Gupta[47], V. Gupta[18], N. Gupte[1], J. Gurs[99], N. Gutierrez[177], N. Guttman[6], F. Guzman[133], D. Haba[220], M. Haberland[1], S. Haino[221], E. D. Hall[36], R. Hamburg[222], E. Z. Hamilton[100], G. Hammond[88], M. Haney[38], J. Hanks[2], C. Hanna[7], M. D. Hannam[34], O. A. Hannuksela[223], A. G. Hanselman[132], H. Hansen[2], J. Hanson[65], S. Hanumasagar[58], R. Harada[43], A. R. Hardison[184], S. Harikumar[189], K. Haris[38,73], I. Harley-Trochimczyk[133], T. Harmark[136], J. Harms[45,46], G. M. Harry[224], I. W. Harry[75], J. Hart[106], B. Haskell[97,225,226], C. J. Haster[216], K. Haughian[88], H. Hayakawa[51], K. Hayama[227], A. Heffernan[100], M. C. Heintze[65], J. Heinze[120], J. Heinzel[36], H. Heitmann[115], F. Hellman[210], A. F. Helmling-Cornell[79], G. Hemming[64], O. Henderson-Sapir[117], M. Hendry[88], I. S. Heng[88], M. H. Hennig[88], C. Henshaw[58], M. Heurs[8,9], A. L. Hewitt[228,229], J. Heynen[15], J. Heyns[36], S. Higginbotham[34], S. Hild[37,38], S. Hill[88], Y. Himemoto[230], N. Hirata[26], C. Hirose[231], D. Hofman[177], B. E. Hogan[67], N. A. Holland[38,109], I. J. Hollows[176], D. E. Holz[132], L. Honet[113], D. J. Horton-Bailey[210], J. Hough[88], S. Hourihane[11], N. T. Howard[146], E. J. Howell[74], C. G. Hoy[75], C. A. Hrishikesh[21], P. Hsi[36], H.-F. Hsieh[144], H.-Y. Hsieh[144], C. Hsiung[232], S.-H. Hsu[148], W.-F. Hsu[111], Q. Hu[88], H. Y. Huang[143], Y. Huang[7], Y. T. Huang[80], A. D. Huddart[233], B. Hughey[67], V. Hui[32], S. Husa[100], R. Huxford[7], L. Iampieri[39,40], G. A. Iandolo[37], M. Ianni[21,22], G. Iannone[134], J. Iascau[79], K. Ide[234], R. Iden[220], A. Ierardi[45,46], S. Ikeda[149], H. Imafuku[43], Y. Inoue[143], G. Iorio[93], P. Iosif[49,187], M. H. Iqbal[35], J. Irwin[88], R. Ishikawa[234], M. Isi[193,194], K. S. Isleif[235], Y. Itoh[207,236], M. Iwaya[206], B. R. Iyer[24], C. Jacquet[102], P.-E. Jacquet[123], T. Jacquot[42], S. J. Jadhav[237], S. P. Jadhav[157], M. Jain[135], T. Jain[228], A. L. James[11], A. Jan[150], K. Jani[146], J. Janquart[15], N. N. Janthalur[237], S. Jaraba[238], P. Jaranowski[239], R. Jaume[100], W. Javed[34], A. Jennings[2], M. Jensen[2], W. Jia[36], J. Jiang[152], H.-B. Jin[240,241], S. J. Jin[74], G. R. Johns[124], N. A. Johnson[47], N. K. Johnson-McDaniel[219], M. C. Johnston[216], R. Johnston[88], N. Johny[8,9], D. H. Jones[35], D. I. Jones[213], R. Jones[88], H. E. Jose[79], P. Joshi[7], S. K. Joshi[81], G. Joubert[57], J. Ju[242], L. Ju[74], K. Jung[243], J. Junker[35], V. Juste[113], H. B. Kabagoz[36,65], T. Kajita[244], I. Kaku[207], V. Kalogera[98], M. Kalomenopoulos[216], M. Kamiizumi[51], N. Kanda[207,236], S. Kandhasamy[81], G. Kang[245], N. C. Kannachel[6], J. B. Kanner[11], S. A. KantiMahanty[18], S. J. Kapadia[81], D. P. Kapasi[55], M. Karthikeyan[135], M. Kasprzack[11], H. Kato[154], T. Kato[206], E. Katsavounidis[36], W. Katzman[65], R. Kaushik[105], K. Kawabe[2], R. Kawamoto[207], D. Keitel[100], L. J. Kemperman[117], J. Kennington[7], F. A. Kerkow[18], R. Kesharwani[81], J. S. Key[246], R. Khadela[8,9],

S. Khadka[91], S. S. Khadkikar[7], F. Y. Khalili[110], F. Khan[8,9], T. Khanam[165], M. Khursheed[105], N. M. Khusid[193,194], W. Kiendrebeogo[115,247], N. Kijbunchoo[117], C. Kim[248], J. C. Kim[249], K. Kim[250], M. H. Kim[242], S. Kim[251], Y.-M. Kim[250], C. Kimball[98], K. Kimes[55], M. Kinnear[34], J. S. Kissel[2], S. Klimenko[47], A. M. Knee[116], E. J. Knox[79], N. Knust[8,9], K. Kobayashi[206], S. M. Koehlenbeck[91], G. Koekoek[37,38], K. Kohri[252,253], K. Kokeyama[34,254], S. Koley[45,167], P. Kolitsidou[120], A. E. Koloniari[255], K. Komori[43], A. K. H. Kong[144], A. Kontos[256], L. M. Koponen[120], M. Korobko[99], X. Kou[18], A. Koushik[23], N. Kouvatsos[69], M. Kovalam[74], T. Koyama[154], D. B. Kozak[11], S. L. Kranzhoff[37,38], V. Kringel[8,9], N. V. Krishnendu[120], S. Kroker[257], A. Królak[189,258], K. Kruska[8,9], J. Kubisz[259], G. Kuehn[8,9], S. Kulkarni[219], A. Kulur Ramamohan[35], Achal Kumar[47], Anil Kumar[237], Praveen Kumar[180], Prayush Kumar[24], Rahul Kumar[2], Rakesh Kumar[95], J. Kume[43,260,261], K. Kuns[36], N. Kuntimaddi[34], S. Kuroyanagi[211,262], S. Kuwahara[43], K. Kwak[243], K. Kwan[35], S. Kwon[43], G. Lacaille[88], D. Laghi[102,191], A. H. Laity[158], E. Lalande[263], M. Lalleman[23], P. C. Lalremruati[264], M. Landry[2], B. B. Lane[36], R. N. Lang[36], J. Lange[150], R. Langgin[216], B. Lantz[91], I. L. Rosa[100], J. Larsen[202], A. Lartaux-Vollard[42], P. D. Lasky[6], J. Lawrence[166], M. Laxen[65], C. Lazarte[140], A. Lazzarini[11], C. Lazzaro[159,160], P. Leaci[39,40], L. Leali[18], Y. K. Lecoeuche[116], H. M. Lee[265], H. W. Lee[266], J. Lee[80], K. Lee[242], R.-K. Lee[144], R. Lee[36], Sungho Lee[250], Sunjae Lee[242], Y. Lee[143], I. N. Legred[11], J. Lehmann[8,9], L. Lehner[185], M. Le Jean[118,177], A. Lemaître[267], M. Lenti[63,179], M. Leonardi[76,77,268], M. Lequime[41], N. Leroy[42], M. Lesovsky[11], N. Letendre[32], M. Lethuillier[57], Y. Levin[6], K. Leyde[75], A. K. Y. Li[11], K. L. Li[269], T. G. F. Li[111], X. Li[151], Y. Li[98], Z. Li[88], A. Lihos[124], E. T. Lin[144], F. Lin[143], L. C.-C. Lin[269], Y.-C. Lin[144], C. Lindsay[203], S. D. Linker[183], A. Liu[223], G. C. Liu[232], Jian Liu[74], F. Llamas Villarreal[166], J. Llobera-Querol[100], R. K. L. Lo[142], J.-P. Locquet[111], S. C. G. Loggins[270], M. R. Loizou[135], L. T. London[69], A. Longo[62,63], D. Lopez[167], M. Lopez Portilla[73], A. Lorenzo-Medina[180], V. Loriette[42], M. Lormand[65], G. Losurdo[82,271], E. Lotti[135], T. P. Lott, IV[58], J. D. Lough[8,9], H. A. Loughlin[36], C. O. Lousto[112], N. Low[125], N. Lu[35], L. Lucchesi[82], H. Lück[8,9], D. Lumaca[22], A. P. Lundgren[272,273], A. W. Lussier[263], R. Macas[75], M. MacInnis[36], D. M. Macleod[34], I. A. O. MacMillan[11], A. Macquet[42], K. Maeda[154], S. Maenaut[111], S. S. Magare[81], R. M. Magee[11], E. Maggio[1], R. Maggiore[38,109], M. Magnozzi[30,31], M. Mahesh[99], M. Maini[158], S. Majhi[81], E. Majorana[39,40], C. N. Makarem[11], D. Malakar[107], J. A. Malaquias-Reis[19], U. Mali[192], S. Maliakal[11], A. Malik[105], L. Mallick[170,192], A.-K. Malz[60], N. Man[115], M. Mancarella[101], V. Mandic[18], V. Mangano[159,172], B. Mannix[79], G. L. Mansell[80], M. Manske[10], M. Mantovani[64], M. Mapelli[93,94,274], C. Marinelli[103], F. Marion[32], A. S. Markosyan[91], A. Markowitz[11], E. Maros[11], S. Marsat[102], F. Martelli[62,63], I. W. Martin[88], R. M. Martin[195], B. B. Martinez[133], D. A. Martinez[55], M. Martinez[44,275], V. Martinez[130], A. Martini[76,77], J. C. Martins[19], D. V. Martynov[120], E. J. Marx[36], L. Massaro[37,38], A. Masserot[32], M. Masso-Reid[88], S. Mastrogiovanni[39], T. Matcovich[52], M. Matiushechkina[8,9], L. Maurin[212], N. Mavalvala[36], N. Maxwell[2], G. McCarrol[65], R. McCarthy[2], D. E. McClelland[35], S. McCormick[65], L. McCuller[11], S. McEachin[124], C. McElhenny[124], G. I. McGhee[88], J. McGinn[88], K. B. M. McGowan[146], J. McIver[116], A. McLeod[74], I. McMahon[191], T. McRae[35], R. McTeague[88], D. Meacher[10], B. N. Meagher[80], R. Mechum[112], Q. Meijer[73], A. Melatos[125], M. Melching[8,9], C. S. Menoni[139], F. Mera[2], R. A. Mercer[10], L. Mereni[177], K. Merfeld[165], E. L. Merilh[65], J. R. Mérou[100], J. D. Merritt[79], M. Merzougui[115], C. Messick[10], B. Mestichelli[45], M. Meyer-Conde[276], F. Meylahn[8,9], A. Mhaske[81], A. Miani[76,77], H. Miao[277], C. Michel[177], Y. Michimura[43], H. Middleton[120], D. P. Mihaylov[106], S. J. Miller[11], M. Millhouse[58], E. Milotti[49,187], V. Milotti[93], Y. Minenkov[22], E. M. Minihan[67], Ll. M. Mir[44], L. Mirasola[159,160], M. Miravet-Tenés[140], C.-A. Miritescu[44], A. Mishra[24], C. Mishra[108], T. Mishra[47], A. L. Mitchell[38,109], J. G. Mitchell[67], S. Mitra[81], V. P. Mitrofanov[110], K. Mitsuhashi[26], R. Mittleman[36], O. Miyakawa[51], S. Miyoki[51], A. Miyoko[67], G. Mo[36], L. Mobilia[62,63], S. R. P. Mohapatra[11], S. R. Mohite[7], M. Molina-Ruiz[210], M. Mondin[183], M. Montani[62,63], C. J. Moore[228], D. Moraru[2], A. More[81], S. More[81], C. Moreno[137], E. A. Moreno[36], G. Moreno[2], A. Moreso Serra[84], S. Morisaki[43,206], Y. Moriwaki[154], G. Morras[211], A. Moscatello[93], M. Mould[36], P. Mourier[100,278], B. Mours[66], C. M. Mow-Lowry[38,109], L. Muccillo[63,179], F. Muciaccia[39,40], D. Mukherjee[120], Samanwaya Mukherjee[24], Soma Mukherjee[166], Subroto Mukherjee[95], Suvodip Mukherjee[13], N. Mukund[36], A. Mullavey[65], H. Mullock[116], J. Mundi[224], C. L. Mungioli[74], M. Murakoshi[234], P. G. Murray[88], D. Nabari[76,77], S. L. Nadji[8,9], A. Nagar[29,279], N. Nagarajan[88], K. Nakagaki[51], K. Nakamura[26], H. Nakano[280], M. Nakano[11], D. Nanadoumgar-Lacroze[44], D. Nandi[12], V. Napolano[64], P. Narayan[219], I. Nardecchia[22], T. Narikawa[206], H. Narola[73], L. Naticchioni[39], R. K. Nayak[264], L. Negri[73], A. Nela[88], C. Nelle[79], A. Nelson[133], T. J. N. Nelson[65], M. Nery[8,9], A. Neunzert[2], S. Ng[55], L. Nguyen Quynh[281], S. A. Nichols[12], A. B. Nielsen[282], Y. Nishino[26,43], A. Nishizawa[283], S. Nissanke[38,284], W. Niu[7], F. Nocera[64], J. Noller[285], M. Norman[34], C. North[34], J. Novak[118,238,286], R. Nowicki[146], J. F. Nuño Siles[211], L. K. Nuttall[75], K. Obayashi[234], J. Oberling[2], J. O'Dell[233], E. Oelker[36], M. Oertel[118,238,286,287], G. Oganesyan[45,46], T. O'Hanlon[65], M. Ohashi[51], F. Ohme[8,9], R. Oliveri[118,286,287], R. Omer[18], B. O'Neal[124], M. Onishi[154], K. Oohara[288], B. O'Reilly[65], R. Oram[65], M. Orselli[52,78], R. O'Shaughnessy[112], S. O'Shea[88], S. Oshino[51], C. Osthelder[11], I. Ota[12], D. J. Ottaway[117], A. Ouzriat[57], H. Overmier[65], B. J. Owen[289], R. Ozaki[234], A. E. Pace[7], R. Pagano[12], M. A. Page[26], A. Pai[197], L. Paiella[45], A. Pal[290], S. Pal[264], M. A. Palaia[82,83], M. Pálfi[205], P. P. Palma[21,22,40], C. Palomba[39], P. Palud[20], H. Pan[144], J. Pan[74], K. C. Pan[144], P. K. Panda[237], Shiksha Pandey[7], Swadha Pandey[36], P. T. H. Pang[38,73], F. Pannarale[39,40], K. A. Pannone[55], B. C. Pant[105], F. H. Panther[74], M. Panzeri[62,63], F. Paoletti[82], A. Paolone[39,291], A. Papadopoulos[88], E. E. Papalexakis[214],

L. Papalini[82,83], G. Papigkiotis[255], A. Paquis[42], A. Parisi[52,78], B.-J. Park[250], J. Park[292], W. Parker[65], G. Pascale[8,9], D. Pascucci[96], A. Pasqualetti[64], R. Passaquieti[82,83], L. Passenger[6], D. Passuello[82], O. Patane[2], A. V. Patel[143], D. Pathak[81], L. Pathak[81], A. Patra[34], B. Patricelli[82,83], B. G. Patterson[34], K. Paul[108], S. Paul[79], E. Payne[11], T. Pearce[34], M. Pedraza[11], A. Pele[11], F. E. Peña Arellano[293], X. Peng[120], Y. Peng[58], S. Penn[294], M. D. Penuliar[55], A. Perego[76,77], Z. Pereira[135], C. Périgois[93,94,295], G. Perna[93], A. Perreca[45,76,77], J. Perret[20], S. Perriès[57], J. W. Perry[38,109], D. Pesios[255], S. Peters[167], S. Petracca[208,209], C. Petrillo[78], H. P. Pfeiffer[1], H. Pham[65], K. A. Pham[18], K. S. Phukon[120], H. Phurailatpam[223], M. Piarulli[102], L. Piccari[39,40], O. J. Piccinni[35], M. Pichot[115], M. Piendibene[82,83], F. Piergiovanni[62,63], L. Pierini[39], G. Pierra[39], V. Pierro[134,296], M. Pietrzak[97], M. Pillas[167], F. Pilo[82], L. Pinard[177], I. M. Pinto[33,134,296,297], M. Pinto[64], B. J. Piotrzkowski[10], M. Pirello[2], M. D. Pitkin[88,228], A. Placidi[52], E. Placidi[39,40], M. L. Planas[100], W. Plastino[22,215], C. Plunkett[36], R. Poggiani[82,83], E. Polini[36], J. Pomper[82,83], L. Pompili[1], J. Poon[223], E. Porcelli[38], E. K. Porter[20], C. Posnansky[7], R. Poulton[64], J. Powell[157], G. S. Prabhu[81], M. Pracchia[167], B. K. Pradhan[81], T. Pradier[66], A. K. Prajapati[95], K. Prasai[298], R. Prasanna[237], P. Prasia[81], G. Pratten[120], G. Principe[49,187], G. A. Prodi[76,77], P. Prosperi[82], P. Prosposito[21,22], A. C. Providence[67], A. Puecher[1], J. Pullin[12], P. Puppo[39], M. Pürrer[158], H. Qi[16], J. Qin[35], G. Quéméner[118,175], V. Quetschke[166], P. J. Quinonez[67], N. Qutob[58], R. Rading[235], I. Rainho[140], S. Raja[105], C. Rajan[105], B. Rajbhandari[112], K. E. Ramirez[65], F. A. Ramis Vidal[100], M. Ramos Arevalo[166], A. Ramos-Buades[38,100], S. Ranjan[58], K. Ransom[65], P. Rapagnani[39,40], B. Ratto[67], A. Ravichandran[135], A. Ray[98], V. Raymond[34], M. Razzano[82,83], J. Read[55], T. Regimbau[32], S. Reid[56], C. Reissel[36], D. H. Reitze[11], A. I. Renzini[11,128], B. Revenu[42,299], A. Revilla Peña[84], R. Reyes[183], L. Ricca[15], F. Ricci[39,40], M. Ricci[39,40], A. Ricciardone[82,83], J. Rice[80], J. W. Richardson[214], M. L. Richardson[117], A. Rijal[67], K. Riles[92], H. K. Riley[34], S. Rinaldi[274], J. Rittmeyer[99], C. Robertson[233], F. Robinet[42], M. Robinson[2], A. Rocchi[22], L. Rolland[32], J. G. Rollins[11], A. E. Romano[300], R. Romano[3,4], A. Romero[32], I. M. Romero-Shaw[228], J. H. Romie[65], S. Ronchini[7], T. J. Roocke[117], L. Rosa[4,33], T. J. Rosauer[214], C. A. Rose[58], D. Rosińska[126], M. P. Ross[54], M. Rossello-Sastre[100], S. Rowan[88], S. K. Roy[193,194], S. Roy[15], D. Rozza[128,129], P. Ruggi[64], N. Ruhama[243], E. Ruiz Morales[211,301], K. Ruiz-Rocha[146], S. Sachdev[58], T. Sadecki[2], P. Saffarieh[38,109], S. Safi-Harb[170], M. R. Sah[13], S. Saha[144], T. Sainrat[66], S. Sajith Menon[39,40,218], K. Sakai[302], Y. Sakai[276], M. Sakellariadou[69], S. Sakon[7], O. S. Salafia[128,129,161], F. Salces-Carcoba[11], L. Salconi[64], M. Saleem[150], F. Salemi[39,40], M. Sallé[38], S. U. Salunkhe[81], S. Salvador[174,175], A. Salvarese[150], A. Samajdar[38,73], A. Sanchez[2], E. J. Sanchez[11], L. E. Sanchez[11], N. Sanchis-Gual[140], J. R. Sanders[184], E. M. Sänger[1], F. Santoliquido[45,46], F. Sarandrea[29], T. R. Saravanan[81], N. Sarin[6], P. Sarkar[8,9], A. Sasli[255], P. Sassi[52,78], B. Sassolas[177], B. S. Sathyaprakash[7,34], R. Sato[231], S. Sato[154], Yukino Sato[154], Yu Sato[154], O. Sauter[47], R. L. Savage[2], T. Sawada[51], H. L. Sawant[81], S. Sayah[177], V. Scacco[21,22], D. Schaetzl[11], M. Scheel[151], A. Schiebelbein[192], M. G. Schiworski[80], P. Schmidt[120], S. Schmidt[73], R. Schnabel[99], M. Schneewind[8,9], R. M. S. Schofield[79], K. Schouteden[111], B. W. Schulte[8,9], B. F. Schutz[8,9,34], E. Schwartz[303], M. Scialpi[304], J. Scott[88], S. M. Scott[35], R. M. Sedas[65], T. C. Seetharamu[88], M. Seglar-Arroyo[44], Y. Sekiguchi[305], D. Sellers[65], N. Sembo[207], A. S. Sengupta[306], E. G. Seo[88], J. W. Seo[111], V. Sequino[4,33], M. Serra[39], A. Sevrin[190], T. Shaffer[2], U. S. Shah[58], M. A. Shaikh[265], L. Shao[307], A. K. Sharma[100], A. Sharma[308], Preeti Sharma[12], Prianka Sharma[105], Ritwik Sharma[18], S. Sharma Chaudhary[107], P. Shawhan[127], N. S. Shcheblanov[267,309], E. Sheridan[146], Z.-H. Shi[144], M. Shikauchi[43], R. Shimomura[310], H. Shinkai[310], S. Shirke[81], D. H. Shoemaker[36], D. M. Shoemaker[150], R. W. Short[2], S. ShyamSundar[105], A. Sider[114], H. Siegel[193,194], D. Sigg[2], L. Silenzi[37,38], L. Silvestri[40,171], M. Simmonds[117], L. P. Singer[311], Amitesh Singh[219], Anika Singh[11], D. Singh[210], M. K. Singh[312], N. Singh[100], S. Singh[61,220], A. M. Sintes[100], V. Sipala[159,172], V. Skliris[34], B. J. J. Slagmolen[35], D. A. Slater[202], T. J. Slaven-Blair[74], J. Smetana[120], J. R. Smith[55], L. Smith[49,88,187], R. J. E. Smith[6], W. J. Smith[146], S. Soares de Albuquerque Filho[62], M. Soares-Santos[191], K. Somiya[220], I. Song[144], S. Soni[36], V. Sordini[57], F. Sorrentino[30], H. Sotani[313], F. Spada[82], V. Spagnuolo[38], A. P. Spencer[88], P. Spinicelli[64], A. K. Srivastava[95], F. Stachurski[88], C. J. Stark[124], D. A. Steer[314], N. Steinle[170], J. Steinlechner[37,38], S. Steinlechner[37,38], N. Stergioulas[255], P. Stevens[42], S. P. Stevenson[157], M. StPierre[158], M. D. Strong[12], A. Strunk[2], A. L. Stuver[104,328], M. Suchenek[97], S. Sudhagar[97], Y. Sudo[234], N. Sueltmann[99], L. Suleiman[55], J. M. Sullivan[315], K. D. Sullivan[12], J. Sun[245], L. Sun[35], S. Sunil[95], J. Suresh[115], B. J. Sutton[69], P. J. Sutton[34], K. Suzuki[220], M. Suzuki[206], B. L. Swinkels[38], A. Syx[118], M. J. Szczepańczyk[316], P. Szewczyk[126], M. Tacca[38], H. Tagoshi[206], K. Takada[206], H. Takahashi[276], R. Takahashi[26], A. Takamori[43], S. Takano[317], H. Takeda[318,319], K. Takeshita[220], I. Takimoto Schmiegelow[45,46], M. Takou-Ayaoh[80], C. Talbot[132], M. Tamaki[206], N. Tamanini[102], D. Tanabe[143], K. Tanaka[51], S. J. Tanaka[234], S. Tanioka[34], D. B. Tanner[47], W. Tanner[8,9], L. Tao[214], R. D. Tapia[7], E. N. Tapia San Martín[38], C. Taranto[21,22], A. Taruya[320], J. D. Tasson[155], J. G. Tau[112], D. Tellez[55], R. Tenorio[100], H. Themann[183], A. Theodoropoulos[140], M. P. Thirugnanasambandam[81], L. M. Thomas[11], M. Thomas[65], P. Thomas[2], J. E. Thompson[213], S. R. Thondapu[105], K. A. Thorne[65], E. Thrane[6], S. Tibrewal[150], J. Tissino[45,46], A. Tiwari[81], Pawan Tiwari[45], Praveer Tiwari[197], S. Tiwari[191], V. Tiwari[120], M. R. Todd[80], M. Toffano[93], A. M. Toivonen[18], K. Toland[88], A. E. Tolley[75], T. Tomaru[26], V. Tommasini[11], T. Tomura[51], H. Tong[6], C. Tong-Yu[143], A. Torres-Forné[140,141], C. I. Torrie[11], I. Tosta e Melo[321], E. Tournefier[32], M. Trad Nery[115], K. Tran[124], A. Trapananti[52,53], R. Travaglini[169], F. Travasso[52,53], G. Traylor[65], M. Trevor[127], M. C. Tringali[64], A. Tripathee[92], G. Troian[49,187], A. Trovato[49,187],

L. Trozzo[4], R. J. Trudeau[11], T. Tsang[34], S. Tsuchida[322], L. Tsukada[216], K. Turbang[23,190], M. Turconi[115], C. Turski[96], H. Ubach[84,85], N. Uchikata[206], T. Uchiyama[51], R. P. Udall[11], T. Uehara[323], K. Ueno[43], V. Undheim[282], L. E. Uronen[223], T. Ushiba[51], M. Vacatello[82,83], H. Vahlbruch[8,9], N. Vaidya[11], G. Vajente[11], A. Vajpeyi[6], J. Valencia[100], M. Valentini[38,109], S. A. Vallejo-Peña[300], S. Vallero[29], V. Valsan[10], M. van Dael[38,324], E. Van den Bossche[190], J. F. J. van den Brand[37,38,109], C. Van Den Broeck[38,73], M. van der Sluys[38,73], A. Van de Walle[42], J. van Dongen[38,109], K. Vandra[104], M. VanDyke[121], H. van Haevermaet[23], J. V. van Heijningen[38,109], P. Van Hove[66], J. Vanier[263], M. VanKeuren[106], J. Vanosky[2], N. van Remortel[23], M. Vardaro[37,38], A. F. Vargas[125], V. Varma[135], A. N. Vazquez[91], A. Vecchio[120], G. Vedovato[94], J. Veitch[88], P. J. Veitch[117], S. Venikoudis[15], R. C. Venterea[18], P. Verdier[57], M. Vereecken[15], D. Verkindt[32], B. Verma[135], Y. Verma[105], S. M. Vermeulen[11], F. Vetrano[62], A. Veutro[39,40], A. Viceré[62,63], S. Vidyant[80], A. D. Viets[90], A. Vijaykumar[192], A. Vilkha[112], N. Villanueva Espinosa[140], V. Villa-Ortega[180], E. T. Vincent[58], J.-Y. Vinet[115], S. Viret[57], S. Vitale[36], H. Vocca[52,78], D. Voigt[99], E. R. G. von Reis[2], J. S. A. von Wrangel[8,9], W. E. Vossius[235], L. Vujeva[142], S. P. Vyatchanin[110], J. Wack[11], L. E. Wade[106], M. Wade[106], K. J. Wagner[112], L. Wallace[11], E. J. Wang[91], H. Wang[220], J. Z. Wang[92], W. H. Wang[166], Y. F. Wang[1], G. Waratkar[197], J. Warner[2], M. Was[32], T. Washimi[26], N. Y. Washington[11], D. Watarai[43], B. Weaver[2], S. A. Webster[88], N. L. Weickhardt[99], M. Weinert[8,9], A. J. Weinstein[11], R. Weiss[36,329], L. Wen[74], K. Wette[35], J. T. Whelan[112], B. F. Whiting[47], C. Whittle[11], E. G. Wickens[75], D. Wilken[8,9], A. T. Wilkin[214], B. M. Williams[121], D. Williams[88], M. J. Williams[75], N. S. Williams[1], J. L. Willis[11], B. Willke[8,9], M. Wils[111], L. Wilson[106], C. W. Winborn[107], J. Winterflood[74], C. C. Wipf[11], G. Woan[88], J. Woehler[37,38], N. E. Wolfe[36], H. T. Wong[143], I. C. F. Wong[111,223], K. Wong[192], T. Wouters[38,73], J. L. Wright[2], M. Wright[73,88], B. Wu[80], C. Wu[144], D. S. Wu[8,9], H. Wu[144], K. Wu[121], Q. Wu[54], Y. Wu[98], Z. Wu[102], E. Wuchner[55], D. M. Wysocki[10], V. A. Xu[210], Y. Xu[100], N. Yadav[29], H. Yamamoto[11], K. Yamamoto[154], T. S. Yamamoto[43], T. Yamamoto[51], R. Yamazaki[234], T. Yan[120], K. Z. Yang[18], Y. Yang[148], Z. Yarbrough[12], J. Yebana[100], S.-W. Yeh[144], A. B. Yelikar[146], X. Yin[36], J. Yokoyama[43,325], T. Yokozawa[51], S. Yuan[74], H. Yuzurihara[51], M. Zanolin[67], M. Zeeshan[112], T. Zelenova[64], J.-P. Zendri[94], M. Zeoli[15], M. Zerrad[41], M. Zevin[98], L. Zhang[11], N. Zhang[58], R. Zhang[152], T. Zhang[120], C. Zhao[74], Yue Zhao[164], Yuhang Zhao[20], Z.-C. Zhao[326], Y. Zheng[107], H. Zhong[18], H. Zhou[80], H. O. Zhu[74], Z.-H. Zhu[326,327], A. B. Zimmerman[150], L. Zimmermann[57], M. E. Zucker[11,36], J. Zweizig[11] The LIGO Scientific Collaboration, the Virgo Collaboration, and the KAGRA Collaboration

[1] Max Planck Institute for Gravitational Physics (Albert Einstein Institute), D-14476 Potsdam, Germany
[2] LIGO Hanford Observatory, Richland, WA 99352, USA
[3] Dipartimento di Farmacia, Università di Salerno, I-84084 Fisciano, Salerno, Italy
[4] INFN, Sezione di Napoli, I-80126 Napoli, Italy
[5] University of Warwick, Coventry CV4 7AL, UK
[6] OzGrav, School of Physics & Astronomy, Monash University, Clayton 3800, Victoria, Australia
[7] The Pennsylvania State University, University Park, PA 16802, USA
[8] Max Planck Institute for Gravitational Physics (Albert Einstein Institute), D-30167 Hannover, Germany
[9] Leibniz Universität Hannover, D-30167 Hannover, Germany
[10] University of Wisconsin-Milwaukee, Milwaukee, WI 53201, USA
[11] LIGO Laboratory, California Institute of Technology, Pasadena, CA 91125, USA
[12] Louisiana State University, Baton Rouge, LA 70803, USA
[13] Tata Institute of Fundamental Research, Mumbai 400005, India
[14] Centre de Physique Théorique, Aix-Marseille Université, Campus de Luminy, 163 Av. de Luminy, 13009 Marseille, France
[15] Université catholique de Louvain, B-1348 Louvain-la-Neuve, Belgium
[16] Queen Mary University of London, London E1 4NS, UK
[17] University of California, Davis, CA 95616, USA
[18] University of Minnesota, Minneapolis, MN 55455, USA
[19] Instituto Nacional de Pesquisas Espaciais, 12227-010 São José dos Campos, São Paulo, Brazil
[20] Université Paris Cité, CNRS, Astroparticule et Cosmologie, F-75013 Paris, France
[21] Università di Roma Tor Vergata, I-00133 Roma, Italy
[22] INFN, Sezione di Roma Tor Vergata, I-00133 Roma, Italy
[23] Universiteit Antwerpen, 2000 Antwerpen, Belgium
[24] International Centre for Theoretical Sciences, Tata Institute of Fundamental Research, Bengaluru 560089, India
[25] University College Dublin, Belfield, Dublin 4, Ireland
[26] Gravitational Wave Science Project, National Astronomical Observatory of Japan, 2-21-1 Osawa, Mitaka City, Tokyo 181-8588, Japan
[27] Advanced Technology Center, National Astronomical Observatory of Japan, 2-21-1 Osawa, Mitaka City, Tokyo 181-8588, Japan
[28] Theoretisch-Physikalisches Institut, Friedrich-Schiller-Universität Jena, D-07743 Jena, Germany
[29] INFN, Sezione di Torino, I-10125 Torino, Italy
[30] INFN, Sezione di Genova, I-16146 Genova, Italy
[31] Dipartimento di Fisica, Università degli Studi di Genova, I-16146 Genova, Italy
[32] Univ. Savoie Mont Blanc, CNRS, Laboratoire d'Annecy de Physique des Particules—IN2P3, F-74000 Annecy, France
[33] Università di Napoli "Federico II," I-80126 Napoli, Italy
[34] Cardiff University, Cardiff CF24 3AA, UK
[35] OzGrav, Australian National University, Canberra, Australian Capital Territory 0200, Australia
[36] LIGO Laboratory, Massachusetts Institute of Technology, Cambridge, MA 02139, USA
[37] Maastricht University, 6200 MD Maastricht, The Netherlands
[38] Nikhef, 1098 XG Amsterdam, The Netherlands
[39] INFN, Sezione di Roma, I-00185 Roma, Italy
[40] Università di Roma "La Sapienza," I-00185 Roma, Italy
[41] Aix Marseille Univ., CNRS, Centrale Med, Institut Fresnel, F-13013 Marseille, France
[42] Université Paris-Saclay, CNRS/IN2P3, IJCLab, 91405 Orsay, France

[43] University of Tokyo, Tokyo, 113-0033, Japan
[44] Institut de Física d'Altes Energies (IFAE), The Barcelona Institute of Science and Technology, Campus UAB, E-08193 Bellaterra (Barcelona), Spain
[45] Gran Sasso Science Institute (GSSI), I-67100 L'Aquila, Italy
[46] INFN, Laboratori Nazionali del Gran Sasso, I-67100 Assergi, Italy
[47] University of Florida, Gainesville, FL 32611, USA
[48] Dipartimento di Scienze Matematiche, Informatiche e Fisiche, Università di Udine, I-33100 Udine, Italy
[49] INFN, Sezione di Trieste, I-34127 Trieste, Italy
[50] Tecnologico de Monterrey, Escuela de Ingeniería y Ciencias, 64849, Monterrey, Nuevo León, Mexico
[51] Institute for Cosmic Ray Research, KAGRA Observatory, The University of Tokyo, 238 Higashi-Mozumi, Kamioka-cho, Hida City, Gifu 506-1205, Japan
[52] INFN, Sezione di Perugia, I-06123 Perugia, Italy
[53] Università di Camerino, I-62032 Camerino, Italy
[54] University of Washington, Seattle, WA 98195, USA
[55] California State University, Fullerton, Fullerton, CA 92831, USA
[56] SUPA, University of Strathclyde, Glasgow G1 1XQ, UK
[57] Université Claude Bernard Lyon 1, CNRS, IP2I Lyon/IN2P3, UMR 5822, F-69622 Villeurbanne, France
[58] Georgia Institute of Technology, Atlanta, GA 30332, USA
[59] Chennai Mathematical Institute, Chennai 603103, India
[60] Royal Holloway, University of London, London TW20 0EX, UK
[61] Astronomical Course, The Graduate University for Advanced Studies (SOKENDAI), 2-21-1 Osawa, Mitaka City, Tokyo 181-8588, Japan
[62] Università degli Studi di Urbino "Carlo Bo," I-61029 Urbino, Italy
[63] INFN, Sezione di Firenze, I-50019 Sesto Fiorentino, Firenze, Italy
[64] European Gravitational Observatory (EGO), I-56021 Cascina, Pisa, Italy
[65] LIGO Livingston Observatory, Livingston, LA 70754, USA
[66] Université de Strasbourg, CNRS, IPHC UMR 7178, F-67000 Strasbourg, France
[67] Embry-Riddle Aeronautical University, Prescott, AZ 86301, USA
[68] Dipartimento di Fisica "E.R. Caianiello," Università di Salerno, I-84084 Fisciano, Salerno, Italy
[69] King's College London, University of London, London WC2R 2LS, UK
[70] Korea Institute of Science and Technology Information, Daejeon 34141, Republic of Korea
[71] International College, Osaka University, 1-1 Machikaneyama-cho, Toyonaka City, Osaka 560-0043, Japan
[72] Accelerator Laboratory, High Energy Accelerator Research Organization (KEK), 1-1 Oho, Tsukuba City, Ibaraki 305-0801, Japan
[73] Institute for Gravitational and Subatomic Physics (GRASP), Utrecht University, 3584 CC Utrecht, The Netherlands
[74] OzGrav, University of Western Australia, Crawley, Western Australia 6009, Australia
[75] University of Portsmouth, Portsmouth, PO1 3FX, UK
[76] Università di Trento, Dipartimento di Fisica, I-38123 Povo, Trento, Italy
[77] INFN, Trento Institute for Fundamental Physics and Applications, I-38123 Povo, Trento, Italy
[78] Università di Perugia, I-06123 Perugia, Italy
[79] University of Oregon, Eugene, OR 97403, USA
[80] Syracuse University, Syracuse, NY 13244, USA
[81] Inter-University Centre for Astronomy and Astrophysics, Pune 411007, India
[82] INFN, Sezione di Pisa, I-56127 Pisa, Italy
[83] Università di Pisa, I-56127 Pisa, Italy
[84] Institut de Ciències del Cosmos (ICCUB), Universitat de Barcelona (UB), c. Martí i Franquès, 1, 08028 Barcelona, Spain
[85] Departament de Física Quàntica i Astrofísica (FQA), Universitat de Barcelona (UB), c. Martí i Franqués, 1, 08028 Barcelona, Spain
[86] Institut d'Estudis Espacials de Catalunya, c. Gran Capità, 2-4, 08034 Barcelona, Spain
[87] Dipartimento di Medicina, Chirurgia e Odontoiatria "Scuola Medica Salernitana," Università di Salerno, I-84081 Baronissi, Salerno, Italy
[88] IGR, University of Glasgow, Glasgow G12 8QQ, UK
[89] HUN-REN Wigner Research Centre for Physics, H-1121 Budapest, Hungary
[90] Concordia University Wisconsin, Mequon, WI 53097, USA
[91] Stanford University, Stanford, CA 94305, USA
[92] University of Michigan, Ann Arbor, MI 48109, USA
[93] Università di Padova, Dipartimento di Fisica e Astronomia, I-35131 Padova, Italy
[94] INFN, Sezione di Padova, I-35131 Padova, Italy
[95] Institute for Plasma Research, Bhat, Gandhinagar 382428, India
[96] Universiteit Gent, B-9000 Gent, Belgium
[97] Nicolaus Copernicus Astronomical Center, Polish Academy of Sciences, 00-716, Warsaw, Poland
[98] Northwestern University, Evanston, IL 60208, USA
[99] Universität Hamburg, D-22761 Hamburg, Germany
[100] IAC3–IEEC, Universitat de les Illes Balears, E-07122 Palma de Mallorca, Spain
[101] Aix-Marseille Université, Université de Toulon, CNRS, CPT, Marseille, France
[102] Laboratoire des 2 Infinis—Toulouse (L2IT-IN2P3), F-31062 Toulouse Cedex 9, France
[103] Università di Siena, Dipartimento di Scienze Fisiche, della Terra e dell'Ambiente, I-53100 Siena, Italy
[104] Villanova University, Villanova, PA 19085, USA
[105] RRCAT, Indore, Madhya Pradesh 452013, India
[106] Kenyon College, Gambier, OH 43022, USA
[107] Missouri University of Science and Technology, Rolla, MO 65409, USA
[108] Indian Institute of Technology Madras, Chennai 600036, India
[109] Department of Physics and Astronomy, Vrije Universiteit Amsterdam, 1081 HV Amsterdam, The Netherlands
[110] Lomonosov Moscow State University, Moscow 119991, Russia
[111] Katholieke Universiteit Leuven, Oude Markt 13, 3000 Leuven, Belgium
[112] Rochester Institute of Technology, Rochester, NY 14623, USA
[113] Université Libre de Bruxelles, 1050 Bruxelles, Belgium
[114] Bar-Ilan University, Ramat Gan, 5290002, Israel
[115] Université Côte d'Azur, Observatoire de la Côte d'Azur, CNRS, Artemis, F-06304 Nice, France
[116] University of British Columbia, Vancouver, BC V6T 1Z4, Canada
[117] OzGrav, University of Adelaide, Adelaide, South Australia 5005, Australia
[118] Centre national de la recherche scientifique, 75016 Paris, France

[119] Univ. Rennes, CNRS, Institut FOTON—UMR 6082, F-35000 Rennes, France
[120] University of Birmingham, Birmingham B15 2TT, UK
[121] Washington State University, Pullman, WA 99164, USA
[122] Cornell University, Ithaca, NY 14850, USA
[123] Laboratoire Kastler Brossel, Sorbonne Université, CNRS, ENS-Université PSL, Collège de France, F-75005 Paris, France
[124] Christopher Newport University, Newport News, VA 23606, USA
[125] OzGrav, University of Melbourne, Parkville, Victoria 3010, Australia
[126] Astronomical Observatory Warsaw University, 00-478 Warsaw, Poland
[127] University of Maryland, College Park, MD 20742, USA
[128] Università degli Studi di Milano-Bicocca, I-20126 Milano, Italy
[129] INFN, Sezione di Milano-Bicocca, I-20126 Milano, Italy
[130] Université de Lyon, Université Claude Bernard Lyon 1, CNRS, Institut Lumière Matière, F-69622 Villeurbanne, France
[131] IGFAE, Campus Sur, Universidade de Santiago de Compostela, 15782, Spain
[132] University of Chicago, Chicago, IL 60637, USA
[133] University of Arizona, Tucson, AZ 85721, USA
[134] INFN, Sezione di Napoli, Gruppo Collegato di Salerno, I-80126 Napoli, Italy
[135] University of Massachusetts Dartmouth, North Dartmouth, MA 02747, USA
[136] Niels Bohr Institute, Copenhagen University, 2100 København, Denmark
[137] Universidad de Guadalajara, 44430, Guadalajara, Jalisco, Mexico
[138] Istituto di Astrofisica e Planetologia Spaziali di Roma, 00133 Roma, Italy
[139] Colorado State University, Fort Collins, CO 80523, USA
[140] Departamento de Astronomía y Astrofísica, Universitat de València, E-46100 Burjassot, València, Spain
[141] Observatori Astronòmic, Universitat de València, E-46980 Paterna, València, Spain
[142] Niels Bohr Institute, University of Copenhagen, 2100 Kóbenhavn, Denmark
[143] National Central University, Taoyuan City 320317, Taiwan
[144] National Tsing Hua University, Hsinchu City 30013, Taiwan
[145] OzGrav, Charles Sturt University, Wagga Wagga, New South Wales 2678, Australia
[146] Vanderbilt University, Nashville, TN 37235, USA
[147] University of the Chinese Academy of Sciences/International Centre for Theoretical Physics Asia-Pacific, Bejing 100049, People's Republic of China
[148] Department of Electrophysics, National Yang Ming Chiao Tung University, 101 Univ. Street, Hsinchu, Taiwan
[149] Kamioka Branch, National Astronomical Observatory of Japan, 238 Higashi-Mozumi, Kamioka-cho, Hida City, Gifu 506-1205, Japan
[150] University of Texas, Austin, TX 78712, USA
[151] CaRT, California Institute of Technology, Pasadena, CA 91125, USA
[152] Northeastern University, Boston, MA 02115, USA
[153] Dipartimento di Ingegneria Industriale (DIIN), Università di Salerno, I-84084 Fisciano, Salerno, Italy
[154] Faculty of Science, University of Toyama, 3190 Gofuku, Toyama City, Toyama 930-8555, Japan
[155] Carleton College, Northfield, MN 55057, USA
[156] University of Szeged, Dóm tér 9, Szeged 6720, Hungary
[157] OzGrav, Swinburne University of Technology, Hawthorn VIC 3122, Australia
[158] University of Rhode Island, Kingston, RI 02881, USA
[159] INFN Cagliari, Physics Department, Università degli Studi di Cagliari, Cagliari 09042, Italy
[160] Università degli Studi di Cagliari, Via Università 40, 09124 Cagliari, Italy
[161] INAF, Osservatorio Astronomico di Brera sede di Merate, I-23807 Merate, Lecco, Italy
[162] Departamento de Matemáticas, Universitat de València, E-46100 Burjassot, València, Spain
[163] Montana State University, Bozeman, MT 59717, USA
[164] The University of Utah, Salt Lake City, UT 84112, USA
[165] Johns Hopkins University, Baltimore, MD 21218, USA
[166] The University of Texas Rio Grande Valley, Brownsville, TX 78520, USA
[167] Université de Liège, B-4000 Liège, Belgium
[168] DIFA—Alma Mater Studiorum Università di Bologna, Via Zamboni, 33—40126 Bologna, Italy
[169] Istituto Nazionale Di Fisica Nucleare—Sezione di Bologna, viale Carlo Berti Pichat 6/2—40127 Bologna, Italy
[170] University of Manitoba, Winnipeg, MB R3T 2N2, Canada
[171] INFN-CNAF—Bologna, Viale Carlo Berti Pichat, 6/2, 40127, Bologna BO, Italy
[172] Università degli Studi di Sassari, I-07100 Sassari, Italy
[173] INFN, Laboratori Nazionali del Sud, I-95125 Catania, Italy
[174] Université de Normandie, ENSICAEN, UNICAEN, CNRS/IN2P3, LPC Caen, F-14000 Caen, France
[175] Laboratoire de Physique Corpusculaire Caen, 6 boulevard du maréchal Juin, F-14050 Caen, France
[176] The University of Sheffield, Sheffield S10 2TN, UK
[177] Université Claude Bernard Lyon 1, CNRS, Laboratoire des Matériaux Avancés (LMA), IP2I Lyon/IN2P3, UMR 5822, F-69622 Villeurbanne, France
[178] International Centre for Theoretical Sciences, Tata Institute of Fundamental Research, Bangalore 560089, India
[179] Università di Firenze, Sesto Fiorentino I-50019, Italy
[180] IGFAE, Universidade de Santiago de Compostela, E-15782 Santiago de Compostela, Spain
[181] Dipartimento di Scienze Matematiche, Fisiche e Informatiche, Università di Parma, I-43124 Parma, Italy
[182] INFN, Sezione di Milano Bicocca, Gruppo Collegato di Parma, I-43124 Parma, Italy
[183] California State University, Los Angeles, Los Angeles, CA 90032, USA
[184] Marquette University, Milwaukee, WI 53233, USA
[185] Perimeter Institute, Waterloo, ON N2L 2Y5, Canada
[186] Corps des Mines, Mines Paris, Université PSL, 60 Bd Saint-Michel, 75272 Paris, France
[187] Dipartimento di Fisica, Università di Trieste, I-34127 Trieste, Italy
[188] Université Côte d'Azur, Observatoire de la Côte d'Azur, CNRS, Lagrange, F-06304 Nice, France
[189] National Center for Nuclear Research, 05-400 Świerk-Otwock, Poland
[190] Vrije Universiteit Brussel, 1050 Brussel, Belgium
[191] University of Zurich, Winterthurerstrasse 190, 8057 Zurich, Switzerland
[192] Canadian Institute for Theoretical Astrophysics, University of Toronto, Toronto, ON M5S 3H8, Canada
[193] Stony Brook University, Stony Brook, NY 11794, USA
[194] Center for Computational Astrophysics, Flatiron Institute, New York, NY 10010, USA

[195] Montclair State University, Montclair, NJ 07043, USA
[196] HUN-REN Institute for Nuclear Research, H-4026 Debrecen, Hungary
[197] Indian Institute of Technology Bombay, Powai, Mumbai 400 076, India
[198] Centro de Física das Universidades do Minho e do Porto, Universidade do Minho, PT-4710-057 Braga, Portugal
[199] Aix Marseille Univ, CNRS/IN2P3, CPPM, Marseille, France
[200] CNR-SPIN, I-84084 Fisciano, Salerno, Italy
[201] Scuola di Ingegneria, Università della Basilicata, I-85100 Potenza, Italy
[202] Western Washington University, Bellingham, WA 98225, USA
[203] SUPA, University of the West of Scotland, Paisley PA1 2BE, UK
[204] Barry University, Miami Shores, FL 33168, USA
[205] Eötvös University, Budapest 1117, Hungary
[206] Institute for Cosmic Ray Research, KAGRA Observatory, The University of Tokyo, 5-1-5 Kashiwa-no-Ha, Kashiwa City, Chiba 277-8582, Japan
[207] Department of Physics, Graduate School of Science, Osaka Metropolitan University, 3-3-138 Sugimoto-cho, Sumiyoshi-ku, Osaka City, Osaka 558-8585, Japan
[208] University of Sannio at Benevento, I-82100 Benevento, Italy
[209] INFN, Sezione di Napoli, I-80100 Napoli, Italy
[210] University of California, Berkeley, CA 94720, USA
[211] Instituto de Fisica Teorica UAM-CSIC, Universidad Autonoma de Madrid, 28049 Madrid, Spain
[212] Laboratoire d'Acoustique de l'Université du Mans, UMR CNRS 6613, F-72085 Le Mans, France
[213] University of Southampton, Southampton SO17 1BJ, UK
[214] University of California, Riverside, CA 92521, USA
[215] Dipartimento di Ingegneria Industriale, Elettronica e Meccanica, Università degli Studi Roma Tre, I-00146 Roma, Italy
[216] University of Nevada, Las Vegas, NV 89154, USA
[217] University of Nottingham, NG7 2RD, UK
[218] Ariel University, Ramat HaGolan St 65, Ari'el, Israel
[219] The University of Mississippi, University, MS 38677, USA
[220] Graduate School of Science, Institute of Science Tokyo, 2-12-1 Ookayama, Meguro-ku, Tokyo 152-8551, Japan
[221] Institute of Physics, Academia Sinica, 128 Sec. 2, Academia Rd., Nankang, Taipei 11529, Taiwan
[222] Science and Technology Institute, Universities Space Research Association, Huntsville, AL 35805, USA
[223] The Chinese University of Hong Kong, Shatin, NT, Hong Kong
[224] American University, Washington, DC 20016, USA
[225] Dipartimento di Fisica, Università degli studi di Milano, Via Celoria 16, I-20133, Milano, Italy
[226] INFN, sezione di Milano, Via Celoria 16, I-20133, Milano, Italy
[227] Department of Applied Physics, Fukuoka University, 8-19-1 Nanakuma, Jonan, Fukuoka City, Fukuoka 814-0180, Japan
[228] University of Cambridge, Cambridge CB2 1TN, UK
[229] University of Lancaster, Lancaster LA1 4YW, UK
[230] College of Industrial Technology, Nihon University, 1-2-1 Izumi, Narashino City, Chiba 275-8575, Japan
[231] Faculty of Engineering, Niigata University, 8050 Ikarashi-2-no-cho, Nishi-ku, Niigata City, Niigata 950-2181, Japan
[232] Department of Physics, Tamkang University, No. 151, Yingzhuan Rd., Danshui Dist., New Taipei City 25137, Taiwan
[233] Rutherford Appleton Laboratory, Didcot OX11 0DE, UK
[234] Department of Physical Sciences, Aoyama Gakuin University, 5-10-1 Fuchinobe, Sagamihara City, Kanagawa 252-5258, Japan
[235] Helmut Schmidt University, D-22043 Hamburg, Germany
[236] Nambu Yoichiro Institute of Theoretical and Experimental Physics (NITEP), Osaka Metropolitan University, 3-3-138 Sugimoto-cho, Sumiyoshi-ku, Osaka City, Osaka 558-8585, Japan
[237] Directorate of Construction, Services & Estate Management, Mumbai 400094, India
[238] Observatoire Astronomique de Strasbourg, 11 Rue de l'Université, 67000 Strasbourg, France
[239] Faculty of Physics, University of Białystok, 15-245 Białystok, Poland
[240] National Astronomical Observatories, Chinese Academic of Sciences, 20A Datun Rd., Chaoyang District, Beijing, People's Republic of China
[241] School of Astronomy and Space Science, University of Chinese Academy of Sciences, 20A Datun Rd., Chaoyang District, Beijing, People's Republic of China
[242] Sungkyunkwan University, Seoul 03063, Republic of Korea
[243] Department of Physics, Ulsan National Institute of Science and Technology (UNIST), 50 UNIST-gil, Ulju-gun, Ulsan 44919, Republic of Korea
[244] Institute for Cosmic Ray Research, The University of Tokyo, 5-1-5 Kashiwa-no-Ha, Kashiwa City, Chiba 277-8582, Japan
[245] Chung-Ang University, Seoul 06974, Republic of Korea
[246] University of Washington Bothell, Bothell, WA 98011, USA
[247] Laboratoire de Physique et de Chimie de l'Environnement, Université Joseph KI-ZERBO, 9GH2+3V5, Ouagadougou, Burkina Faso
[248] Ewha Womans University, Seoul 03760, Republic of Korea
[249] National Institute for Mathematical Sciences, Daejeon 34047, Republic of Korea
[250] Korea Astronomy and Space Science Institute, Daejeon 34055, Republic of Korea
[251] Department of Astronomy and Space Science, Chungnam National University, 9 Daehak-ro, Yuseong-gu, Daejeon 34134, Republic of Korea
[252] Institute of Particle and Nuclear Studies (IPNS), High Energy Accelerator Research Organization (KEK), 1-1 Oho, Tsukuba City, Ibaraki 305-0801, Japan
[253] Division of Science, National Astronomical Observatory of Japan, 2-21-1 Osawa, Mitaka City, Tokyo 181-8588, Japan
[254] Nagoya University, Nagoya, 464-8601, Japan
[255] Department of Physics, Aristotle University of Thessaloniki, 54124 Thessaloniki, Greece
[256] Bard College, Annandale-On-Hudson, NY 12504, USA
[257] Technical University of Braunschweig, D-38106 Braunschweig, Germany
[258] Institute of Mathematics, Polish Academy of Sciences, 00656 Warsaw, Poland
[259] Astronomical Observatory, Jagiellonian University, 31-007 Cracow, Poland
[260] Department of Physics and Astronomy, University of Padova, Via Marzolo, 8-35151 Padova, Italy
[261] Sezione di Padova, Istituto Nazionale di Fisica Nucleare (INFN), Via Marzolo, 8-35131 Padova, Italy
[262] Department of Physics, Nagoya University, ES building, Furocho, Chikusa-ku, Nagoya, Aichi 464-8602, Japan
[263] Université de Montréal/Polytechnique, Montreal, Quebec H3T 1J4, Canada
[264] Indian Institute of Science Education and Research, Kolkata, Mohanpur, West Bengal 741252, India
[265] Seoul National University, Seoul 08826, Republic of Korea
[266] Department of Computer Simulation, Inje University, 197 Inje-ro, Gimhae, Gyeongsangnam-do 50834, Republic of Korea
[267] NAVIER, École des Ponts, Univ. Gustave Eiffel, CNRS, Marne-la-Vallée, France
[268] Gravitational Wave Science Project, National Astronomical Observatory of Japan (NAOJ), Mitaka City, Tokyo 181-8588, Japan
[269] Department of Physics, National Cheng Kung University, No.1, University Rd., Tainan City 701, Taiwan

[270] St. Thomas University, Miami Gardens, FL 33054, USA
[271] Scuola Normale Superiore, I-56126 Pisa, Italy
[272] Institució Catalana de Recerca i Estudis Avançats, E-08010 Barcelona, Spain
[273] Institut de Física d'Altes Energies, E-08193 Barcelona, Spain
[274] Institut für Theoretische Astrophysik, Zentrum für Astronomie Heidelberg, Universitaet Heidelberg, Albert Ueberle Str. 2, 69120 Heidelberg, Germany
[275] Institucio Catalana de Recerca i Estudis Avançats (ICREA), Passeig de Lluís Companys, 23, 08010 Barcelona, Spain
[276] Research Center for Space Science, Advanced Research Laboratories, Tokyo City University, 3-3-1 Ushikubo-Nishi, Tsuzuki-Ku, Yokohama, Kanagawa 224-8551, Japan
[277] Tsinghua University, Beijing 100084, People's Republic of China
[278] School of Physical & Chemical Sciences, University of Canterbury, Private Bag 4800, Christchurch 8041, New Zealand
[279] Institut des Hautes Etudes Scientifiques, F-91440 Bures-sur-Yvette, France
[280] Faculty of Law, Ryukoku University, 67 Fukakusa Tsukamoto-cho, Fushimi-ku, Kyoto City, Kyoto 612-8577, Japan
[281] Phenikaa Institute for Advanced Study (PIAS), Phenikaa University, Yen Nghia, Ha Dong, Hanoi, Vietnam
[282] University of Stavanger, 4021 Stavanger, Norway
[283] Physics Program, Graduate School of Advanced Science and Engineering, Hiroshima University, 1-3-1 Kagamiyama, Higashihiroshima City, Hiroshima 739-8526, Japan
[284] GRAPPA, Anton Pannekoek Institute for Astronomy and Institute for High-Energy Physics, University of Amsterdam, 1098 XH Amsterdam, The Netherlands
[285] University College London, London WC1E 6BT, UK
[286] Observatoire de Paris, 75014 Paris, France
[287] Laboratoire Univers et Théories, Observatoire de Paris, 92190 Meudon, France
[288] Graduate School of Science and Technology, Niigata University, 8050 Ikarashi-2-no-cho, Nishi-ku, Niigata City, Niigata 950-2181, Japan
[289] University of Maryland, Baltimore County, Baltimore, MD 21250, USA
[290] CSIR-Central Glass and Ceramic Research Institute, Kolkata, West Bengal 700032, India
[291] Consiglio Nazionale delle Ricerche—Istituto dei Sistemi Complessi, I-00185 Roma, Italy
[292] Department of Astronomy, Yonsei University, 50 Yonsei-Ro, Seodaemun-Gu, Seoul 03722, Republic of Korea
[293] Department of Physics, University of Guadalajara, Av. Revolucion 1500, Colonia Olimpica C.P. 44430, Guadalajara, Jalisco, Mexico
[294] Hobart and William Smith Colleges, Geneva, NY 14456, USA
[295] INAF, Osservatorio Astronomico di Padova, I-35122 Padova, Italy
[296] Dipartimento di Ingegneria, Università del Sannio, I-82100 Benevento, Italy
[297] Museo Storico della Fisica e Centro Studi e Ricerche "Enrico Fermi," I-00184 Roma, Italy
[298] Kennesaw State University, Kennesaw, GA 30144, USA
[299] Subatech, CNRS/IN2P3—IMT Atlantique—Nantes Université, 4 rue Alfred Kastler BP 20722 44307 Nantes CÉDEX 03, France
[300] Universidad de Antioquia, Medellín, Colombia
[301] Departamento de Física—ETSIDI, Universidad Politécnica de Madrid, 28012 Madrid, Spain
[302] Department of Electronic Control Engineering, National Institute of Technology, Nagaoka College, 888 Nishikatakai, Nagaoka City, Niigata 940-8532, Japan
[303] Trinity College, Hartford, CT 06106, USA
[304] Dipartimento di Fisica e Scienze della Terra, Università Degli Studi di Ferrara, Via Saragat, 1, 44121, Ferrara FE, Italy
[305] Faculty of Science, Toho University, 2-2-1 Miyama, Funabashi City, Chiba 274-8510, Japan
[306] Indian Institute of Technology, Palaj, Gandhinagar, Gujarat 382355, India
[307] Kavli Institute for Astronomy and Astrophysics, Peking University, Yiheyuan Rd. 5, Haidian District, Beijing 100871, People's Republic of China
[308] Department of Physics, Indian Institute of Technology Gandhinagar, Gujarat 382055, India
[309] Laboratoire MSME, Cité Descartes, 5 Boulevard Descartes, Champs-sur-Marne, 77454, Marne-la-Vallée Cedex 2, France
[310] Faculty of Information Science and Technology, Osaka Institute of Technology, 1-79-1 Kitayama, Hirakata City, Osaka 573-0196, Japan
[311] NASA Goddard Space Flight Center, Greenbelt, MD 20771, USA
[312] Gravity Exploration Institute, Cardiff School of Physics and Astronomy, Cardiff University, Cardiff, CF24 3AA, UK
[313] Faculty of Science and Technology, Kochi University, 2-5-1 Akebono-cho, Kochi-shi, Kochi 780-8520, Japan
[314] Laboratoire de Physique de l'École Normale Supérieure, ENS, (CNRS, Université PSL, Sorbonne Université, Université Paris Cité), F-75005 Paris, France
[315] School of Physics, Georgia Institute of Technology, Atlanta, GA 30332, USA
[316] Faculty of Physics, University of Warsaw, Ludwika Pasteura 5, 02-093 Warszawa, Poland
[317] Laser Interferometry and Gravitational Wave Astronomy, Max Planck Institute for Gravitational Physics, Callinstrasse 38, 30167 Hannover, Germany
[318] The Hakubi Center for Advanced Research, Kyoto University, Yoshida-honmachi, Sakyou-ku, Kyoto City, Kyoto 606-8501, Japan
[319] Department of Physics, Kyoto University, Kita-Shirakawa Oiwake-cho, Sakyou-ku, Kyoto City, Kyoto 606-8502, Japan
[320] Yukawa Institute for Theoretical Physics (YITP), Kyoto University, Kita-Shirakawa Oiwake-cho, Sakyou-ku, Kyoto City, Kyoto 606-8502, Japan
[321] University of Catania, Department of Physics and Astronomy, Via S. Sofia, 64, 95123, Catania CT, Italy
[322] National Institute of Technology, Fukui College, Geshi-cho, Sabae-shi, Fukui 916-8507, Japan
[323] Department of Communications Engineering, National Defense Academy of Japan, 1-10-20 Hashirimizu, Yokosuka City, Kanagawa 239-8686, Japan
[324] Eindhoven University of Technology, 5600 MB Eindhoven, The Netherlands
[325] Kavli Institute for the Physics and Mathematics of the Universe (Kavli IPMU), WPI, The University of Tokyo, 5-1-5 Kashiwa-no-Ha, Kashiwa City, Chiba 277-8583, Japan
[326] Department of Astronomy, Beijing Normal University, Xinjiekouwai St. 19, Haidian District, Beijing 100875, People's Republic of China
[327] School of Physics and Technology, Wuhan University, Bayi Rd. 299, Wuchang District, Wuhan, Hubei 430072, People's Republic of China

[328] Deceased, September 2024.
[329] Deceased, August 2025.

## Abstract

Version 4.0 of the Gravitational-Wave Transient Catalog (GWTC-4.0) adds new candidates detected by the LIGO, Virgo, and KAGRA observatories through the first part of the fourth observing run (O4a: 2023 May 24 15:00:00 to 2024 January 16 16:00:00 UTC) and a preceding engineering run. In these new data, we find 128 compact binary coalescence candidates that are identified by at least one of our search algorithms with a probability of astrophysical origin $p_{\mathrm{astro}} \geqslant 0.5$ and that are not vetoed during event validation. We also provide detailed source property measurements for 86 of these that have a false-alarm rate <1 $\mathrm{yr}^{-1}$. Based on the inferred component masses, these candidates are consistent with signals from binary black holes (BBHs) and neutron star–black hole binaries (GW230518_125908 and GW230529_181500). Median-inferred component masses of BBHs in the catalog now range from 5.79M⊙ (GW230627_015337) to 137$M_{\odot}$ (GW231123_135430), while GW231123_135430 was probably produced by the most massive binary observed in the catalog. For the first time, we have discovered BBH signals with network signal-to-noise ratio exceeding 30, GW230814_230901 and GW231226_101520, enabling high-fidelity studies of the waveforms and astrophysical properties of these systems. Combined with the 90 candidates included in GWTC-3.0, the catalog now contains 218 candidates with $p_{\mathrm{astro}} \geqslant 0.5$ and not otherwise vetoed, more than doubling the size of the catalog and further opening our view of the gravitational-wave Universe.

## 1. Introduction

An increasingly diverse population of gravitational-wave (GW) sources in our Universe is now being uncovered by the Laser Interferometer Gravitational-Wave Observatory (LIGO; J. Aasi et al. 2015) and the Virgo (F. Acernese et al. 2015) and KAGRA (B. P. Abbott et al. 2020a) observatories. In the past decade, these detectors have seen transient GWs from binary black holes (BBHs; B. P. Abbott et al. 2016a), binary neutron stars (BNSs; B. P. Abbott et al. 2017a), and neutron star (NS)–black hole (BH) binaries (NSBHs; R. Abbott et al. 2021a) with a significant impact on physics, astrophysics, and cosmology. This work presents the observations and results of version 4.0 of the LIGO–Virgo–KAGRA Collaboration (LVK) Gravitational-Wave Transient Catalog (GWTC-4.0) and is part of a collection of articles that also includes an introduction (A. G. Abac et al. 2025a) and a description of the data-analysis methods (A. G. Abac et al. 2026a); we refer readers to these articles for contextual information on the results presented here.

GWTC-4.0 updates the previous GWTC-3.0 (R. Abbott et al. 2023a) by including the results of searches for compact binary coalescences (CBCs) in data collected through to the end of the first part of the fourth observing run (O4a). This period consists of data collected by the two-detector network of LIGO Hanford Observatory (LHO) and LIGO Livingston Observatory (LLO) between 2023 May 24 15:00:00 and 2024 January 16 16:00:00 UTC. In the analysis, we also include 5 days of data (between 2023 May 15 and 2023 May 19) from a pre-O4a engineering run that was conducted between 2023 April 26 and 2023 May 24 UTC (A. G. Abac et al. 2026b). The KAGRA detector joined the beginning of O4a, collecting data until 2023 June 20; however, we do not include this data in the analyses presented in this Letter as it is substantially less sensitive than the data from LHO and LLO.

We identify 1382 candidates with false-alarm rate (FAR) < 2 $\mathrm{day}^{-1}$ in at least one of our four search pipelines. Of these candidates, 128 have a probability of astrophysical CBC origin of $p_{\mathrm{astro}} \geqslant 0.5$ in at least one of our four search pipelines and are not vetoed during event validation (A. G. Abac et al. 2026a), bringing the total number of transients in the cumulative GWTC fulfilling these criteria to 218.

We analyze in detail the properties of a smaller, higher-purity subset of 86 candidates with FAR < 1 $\mathrm{yr}^{-1}$ and which pass our event-validation criteria. The bulk of new candidates reported here have properties that span a similar range of values as those reported in previous GWTC versions (B. P. Abbott et al. 2019a, 2021b, 2023a, 2024); however, a number of candidates exhibit new extremes.

The source of GW231123_135430 (A. G. Abac et al. 2025b) is probably more massive than that of any of our previously detected BBH candidates, with an inferred total mass $M = 236^{+29}_{-48}\,M_{\odot}$. GWTC-4.0 also includes the highest signal-to-noise ratio (SNR; throughout, we define the SNR to be the matched-filter network SNR) signal detected through to the end of O4a, GW230814_230901 (A. G. Abac et al. 2026c). With an SNR of 42.1, it is significantly louder than the previous record-holder, GW170817, which had an SNR of 32.4 (B. P. Abbott et al. 2017a).

In addition, the catalog update includes two NSBH candidates: GW230529_181500 (A. G. Abac et al. 2024b) and GW230518_125908, observed during the pre-O4a engineering run. All other O4a candidates have inferred component masses above the theoretical upper limit of the NS maximum mass (C. E. Rhoades & R. Ruffini 1974; V. Kalogera & G. Baym 1996) and are thus consistent with BBH systems.

As our catalog grows in tandem with the increasing sensitivity of the GW detector network, we detect a larger number of remarkable candidates as well as a larger number of candidates whose source properties we measure well. Several BBH sources reported here have total masses above 100 $M_{\odot}$, including the source of GW231028_153006, which has a total mass $152^{+29}_{-14}\,M_{\odot}$(here, and throughout this work, we present the median value and uncertainties based on the 90% credible interval). This BBH probably has a large, positive effective inspiral spin $\chi_{\mathrm{eff}} = 0.44^{+0.16}_{-0.20}$, a measure of the total spin angular momentum aligned with the orbital angular momentum (see A. G. Abac et al. 2025a for a definition). Other BBH sources are notable for having measured masses that rule out equal component masses, for example, that of GW231114_043211 which has mass ratio $q < 0.55$ with 90% probability. We also add a number of candidates with support for nonnegligible spins to GWTC-4.0. One such example is GW231118_005626, whose source has asymmetric masses, with a mass ratio $q = 0.55^{+0.37}_{-0.22}$ and a large $\chi_{\rm eff}$, with a primary spin magnitude $\chi_1 = 0.65^{+0.28}_{-0.38}$.

As in past observing runs (B. P. Abbott et al. 2019b, 2021b, 2023a), GW candidates identified by initial analysis of the data were publicly announced to enable searches for multimessenger counterparts. In O4a, these announcements were in the form of Notices distributed via General Coordinates Network (GCN; NASA 2025) and SCiMMA Hopskotch (SCiMMA 2025) and through GCN Circulars. There were 1702 candidates assigned an FAR $< 2\,{\rm day}^{-1}$ by at least one analysis in low latency. Of these candidates, 93 had an FAR $<$ 1 per 30 days after applying a trials factor to account for the number of simultaneously observing analyses and were reported as significant detection candidates. No confident multimessenger counterparts have been reported for any O4a candidates. Our offline analyses recover 77 of the candidates identified as significant in online search analyses in our higher-purity subset with both $p_{\rm astro} \geqslant 0.5$ and FAR $< 1\,{\rm yr}^{-1}$. We identify eight new candidates with $p_{\rm astro} \geqslant 0.5$ and which were not part of the 1702 candidates identified online as either significant or of low significance; of these, only one has FAR $< 1\,{\rm yr}^{-1}$. GWTC-4.0 is the most comprehensive set of GW observations to date.

The remainder of this Letter is structured as follows. In Section 2, we present the new candidates identified by search pipelines and discuss similarities and differences with the set identified by low-latency analyses. For a high-purity subset of these candidates, we present measurements of the inferred astrophysical source properties and discuss the impact of systematic differences between waveform models in Section 3. Finally, we summarize the results and discuss future prospects for GW astronomy in Section 4.

## 2. Candidate List

Search algorithms operate in two different modes: online (low latency) and offline. Online searches analyze data in near-real time as they are collected. The rapid identification of candidate astrophysical transients in low latency enables public alerts and facilitates searches for multimessenger counterparts. Offline analyses can be run at a higher latency using data that have undergone final calibration and benefit from noise subtraction (G. Vajente et al. 2020; S. Soni et al. 2025) and the identification of transient noise artifacts, known as *glitches* (L. K. Nuttall 2018; J. Glanzer et al. 2023; S. Soni et al. 2025), based on detector-monitor channel information (R. Essick et al. 2020; R. Huxford et al. 2024). Additionally, online analyses may be subject to occasional data dropout due to network instability or computing outages, while offline analyses have access to the complete dataset. For these reasons, offline analyses are more sensitive than their online counterparts, which leads to differences between the final candidate list and the initial online results. The majority of low-FAR online candidates (e.g., with FAR $< 1\,{\rm yr}^{-1}$) are expected to remain significant in offline analyses, but candidates that were initially identified with a higher FAR can change in significance when later reevaluated.

In this Letter, we describe the offline search results in O4a data from four search pipelines: Coherent WaveBurst (cWB)-BBH (S. Klimenko et al. 2005, 2008, 2016; T. Mishra et al. 2025), GstLAL (C. Messick et al. 2017; S. Sachdev et al. 2019; L. Tsukada et al. 2023; S. Sakon et al. 2024; P. Joshi et al. 2025), MBTA (C. Alléné et al. 2025), and PyCBC (T. Dal Canton et al. 2014; S. A. Usman et al. 2016; A. H. Nitz et al. 2017). Each of these analyses also searched for GW transients in low latency (K. Cannon et al. 2012; T. Adams et al. 2016; A. H. Nitz et al. 2018), as did the SPIIR analysis (Q. Chu et al. 2022); the online candidates are discussed more in Section 2.1.1. The cWB-BBH analysis is a minimally modeled search that coherently analyzes the data from the network to identify transient signals, while GstLAL, MBTA, PyCBC, and SPIIR use matched filtering to correlate the data with CBC waveform templates. Additional differences are detailed in A. G. Abac et al. (2026a).

### 2.1. Search Results

While there are many potential sources of transient GWs, only CBCs have been confidently identified (B. P. Abbott et al. 2017b, 2019c, 2021g; A. G. Abac et al. 2025c) to date. We therefore limit the GWTC-4.0 candidate list to potential BBHs, NSBHs, and BNSs. In GWTC-3.0 (R. Abbott et al. 2023a), candidates found by at least one pipeline with a probability of astrophysical origin $p_{\rm astro} \geqslant 0.5$ were selected for detailed analysis. Here, we additionally impose a threshold on the FAR of a given candidate, which is a measure of its detection significance defined by the estimated rate of nonastrophysical (noise) events found by a pipeline with a rank at least as high as the candidate. Both $p_{\rm astro}$ and FAR depend on the specific noise background seen by each pipeline and also incorporate assumptions on the (absolute or relative) astrophysical rates of signals for different binary sources. The $p_{\rm astro}$ calculation also depends on the sensitivity of the analyses to CBC signals, which varies across the source-parameter space as seen in Figure 1. This, coupled with methodological differences in estimating the noise background, can lead to significant differences in estimated FAR and $p_{\rm astro}$ across pipelines. Such differences are more pronounced for candidates of marginal significance, as well as those with inferred properties (under the assumption of astrophysical origin) lying outside the range of previously observed sources.

Given such systematic variability over pipelines, we sort candidates into three disjoint sets. First, candidates with both FAR $< 1\,{\rm yr}^{-1}$ and $p_{\rm astro} \geqslant 0.5$ in one or more pipelines. Second, candidates with $p_{\rm astro} \geqslant 0.5$ in one or more pipelines but FAR $\geqslant 1\,{\rm yr}^{-1}$ in all pipelines. Third, subthreshold candidates with $p_{\rm astro} < 0.5$ in all pipelines. There are no candidates with FAR $< 1\,{\rm yr}^{-1}$ for which $p_{\rm astro} < 0.\,5$ in all pipelines.

All candidates identified in GW data by the search pipelines are named with a GW prefix (see also A. G. Abac et al. 2025a). Those GW candidates found in O4a with an FAR $< 1\,{\rm yr}^{-1}$ in at least one analysis, and for which $p_{\rm astro} > 0.5$, can be found in Table 1 alongside their FAR, SNR, and $p_{\rm astro}$. Their individual-detector SNRs are reported in Table 6 of Appendix A. We carry out event validation for this high-purity subset of candidates (A. G. Abac et al. 2026a), and for those that pass validation, we estimate their source properties, which we report in Section 3. In addition, we list all other candidates assigned $p_{\rm astro} \geqslant 0.5$ by at least one pipeline in Table 2 and discuss the remaining subthreshold candidates in Section 2.1.4 below.

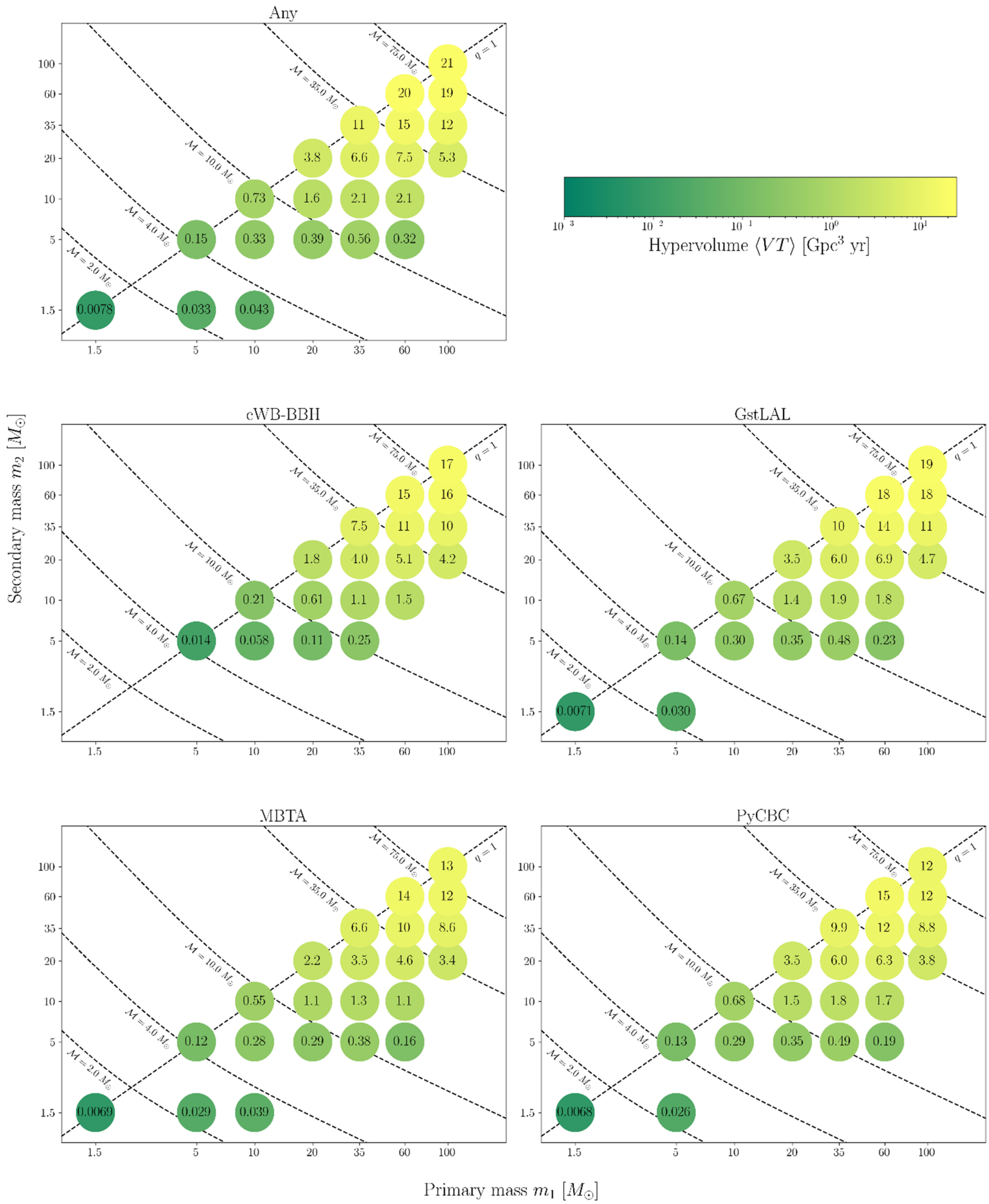

**Figure 1.** The sensitive hypervolume ⟨*VT*⟩ for searches of O4a data applying a significance threshold FAR $< 1$ yr$^{-1}$, evaluated at points in component mass space. The results labeled "Any," presented in the upper left-hand panel, come from calculating the ⟨*VT*⟩ for injections found by at least one search analysis. The color of each circle corresponds to the ⟨*VT*⟩ value. The plotted points correspond to the central points of log-normal distributions with width 0.1 used to estimate ⟨*VT*⟩.

#### 2.1.1. Online Candidates

In O4a, five pipelines conducted online searches for CBCs: cWB-BBH (T. Mishra et al. 2022), GstLAL (B. Ewing et al. 2024), MBTA (C. Alléné et al. 2025), PyCBC (T. Dal Canton et al. 2021), and SPIIR (Q. Chu et al. 2022). These searches distributed Notices via GCN (NASA 2025) and SCiMMA Hopskotch (SCiMMA 2025) for 1702 GW candidates with FAR $< 2$ day$^{-1}$. Of these, 93 passed a stricter threshold of FAR $< 1$ per 30 days after applying a trials factor to account for the number of simultaneous searches (LIGO–Virgo–KAGRA Collaboration 2025a); these candidates were reported as high-significance candidates. The high-significance

**Table 1**
Candidate GW Signals from O4a with an FAR $\leqslant 1\ \mathrm{yr}^{-1}$ in at Least One Analysis and for Which $p_{\mathrm{astro}} > 0.5$

| Candidate | Inst. | cWB-BBH | | | GstLAL | | | MBTA | | | PyCBC | | |
|---|---|---|---|---|---|---|---|---|---|---|---|---|---|
| | | FAR ($\mathrm{yr}^{-1}$) | SNR | $p_{\mathrm{astro}}$ | FAR ($\mathrm{yr}^{-1}$) | SNR | $p_{\mathrm{astro}}$ | FAR ($\mathrm{yr}^{-1}$) | SNR | $p_{\mathrm{astro}}$ | FAR ($\mathrm{yr}^{-1}$) | SNR | $p_{\mathrm{astro}}$ |
| GW230518_125908 | HL | … | … | … | $<1.0 \times 10^{-5}$ | 13.7 | >0.99 | $<1.0 \times 10^{-5}$ | 14.1 | >0.99 | $7.1 \times 10^{-4}$ | 13.6 | >0.99 |
| GW230529_181500 | L | … | … | … | 0.0058 | 11.8 | 0.85 | $2.2 \times 10^{-4}$ | 11.4 | >0.99 | $1.0 \times 10^{-3}$ | 11.7 | >0.99 |
| GW230601_224134 | HL | 0.0013 | 13.4 | >0.99 | $<1.0 \times 10^{-5}$ | 11.8 | >0.99 | 0.0082 | 12.4 | >0.99 | 0.001 | 12.2 | >0.99 |
| GW230605_065343 | HL | *560* | *7.5* | *<0.01* | $2.4 \times 10^{-5}$ | 10.7 | >0.99 | $<1.0 \times 10^{-5}$ | 11.1 | >0.99 | $1.3 \times 10^{-5}$ | 11.4 | >0.99 |
| GW230606_004305 | HL | 0.0067 | 11.1 | >0.99 | 0.0013 | 10.9 | >0.99 | *1.9* | *10.9* | *0.83* | $4.1 \times 10^{-4}$ | 10.7 | >0.99 |
| GW230608_205047 | HL | 0.032 | 9.9 | >0.99 | 0.0012 | 10.2 | >0.99 | 0.27 | 10.2 | 0.96 | … | … | … |
| GW230609_064958 | HL | 0.0013 | 10.6 | >0.99 | $1.4 \times 10^{-4}$ | 10.0 | >0.99 | *3.6* | *10.5* | *0.73* | 0.0011 | 9.6 | >0.99 |
| GW230624_113103 | HL | 0.0022 | 11.4 | >0.99 | $1.8 \times 10^{-4}$ | 10.0 | >0.99 | 0.018 | 10.3 | >0.99 | 0.017 | 10.2 | >0.99 |
| GW230627_015337 | HL | 0.0011 | 27.8 | >0.99 | $<1.0 \times 10^{-5}$ | 28.3 | >0.99 | $<1.0 \times 10^{-5}$ | 28.4 | >0.99 | $<1.0 \times 10^{-5}$ | 28.7 | >0.99 |
| GW230628_231200 | HL | 0.0011 | 16.4 | >0.99 | $<1.0 \times 10^{-5}$ | 15.3 | >0.99 | $<1.0 \times 10^{-5}$ | 15.9 | >0.99 | $<1.0 \times 10^{-5}$ | 15.9 | >0.99 |
| **GW230630_070659*** | HL | … | … | … | 0.47 | 9.8 | 0.88 | … | … | … | … | … | … |
| GW230630_125806 | HL | 0.16 | 9.0 | 0.97 | 0.25 | 8.1 | 0.93 | *1.3* | *8.2* | *0.87* | 0.48 | 8.1 | >0.99 |
| GW230630_234532 | HL | … | … | … | 0.028 | 9.8 | 0.99 | $4.2 \times 10^{-4}$ | 9.9 | >0.99 | 0.25 | 9.8 | >0.99 |
| GW230702_185453 | HL | 0.0089 | 10.1 | >0.99 | $<1.0 \times 10^{-5}$ | 9.8 | >0.99 | 0.21 | 9.9 | 0.97 | 0.031 | 9.2 | >0.99 |
| GW230704_021211 | HL | … | … | … | 0.21 | 9.4 | 0.94 | *2.7* | *9.2* | *0.78* | 0.38 | 9.2 | >0.99 |
| GW230704_212616 | HL | *43* | *8.3* | *0.14* | *11* | *8.3* | *0.30* | 0.51 | 8.7 | 0.93 | … | … | … |
| GW230706_104333 | HL | … | … | … | 0.23 | 9.2 | 0.94 | … | … | … | *1.5* | *8.8* | *0.98* |
| GW230707_124047 | HL | 0.0011 | 11.9 | >0.99 | 0.0026 | 10.1 | >0.99 | 0.072 | 10.3 | 0.99 | 0.0055 | 10.5 | >0.99 |
| GW230708_053705 | HL | … | … | … | *2.5* | *8.6* | *0.63* | *54* | *8.9* | *0.15* | 0.22 | 8.9 | >0.99 |
| GW230708_230935 | HL | *1.2* | *10.0* | *0.80* | 0.0037 | 9.6 | >0.99 | 0.26 | 9.7 | 0.96 | 0.012 | 9.4 | >0.99 |
| GW230709_122727 | HL | 0.071 | 10.2 | >0.99 | 0.16 | 9.9 | 0.95 | *12* | *10.1* | *0.48* | 0.011 | 10.0 | >0.99 |
| GW230712_090405 | HL | 0.018 | 9.5 | >0.99 | *99* | *8.2* | *0.05* | … | … | … | *260* | *8.2* | *0.19* |
| GW230723_101834 | HL | … | … | … | 0.0053 | 9.9 | >0.99 | 0.0034 | 10.0 | >0.99 | 0.0038 | 10.1 | >0.99 |
| GW230726_002940 | L | … | … | … | $<1.0 \times 10^{-5}$ | 10.5 | >0.99 | … | … | … | *4.5* | *10.0* | *0.58* |
| GW230729_082317 | HL | … | … | … | 0.18 | 9.5 | 0.95 | … | … | … | *31* | *9.4* | *0.77* |
| GW230731_215307 | HL | … | … | … | $<1.0 \times 10^{-5}$ | 12.2 | >0.99 | $<1.0 \times 10^{-5}$ | 11.9 | >0.99 | $<1.0 \times 10^{-5}$ | 11.9 | >0.99 |
| GW230803_033412 | HL | *3.0* | *9.4* | *0.68* | *3.0* | *8.0* | *0.59* | *19* | *8.6* | *0.35* | 0.31 | 8.2 | >0.99 |
| GW230805_034249 | HL | *7.5* | *9.5* | *0.49* | 0.0065 | 9.3 | >0.99 | *6.4* | *9.4* | *0.62* | 0.0037 | 9.4 | >0.99 |
| GW230806_204041 | HL | 0.0065 | 9.4 | >0.99 | 0.0037 | 9.1 | >0.99 | 0.20 | 9.4 | 0.97 | 0.035 | 9.1 | >0.99 |
| GW230811_032116 | HL | 0.0013 | 13.6 | >0.99 | $<1.0 \times 10^{-5}$ | 12.9 | >0.99 | $8.6 \times 10^{-5}$ | 13.3 | >0.99 | $<1.0 \times 10^{-5}$ | 12.4 | >0.99 |
| GW230814_061920 | HL | 0.0039 | 11.2 | >0.99 | $6.3 \times 10^{-4}$ | 10.2 | >0.99 | 0.041 | 10.0 | >0.99 | 0.0081 | 9.6 | >0.99 |
| GW230814_230901 | L | … | … | … | $<1.0 \times 10^{-5}$ | 42.3 | >0.99 | … | … | … | $1.0 \times 10^{-3}$ | 43.0 | >0.99 |
| GW230819_171910 | HL | 0.011 | 9.9 | >0.99 | 0.013 | 9.0 | >0.99 | … | … | … | *11* | *8.9* | *0.81* |
| GW230820_212515 | HL | *68* | *7.9* | *0.10* | 0.24 | 9.1 | 0.93 | 0.30 | 9.3 | 0.96 | 0.96 | 9.0 | 0.97 |
| GW230824_033047 | HL | 0.0035 | 11.1 | >0.99 | $<1.0 \times 10^{-5}$ | 10.5 | >0.99 | 0.017 | 10.6 | >0.99 | $<1.0 \times 10^{-5}$ | 10.7 | >0.99 |
| GW230825_041334 | HL | *1.3* | *8.8* | *0.83* | 0.10 | 8.7 | 0.97 | *1.8* | *8.5* | *0.84* | 0.83 | 8.7 | 0.98 |
| GW230831_015414 | HL | *27* | *7.8* | *0.20* | 0.63 | 8.6 | 0.86 | *1.4* | *8.6* | *0.87* | 0.29 | 8.5 | >0.99 |
| GW230904_051013 | HL | … | … | … | $3.9 \times 10^{-5}$ | 10.5 | >0.99 | $4.3 \times 10^{-5}$ | 10.4 | >0.99 | 0.0042 | 10.2 | >0.99 |
| GW230911_195324 | H | … | … | … | 0.014 | 10.7 | >0.99 | … | … | … | $1.0 \times 10^{-3}$ | 11.1 | >0.99 |
| GW230914_111401 | HL | 0.0012 | 17.2 | >0.99 | $<1.0 \times 10^{-5}$ | 15.9 | >0.99 | $<1.0 \times 10^{-5}$ | 16.6 | >0.99 | $<1.0 \times 10^{-5}$ | 16.0 | >0.99 |
| GW230919_215712 | HL | 0.0012 | 16.8 | >0.99 | $<1.0 \times 10^{-5}$ | 16.3 | >0.99 | $<1.0 \times 10^{-5}$ | 16.1 | >0.99 | $<1.0 \times 10^{-5}$ | 16.5 | >0.99 |
| GW230920_071124 | HL | 0.0012 | 11.1 | >0.99 | $<1.0 \times 10^{-5}$ | 10.1 | >0.99 | 0.11 | 10.2 | 0.98 | $3.4 \times 10^{-4}$ | 9.6 | >0.99 |
| GW230922_020344 | HL | 0.013 | 13.4 | >0.99 | $<1.0 \times 10^{-5}$ | 12.3 | >0.99 | $4.1 \times 10^{-5}$ | 12.2 | >0.99 | $3.6 \times 10^{-4}$ | 11.9 | >0.99 |
| GW230922_040658 | HL | 0.0012 | 12.5 | >0.99 | $<1.0 \times 10^{-5}$ | 11.6 | >0.99 | $3.0 \times 10^{-4}$ | 11.6 | >0.99 | $6.3 \times 10^{-4}$ | 11.6 | >0.99 |
| GW230924_124453 | HL | 0.0012 | 13.5 | >0.99 | $<1.0 \times 10^{-5}$ | 13.3 | >0.99 | $<1.0 \times 10^{-5}$ | 13.3 | >0.99 | $<1.0 \times 10^{-5}$ | 13.0 | >0.99 |
| GW230927_043729 | HL | 0.0012 | 12.1 | >0.99 | $<1.0 \times 10^{-5}$ | 11.3 | >0.99 | $8.6 \times 10^{-4}$ | 11.1 | >0.99 | $1.1 \times 10^{-4}$ | 11.1 | >0.99 |

**Table 1**
(Continued)

| Candidate | Inst. | cWB-BBH | | | GstLAL | | | MBTA | | | PyCBC | | |
|---|---|---|---|---|---|---|---|---|---|---|---|---|---|
| | | FAR ($yr^{-1}$) | SNR | $p_{astro}$ | FAR ($yr^{-1}$) | SNR | $p_{astro}$ | FAR ($yr^{-1}$) | SNR | $p_{astro}$ | FAR ($yr^{-1}$) | SNR | $p_{astro}$ |
| GW230927_153832 | HL | 0.0012 | 20.3 | >0.99 | $<1.0 \times 10^{-5}$ | 19.8 | >0.99 | $<1.0 \times 10^{-5}$ | 20.2 | >0.99 | $<1.0 \times 10^{-5}$ | 19.6 | >0.99 |
| GW230928_215827 | HL | 0.0035 | 10.5 | >0.99 | $1.5 \times 10^{-5}$ | 9.5 | >0.99 | *1.3* | *9.3* | *0.88* | 0.0092 | 9.5 | >0.99 |
| GW230930_110730 | HL | *5.4* | *9.0* | *0.58* | 0.17 | 8.5 | 0.95 | *1.1* | *8.6* | *0.89* | 0.73 | 8.3 | >0.99 |
| GW231001_140220 | HL | 0.0012 | 11.5 | >0.99 | $1.6 \times 10^{-5}$ | 10.3 | >0.99 | $1.8 \times 10^{-4}$ | 10.6 | >0.99 | 0.0031 | 9.9 | >0.99 |
| GW231004_232346 | HL | 0.16 | 8.9 | 0.97 | *6.5* | *7.9* | *0.41* | ⋯ | ⋯ | ⋯ | *420* | *7.0* | *0.01* |
| GW231005_021030 | HL | 0.010 | 10.4 | >0.99 | 0.17 | 9.3 | 0.95 | 0.019 | 9.7 | >0.99 | 0.21 | 9.8 | >0.99 |
| GW231005_091549 | HL | *54* | *11.0* | *0.13* | 0.040 | 8.9 | 0.99 | *2.6* | *8.6* | *0.79* | *3.6* | *8.5* | *0.97* |
| GW231008_142521 | HL | ⋯ | ⋯ | ⋯ | 0.0016 | 9.3 | >0.99 | *1.6* | *9.1* | *0.86* | 0.17 | 8.7 | >0.99 |
| GW231014_040532 | HL | *29* | *8.6* | *0.23* | 0.21 | 9.0 | 0.94 | *1.2* | *8.8* | *0.88* | *3.2* | *8.7* | *0.96* |
| GW231018_233037 | HL | ⋯ | ⋯ | ⋯ | *130* | *8.7* | *0.04* | 0.68 | 9.1 | 0.93 | *170* | *8.6* | *0.29* |
| GW231020_142947 | HL | ⋯ | ⋯ | ⋯ | $<1.0 \times 10^{-5}$ | 11.8 | >0.99 | $<1.0 \times 10^{-5}$ | 12.0 | >0.99 | $<1.0 \times 10^{-5}$ | 11.8 | >0.99 |
| GW231028_153006 | HL | 0.0012 | 22.4 | >0.99 | $<1.0 \times 10^{-5}$ | 21.0 | >0.99 | $<1.0 \times 10^{-5}$ | 21.9 | >0.99 | $<1.0 \times 10^{-5}$ | 21.9 | >0.99 |
| GW231029_111508 | L | ⋯ | ⋯ | ⋯ | $5.2 \times 10^{-5}$ | 10.8 | >0.99 | ⋯ | ⋯ | ⋯ | ⋯ | ⋯ | ⋯ |
| GW231102_071736 | HL | 0.0012 | 15.6 | >0.99 | $<1.0 \times 10^{-5}$ | 13.8 | >0.99 | $<1.0 \times 10^{-5}$ | 14.8 | >0.99 | $<1.0 \times 10^{-5}$ | 13.4 | >0.99 |
| GW231104_133418 | HL | ⋯ | ⋯ | ⋯ | $<1.0 \times 10^{-5}$ | 11.3 | >0.99 | $<1.0 \times 10^{-5}$ | 11.4 | >0.99 | $<1.0 \times 10^{-5}$ | 11.8 | >0.99 |
| GW231108_125142 | HL | $2.1 \times 10^{-4}$ | 12.6 | >0.99 | $<1.0 \times 10^{-5}$ | 12.6 | >0.99 | $7.5 \times 10^{-5}$ | 12.5 | >0.99 | $<1.0 \times 10^{-5}$ | 12.3 | >0.99 |
| GW231110_040320 | HL | ⋯ | ⋯ | ⋯ | $<1.0 \times 10^{-5}$ | 11.4 | >0.99 | $8.7 \times 10^{-4}$ | 11.5 | >0.99 | $<1.0 \times 10^{-5}$ | 11.1 | >0.99 |
| GW231113_122623 | HL | ⋯ | ⋯ | ⋯ | 0.75 | 8.3 | 0.83 | *38* | *8.6* | *0.16* | 0.28 | 8.6 | >0.99 |
| GW231113_200417 | HL | ⋯ | ⋯ | ⋯ | $8.0 \times 10^{-4}$ | 10.3 | >0.99 | $3.8 \times 10^{-5}$ | 10.1 | >0.99 | $3.7 \times 10^{-4}$ | 10.5 | >0.99 |
| GW231114_043211 | HL | ⋯ | ⋯ | ⋯ | $1.3 \times 10^{-4}$ | 10.0 | >0.99 | $2.0 \times 10^{-4}$ | 9.9 | >0.99 | 0.0059 | 9.6 | >0.99 |
| GW231118_005626 | HL | ⋯ | ⋯ | ⋯ | $1.2 \times 10^{-5}$ | 10.4 | >0.99 | $<1.0 \times 10^{-5}$ | 10.7 | >0.99 | $8.4 \times 10^{-5}$ | 10.5 | >0.99 |
| GW231118_071402 | HL | 0.078 | 9.2 | >0.99 | 0.0047 | 9.2 | >0.99 | 0.5 | 9.2 | 0.93 | 0.0028 | 9.2 | >0.99 |
| GW231118_090602 | HL | ⋯ | ⋯ | ⋯ | $<1.0 \times 10^{-5}$ | 10.8 | >0.99 | $<1.0 \times 10^{-5}$ | 11.0 | >0.99 | $7.1 \times 10^{-5}$ | 10.8 | >0.99 |
| GW231119_075248 | HL | *22* | *7.9* | *0.30* | 0.51 | 8.1 | 0.88 | *1.9* | *8.0* | *0.83* | 0.019 | 8.3 | >0.99 |
| GW231123_135430 | HL | $1.0 \times 10^{-4}$ | 21.8 | >0.99 | $<1.0 \times 10^{-5}$ | 20.1 | >0.99 | 0.016 | 19.0 | >0.99 | 0.0063 | 19.9 | >0.99 |
| GW231127_165300 | HL | 0.010 | 9.9 | 0.99 | 0.032 | 9.8 | 0.99 | 0.24 | 9.5 | 0.96 | 0.73 | 9.6 | 0.98 |
| GW231129_081745 | HL | 0.056 | 9.4 | >0.99 | 0.23 | 8.5 | 0.93 | *2.3* | *8.4* | *0.80* | *1.1* | *8.5* | *0.97* |
| GW231206_233134 | HL | 0.0012 | 12.8 | >0.99 | $<1.0 \times 10^{-5}$ | 11.9 | >0.99 | 0.074 | 11.7 | 0.98 | $1.6 \times 10^{-5}$ | 11.5 | >0.99 |
| GW231206_233901 | HL | 0.0012 | 21.9 | >0.99 | $<1.0 \times 10^{-5}$ | 20.7 | >0.99 | $<1.0 \times 10^{-5}$ | 21.4 | >0.99 | $1.6 \times 10^{-5}$ | 21.0 | >0.99 |
| GW231213_111417 | HL | 0.0046 | 10.0 | >0.99 | $<1.0 \times 10^{-5}$ | 10.2 | >0.99 | 0.029 | 10.4 | >0.99 | $7.7 \times 10^{-5}$ | 10.1 | >0.99 |
| GW231221_135041 | HL | 0.54 | 10.0 | 0.96 | *8.7* | *8.4* | *0.34* | *520* | *8.1* | *<0.01* | *240* | *8.3* | *0.11* |
| GW231223_032836 | HL | 0.0046 | 10.2 | >0.99 | $3.8 \times 10^{-4}$ | 9.4 | >0.99 | *13* | *9.1* | *0.42* | 0.0015 | 9.0 | >0.99 |
| GW231223_075055 | HL | ⋯ | ⋯ | ⋯ | *9.7* | *9.3* | *0.32* | *1.6* | *9.4* | *0.85* | 0.55 | 9.4 | 0.98 |
| GW231223_202619 | H | ⋯ | ⋯ | ⋯ | *7.1* | *10.0* | *0.39* | ⋯ | ⋯ | ⋯ | 0.0020 | 10.0 | >0.99 |
| GW231224_024321 | HL | ⋯ | ⋯ | ⋯ | $<1.0 \times 10^{-5}$ | 13.0 | >0.99 | $<1.0 \times 10^{-5}$ | 14.0 | >0.99 | $<1.0 \times 10^{-5}$ | 13.3 | >0.99 |
| GW231226_101520 | HL | 0.0012 | 34.7 | >0.99 | $<1.0 \times 10^{-5}$ | 34.2 | >0.99 | $<1.0 \times 10^{-5}$ | 33.6 | >0.99 | $<1.0 \times 10^{-5}$ | 33.2 | >0.99 |
| GW231230_170116 | HL | 0.42 | 8.2 | 0.96 | *70* | *8.0* | *0.07* | ⋯ | ⋯ | ⋯ | ⋯ | ⋯ | ⋯ |
| GW231231_154016 | H | ⋯ | ⋯ | ⋯ | $<1.0 \times 10^{-5}$ | 13.4 | >0.99 | ⋯ | ⋯ | ⋯ | $1.0 \times 10^{-3}$ | 13.4 | >0.99 |
| GW240104_164932 | H | ⋯ | ⋯ | ⋯ | $<1.0 \times 10^{-5}$ | 14.8 | >0.99 | ⋯ | ⋯ | ⋯ | 0.042 | 12.2 | 0.99 |
| GW240107_013215 | HL | 0.37 | 9.4 | 0.95 | 0.24 | 9.1 | 0.93 | 0.24 | 9.6 | 0.96 | 0.028 | 9.1 | >0.99 |
| GW240109_050431 | H | ⋯ | ⋯ | ⋯ | $2.3 \times 10^{-4}$ | 10.4 | >0.99 | ⋯ | ⋯ | ⋯ | $1.0 \times 10^{-3}$ | 10.0 | >0.99 |

**Note.** The date and time of each candidate is encoded in the name as GWYYMMDD_hhmmss. The names of candidates not previously reported are given in bold. The detectors that were observing at the time of each transient are denoted by a single letter (e.g., H for LIGO Hanford, L for LIGO Livingston). This does not necessarily indicate that the same detectors contributed triggers for a given candidate. We include results from analyses that observe a candidate with FAR $> 1$ $yr^{-1}$ in italics. A dash (–) indicates that a candidate was not found by an analysis. There is evidence that the candidate labeled with an asterisk (*) is of instrumental origin. FARs have been capped at $1 \times 10^{-5}$ $yr^{-1}$ to maintain a consistent limiting FAR across pipelines.

**Table 2**
Candidate GW Signals from O4a with $p_{astro} \geqslant 0.5$ in at Least One Analysis and an FAR $> 1$ yr$^{-1}$

| Candidate | Inst. | cWB-BBH | | | GstLAL | | | MBTA | | | PyCBC | | |
|---|---|---|---|---|---|---|---|---|---|---|---|---|---|
| | | FAR (yr$^{-1}$) | SNR | $p_{astro}$ | FAR (yr$^{-1}$) | SNR | $p_{astro}$ | FAR (yr$^{-1}$) | SNR | $p_{astro}$ | FAR (yr$^{-1}$) | SNR | $p_{astro}$ |
| **GW230531_141100** | L | ⋯ | ⋯ | ⋯ | ⋯ | ⋯ | ⋯ | ⋯ | ⋯ | ⋯ | *3.5* | *8.0* | *0.69* |
| **GW230603_174756** | HL | ⋯ | ⋯ | ⋯ | ⋯ | ⋯ | ⋯ | *210* | *8.0* | *0.03* | *8.4* | *7.9* | *0.73* |
| **GW230606_024545** | HL | ⋯ | ⋯ | ⋯ | *450* | *7.4* | *<0.01* | ⋯ | ⋯ | ⋯ | *3.2* | *7.5* | *0.88* |
| GW230609_010824 | HL | *46* | *8.0* | *0.16* | *2.8* | *7.9* | *0.60* | *16* | *7.9* | *0.40* | *14* | *7.7* | *0.81* |
| GW230615_160825 | HL | *320* | *8.9* | *0.02* | *4.3* | *8.3* | *0.50* | ⋯ | ⋯ | ⋯ | ⋯ | ⋯ | ⋯ |
| GW230618_102550 | HL | ⋯ | ⋯ | ⋯ | *190* | *7.8* | *0.03* | ⋯ | ⋯ | ⋯ | *49* | *7.6* | *0.59* |
| GW230624_214944 | H | ⋯ | ⋯ | ⋯ | *570* | *10.1* | *<0.01* | ⋯ | ⋯ | ⋯ | *2.2* | *10.6* | *0.69* |
| GW230625_211655 | HL | ⋯ | ⋯ | ⋯ | *42* | *7.9* | *0.11* | *280* | *8.1* | *0.02* | *50* | *7.9* | *0.61* |
| **GW230702_162025** | HL | ⋯ | ⋯ | ⋯ | *640* | *8.5* | *<0.01* | *6.5* | *9.2* | *0.60* | ⋯ | ⋯ | ⋯ |
| GW230708_071859 | HL | *5.3* | *9.1* | *0.54* | *1.5* | *8.1* | *0.73* | *74* | *7.8* | *0.10* | *470* | *7.0* | *0.11* |
| GW230709_063445 | HL | *310* | *7.9* | *0.02* | *40* | *7.3* | *0.11* | *91* | *7.3* | *0.08* | *15* | *7.5* | *0.83* |
| GW230717_102139 | HL | ⋯ | ⋯ | ⋯ | *130* | *7.8* | *0.04* | *67* | *8.2* | *0.12* | *29* | *8.3* | *0.70* |
| GW230721_222634 | HL | *290* | *7.9* | *0.02* | ⋯ | ⋯ | ⋯ | *4.1* | *7.6* | *0.73* | ⋯ | ⋯ | ⋯ |
| GW230723_084820 | HL | *3.9* | *8.4* | *0.65* | ⋯ | ⋯ | ⋯ | ⋯ | ⋯ | ⋯ | *210* | *6.9* | *0.01* |
| GW230728_083628 | HL | ⋯ | ⋯ | ⋯ | *1.5* | *13.1* | *0.73* | ⋯ | ⋯ | ⋯ | ⋯ | ⋯ | ⋯ |
| GW230807_205045 | HL | *30* | *7.9* | *0.20* | *2.9* | *8.1* | *0.59* | ⋯ | ⋯ | ⋯ | *$1.8 \times 10^3$* | *8.7* | *<0.01* |
| GW230817_212349 | HL | ⋯ | ⋯ | ⋯ | *130* | *7.6* | *0.04* | *590* | *7.4* | *<0.01* | *6.2* | *7.8* | *0.86* |
| GW230822_230337 | HL | *95* | *8.3* | *0.07* | *1.5* | *8.2* | *0.72* | *85* | *7.5* | *0.08* | *6.1* | *8.0* | *0.86* |
| GW230823_142524 | HL | ⋯ | ⋯ | ⋯ | ⋯ | ⋯ | ⋯ | *1.6* | *8.8* | *0.86* | *400* | *8.6* | *0.10* |
| GW230824_135331 | HL | *6.2* | *9.7* | *0.58* | ⋯ | ⋯ | ⋯ | ⋯ | ⋯ | ⋯ | ⋯ | ⋯ | ⋯ |
| GW230830_064744 | HL | ⋯ | ⋯ | ⋯ | *3.5* | *8.5* | *0.55* | ⋯ | ⋯ | ⋯ | ⋯ | ⋯ | ⋯ |
| GW230831_134621 | HL | ⋯ | ⋯ | ⋯ | *42* | *8.8* | *0.11* | *20* | *8.8* | *0.34* | *12* | *8.5* | *0.80* |
| **GW230902_122814** | HL | *220* | *8.1* | *0.03* | ⋯ | ⋯ | ⋯ | *2.0* | *8.1* | *0.84* | ⋯ | ⋯ | ⋯ |
| GW230902_172430 | HL | ⋯ | ⋯ | ⋯ | *280* | *8.4* | *0.02* | *98* | *8.4* | *0.07* | *39* | *8.5* | *0.59* |
| GW230902_224555 | HL | ⋯ | ⋯ | ⋯ | *82* | *7.4* | *0.06* | ⋯ | ⋯ | ⋯ | *51* | *7.5* | *0.52* |
| GW230904_152545 | HL | ⋯ | ⋯ | ⋯ | ⋯ | ⋯ | ⋯ | *37* | *9.2* | *0.19* | *18* | *9.0* | *0.74* |
| GW230920_064709 | HL | ⋯ | ⋯ | ⋯ | *490* | *9.5* | *<0.01* | *200* | *9.2* | *0.03* | *4.8* | *9.2* | *0.81* |
| GW230925_143957 | HL | *35* | *8.0* | *0.20* | ⋯ | ⋯ | ⋯ | *8.2* | *7.4* | *0.55* | ⋯ | ⋯ | ⋯ |
| GW231002_143916 | HL | ⋯ | ⋯ | ⋯ | *78* | *8.8* | *0.06* | *1.0* | *9.4* | *0.90* | ⋯ | ⋯ | ⋯ |
| **GW231005_144455** | HL | ⋯ | ⋯ | ⋯ | ⋯ | ⋯ | ⋯ | ⋯ | ⋯ | ⋯ | *60* | *7.3* | *0.55* |
| GW231013_135504 | HL | ⋯ | ⋯ | ⋯ | ⋯ | ⋯ | ⋯ | ⋯ | ⋯ | ⋯ | *64* | *8.4* | *0.54* |
| GW231026_130704 | HL | ⋯ | ⋯ | ⋯ | *1.7* | *8.1* | *0.70* | *340* | *8.3* | *0.02* | *1.8* | *8.1* | *0.93* |
| GW231102_052214 | HL | *75* | *7.7* | *0.11* | ⋯ | ⋯ | ⋯ | *4.0* | *7.9* | *0.72* | ⋯ | ⋯ | ⋯ |
| GW231102_232433 | HL | ⋯ | ⋯ | ⋯ | ⋯ | ⋯ | ⋯ | *2.5* | *7.6* | *0.80* | ⋯ | ⋯ | ⋯ |
| GW231113_150041 | HL | *2.4* | *8.7* | *0.81* | *4.7* | *7.9* | *0.48* | *130* | *8.0* | *0.04* | *1.8* | *7.9* | *0.96* |
| GW231120_022103 | HL | ⋯ | ⋯ | ⋯ | *3.8* | *9.6* | *0.53* | *12* | *10.0* | *0.45* | *5.4* | *9.6* | *0.90* |
| GW231126_010928 | HL | *4.6* | *9.1* | *0.67* | *8.7* | *8.5* | *0.34* | *4.2* | *8.4* | *0.71* | *6.2* | *8.6* | *0.86* |
| GW231204_090648 | HL | ⋯ | ⋯ | ⋯ | *3.6* | *8.4* | *0.54* | ⋯ | ⋯ | ⋯ | ⋯ | ⋯ | ⋯ |
| GW231206_010629 | HL | *17* | *9.7* | *0.46* | *59* | *7.9* | *0.08* | *6.9* | *8.2* | *0.58* | *65* | *7.6* | *0.30* |
| GW231220_173406 | HL | *68* | *7.9* | *0.16* | *35* | *7.6* | *0.13* | ⋯ | ⋯ | ⋯ | *6.2* | *7.4* | *0.77* |
| GW231231_120147 | HL | *7.5* | *11.3* | *0.61* | *61* | *9.0* | *0.08* | *150* | *8.6* | *0.03* | *4.5* | *9.2* | *0.86* |
| **GW240105_151143** | H | ⋯ | ⋯ | ⋯ | ⋯ | ⋯ | ⋯ | ⋯ | ⋯ | ⋯ | *3.3* | *25.9* | *0.70* |

**Note.** These candidates do not meet our criterion for source- property estimation but are likely astrophysical in origin. The names of candidates not previously reported are given in bold. The date and time of each candidate are encoded in the name as GWYYMMDD_hhmmss. The detectors that were observing at the time of each transient are denoted by a single letter (e.g., H for LIGO Hanford). This does not necessarily indicate that the same detectors contributed triggers for a given candidate. We include results from analyses that observe a candidate with $p_{astro} < 0.5$. Italics denote that the candidate was found with FAR $> 1$ yr$^{-1}$. A dash (–) indicates that a candidate was not found by an analysis.

candidates underwent human vetting (LIGO–Virgo–KAGRA Collaboration 2025a), and additional information was disseminated via GCN Circulars. These alerts facilitated rapid searches for potential electromagnetic counterparts to transient GWs.

The online searches run in near-real time and are only able to use data collected up to the current time. Data quality can vary suddenly and noise transients in the detectors can be mistakenly identified as astrophysical in origin. Of the 93 high-significance candidates identified in low latency, 11 were later retracted. Of these 11 retractions, 4 were due to triggers from early-warning search pipelines (A. H. Nitz et al. 2020b; S. Sachdev et al. 2020; M. Kovalam et al. 2022; C. Alléné et al. 2025) that did not have corresponding triggers in the full-bandwidth analyses. The other 7 retracted candidates were all found to be significant by only one analysis. None of the 11 retracted online candidates were recovered with an FAR $< 2$ day$^{-1}$ in the offline analyses.

There were five candidates identified as significant online candidates that were not retracted but are also not recovered with an FAR $< 1$ yr$^{-1}$ in the offline analyses:

1. S230708z, S230807f, and S230822bm were all found in low latency by GstLAL in both LHO and LLO. All of

these candidates were found with an SNR $< 10$. These BBH candidates were all identified offline by multiple pipelines with an FAR $> 1$ yr$^{-1}$. They each have $p_{\rm astro} \geqslant 0.5$ and are reported in Table 2 as GW230708_071859, GW230807_205045, and GW230822_230337.
2. S230802aq was a single-detector BBH candidate in LHO found by GstLAL with SNR $< 10$. No pipelines recover an offline trigger at this time with an FAR $< 2$ day$^{-1}$.
3. S231020bw was identified by GstLAL, while both LHO and LLO were observing, but with disparate SNRs in each detector. No pipelines recover an offline trigger at this time with an FAR $< 2$ day$^{-1}$.

#### 2.1.2. New O4a Candidates

There are eight candidates identified with $p_{\rm astro} \geqslant 0.5$ by at least one offline analysis that were not identified in low latency and not previously shared via GCN or SCiMMA Hopskotch Notices or GCN Circulars. These new candidates are indicated in bold in Tables 1 and 2.

The eight new candidates were all identified by a single pipeline. The majority are of low significance (FAR $> 1$ yr$^{-1}$) and are listed in Table 2. All of the new candidates have moderate network SNRs ($\lesssim$10), with the exception of GW240105_151143, which has an SNR $> 25$. However, there are multiple regions of significant excess power inconsistent with a CBC signal in the time–frequency spectrogram of GW240105_151143, as shown in Appendix C, which may contribute to its identification by only one pipeline despite its large SNR. Of the eight new candidates, six are coincident triggers involving both LHO and LLO. Only GW230531_141100 and GW240105_151143 are observed using data from a single detector.

GW230630_070659 is the only new candidate found with FAR $< 1$ yr$^{-1}$. However, offline follow-up (S. Soni et al. 2025) of the data quality for this event indicated evidence of instrumental origin. Specifically, excess power in strain data in both detectors was found to be inconsistent with a CBC signal. This conclusion was based on both calculating the residual power in the strain data (L. Vazsonyi & D. Davis 2023) and machine learning–based image classification (S. Álvarez-López et al. 2024). A spectrogram for this event is shown in Appendix C. In contrast to GWTC-3.0 (R. Abbott et al. 2023a), here, we do not remove the GW prefix for candidates where there is evidence of terrestrial or instrumental noise origin; we instead assign the prefix to all candidates identified by a search pipeline. We thus report GW230630_070659 with the other candidates with FAR $< 1$ yr$^{-1}$ and $p_{\rm astro} \geqslant 0.5$ (Table 1) according to the search pipelines. We do not estimate source properties for GW230630_070659.

#### 2.1.3. Pipeline Consistency

The search algorithms, methods, and configurations used differ between our analyses (A. G. Abac et al. 2026a), which cause them to have different responses to both noise and astrophysical transients. As a result, we expect candidate lists to differ between pipelines. There is less disagreement between pipelines for high-SNR candidates. Lower-SNR candidates, however, may be identified by only a subset of pipelines. Some candidates are observed in only one detector, which increases the uncertainty in significance estimation and can lead to additional disagreement across pipelines in both the estimated FAR and $p_{\rm astro}$.

Not all of the 129 candidates with $p_{\rm astro} \geqslant 0.5$ were observed with $p_{\rm astro} \geqslant 0.5$ by all pipelines. Of these candidates, 54 were found by cWB-BBH, 89 were found by GstLAL, 76 were found by MBTA, and 103 were found by PyCBC. Only 39 of these candidates were found by all pipelines, while 62 were found by all matched-filter-based pipelines, 68 by three or more pipelines, and 86 by two or more pipelines. Of the 129 candidates for which $p_{\rm astro} \geqslant 0.5$ in any pipeline, several meet this criterion uniquely in only a single pipeline:

1. The cWB-BBH analysis found six unique candidates, four of which were assigned an FAR $< 1$ yr$^{-1}$: GW230712_090405, GW231004_232346, GW231221_135041, and GW231230_170116. In each case, the coherent SNR reported by cWB-BBH is higher than the matched-filter SNR obtained by the other analyses. If these candidates are astrophysical in origin, this disagreement in SNRs could result from physics not captured by the matched-filter templates, such as spin effects or orbital eccentricity (T. Mishra et al. 2025) or the difference in SNR definition (A. G. Abac et al. 2026a). GW230824_135331 has no triggers from any other pipeline. Here, we do not require independent support from a matched-filter based pipeline for a cWB-BBH-only candidate to remain in the candidate list. Further discussion and a spectrogram of GW230824_135331 can be found in Appendix C.
2. The GstLAL analysis found seven unique candidates, all of which are classified by the GstLAL $p_{\rm astro}$ estimate as BBH candidates. GW230807_205045 and GW231029_111508 were both identified with FAR $< 1$ per 30 days in low latency; the former was demoted in significance after offline analysis and can be found in Table 2. GW231029_111508 is a single-detector candidate found by a high-mass template. MBTA and PyCBC do not produce single-detector triggers for candidates with detector-frame chirp masses $(1+z)\mathcal{M} > 7 M_{\odot}$ or durations $\lesssim$0.3 s, respectively (A. G. Abac et al. 2026a).
3. The MBTA analysis found 11 unique candidates. These candidates are all inferred to be from BBHs via the multicomponent $p_{\rm astro}$ measurement. GW230704_212616 and GW231018_233037 were found with a lower significance in low latency by multiple pipelines. These are found offline by MBTA with FAR $< 1$ yr$^{-1}$ and are listed in Table 1. The remaining unique MBTA candidates can be found in Table 2.
4. The PyCBC analysis found 19 unique candidates. GW230904_152545, GW230920_064709, GW231013_135504, and GW240105_151143 have greater support from PyCBC for being NSBH candidates, more than that of GstLAL or MBTA; their multicomponent $p_{\rm astro}$ can be found in Table 7 in Appendix A. The remaining candidates are inferred to be from BBHs. GW231223_202619 was identified in low latency with a lower significance. This candidate is found offline by PyCBC with FAR $< 1$ yr$^{-1}$ and is listed in Table 1. The other unique PyCBC candidates are all listed in Table 2.

#### 2.1.4. Subthreshold Candidates

We have highlighted 129 GW candidates in O4a with $p_{\rm astro} \geqslant 0.5$. Of those, 42 with FAR $> 1$ yr$^{-1}$ are listed in Table 2.

The 87 candidates assigned an FAR $< 1\ \mathrm{yr}^{-1}$, listed in Table 1, underwent additional analyses (excluding GW230630_070659) and are discussed in more detail in Section 3. These candidates are a subset of the O4a candidates in GWTC-4.0, which is comprised of 1382 triggers. The remaining 1253 subthreshold candidates in O4a have an FAR $< 2\ \mathrm{day}^{-1}$ and $p_{\mathrm{astro}} < 0.5$. These candidates are made publicly available at Gravitational Wave Open Science Center (GWOSC; A. G. Abac et al. 2026b) for completeness, but they have not been examined for possibly being instrumental in origin. The purity of this sample is expected to be low: 0.013 when considering all of the subthreshold candidates as estimated in Section 2.2.

### 2.2. Search Sensitivity

Here, we describe the estimated sensitivity of each analysis, calculated by simulating astrophysical signals in the data and running search pipelines to recover them. These simulated signals are referred to as *injections*. We parameterize the sensitivity of the search analyses via the estimated time–volume product or hypervolume $\langle VT \rangle$ (A. G. Abac et al. 2025a). As discussed further in A. G. Abac et al. (2026a), the number of astrophysical signals, $\hat{N}$, that a pipeline is expected to detect can be estimated as

$$\hat{N} = \langle \mathrm{VT} \rangle \mathcal{R}, \tag{1}$$

where $\mathcal{R}$ is the volumetric rate of mergers per unit (source-frame) time. To estimate $\langle VT \rangle$ for each pipeline, we simulate a distribution of signals that approximates the detected population of BBHs, NSBHs, and BNSs (R. Essick et al. 2025). To estimate the overall catalog sensitivity, each pipeline analyzes the same set of simulated signals: the injections are added into the collected data, and we record how many are recovered significantly.

As in R. Abbott et al. (2023a), several combinations of masses are used to assess our sensitivity to BBH, NSBH, and BNS systems. We thus estimate $\langle VT \rangle$ at a set of fiducial points in the component mass space:

1. BHs at $35 M_{\odot}$. These correspond to GW150914-like systems (B. P. Abbott et al. 2016b, 2024). This is also approximately where we see a peak in the BH mass spectrum (A. G. Abac et al. 2026d).
2. BHs at 100 M⊙ , 60 M⊙ , 20 M⊙ , 10 M⊙ , and 5 $M_{\odot}$. These lie within the range of previously detected BH masses (R. Abbott et al. 2023a).
3. NSs at 1.5 $M_{\odot}$, which is consistent with the distribution of known NS masses (J. Antoniadis et al. 2016; F. Özel & P. Freire 2016; J. Alsing et al. 2018; N. Farrow et al. 2019; P. Landry & J. S. Read 2021; R. Abbott et al. 2023b).

For each given point, injections are weighted so that they follow a log-normal distribution about the central mass with a width of 0.1 (R. Essick et al. 2025). Figure 1 shows the resulting variation in the O4a $\langle VT \rangle$ with a detection threshold of FAR $< 1\ \mathrm{yr}^{-1}$ across the component mass parameter space for each search.

The results at the top of Figure 1, labeled as "Any," represent our overall sensitivity to CBCs in the specified region. They come from taking the minimum FAR for an injection from all of the analyses. The sensitivity is greatest for the "Any" pipeline for $100 M_{\odot} + 100 M_{\odot}$ binaries, though different pipelines are more or less sensitive to different regions of the binary parameter space. Only sensitivity estimates that have an uncertainty smaller than 50% are shown in Figure 1.[330]

We also estimate the number of astrophysical signals among the set of 1253 subthreshold candidates. If the source population distribution assumed for $p_{\mathrm{astro}}$ calculations is close to the true astrophysical distribution, then for each individual pipeline, the sum of $p_{\mathrm{astro}}$ values provides the expected number of true signals. The expected signal count for a given pipeline is also proportional to that pipeline's $\langle VT \rangle$ for the true signal population; thus, if we knew the true $\langle VT \rangle$, we could scale the sum of $p_{\mathrm{astro}}$ values accordingly to obtain an estimate of the signal count for the "Any" pipeline (R. Abbott et al. 2023a). Due to the larger differences in $\langle VT \rangle$ across pipelines in GWTC-4.0 relative to GWTC-3.0, we compute the sum of $p_{\mathrm{astro}}$ of subthreshold candidates and take the average over the number of pipelines as a conservative lower bound on the expected signal count. From this bound, we estimate $\gtrsim 16$ signals among the subthreshold candidates.

## 3. Source Properties

After identifying GW candidates, we coherently analyze the data from the network of GW detectors to infer the properties of the source of each signal (A. G. Abac et al. 2026a). These inferences are in turn used in companion papers to understand the population of compact objects (A. G. Abac et al. 2026d), measure cosmic expansion history (A. G. Abac et al. 2026e), and as a baseline for analyses that extend beyond our standard assumptions, such as for tests of general relativity carried out on new candidates from O4a (A. G. Abac et al. 2026f, 2026g, 2026h). These downstream analyses set stricter thresholds than used in Section 2 in order to mitigate the impact of differing methods of computing $p_{\mathrm{astro}}$ across search pipelines and attain a higher-purity subset of candidates. We restrict our estimation of the source properties to those that have both $p_{\mathrm{astro}} \geqslant 0.5$ and FARs $< 1\ \mathrm{yr}^{-1}$. Additionally, we exclude from consideration GW230630_070659, which we find likely to be of instrumental origin (Section 2.1.2 and Appendix C).

We use Bayesian parameter estimation in order to infer the posterior probability distributions over the source parameters given a segment of data around each candidate. The posteriors are derived assuming models of GW emission and under the assumption of stationary, Gaussian noise that is uncorrelated across detectors (e.g., J. Veitch et al. 2015; B. P. Abbott et al. 2016b, 2020b; E. Thrane & C. Talbot 2019; N. Christensen & R. Meyer 2022; A. G. Abac et al. 2026a). When we identify the presence of transient, non-Gaussian noise around the time of a candidate, we exclude the affected frequency band or model and coherently remove the noise transient using the BayesWave algorithm (N. J. Cornish & T. B. Littenberg 2015; T. B. Littenberg & N. J. Cornish 2015; N. J. Cornish et al. 2021; S. Hourihane et al. 2022; T. Gupta & N. J. Cornish 2024). We discuss these cases further in Appendix B. Our default priors are chosen to be agnostic and sufficiently wide to cover the region of the parameter space where the posteriors have support (A. G. Abac et al. 2026a). Specifically, they are uniform in (redshifted) component masses, uniform in spin magnitudes, isotropic in spin orientations, isotropic in binary orientation, uniform in merger time and coalescence phase, isotropic in sky location, and our distance prior corresponds to a uniform merger rate in comoving volume and time. Our inferences are given in terms of samples from the posteriors, from which we derive point estimates and uncertainties for the binary parameters (in the form of median values and 90% credible intervals) after marginalizing over the remaining parameters.

[330] Earlier versions of this work (v1 and v2), made available on the arXiv at https://arxiv.org/abs/2508.18082v1 and https://arxiv.org/abs/2508.18082v2, contained a version of this plot produced using an incorrect calculation, which resulted in the inflation of sensitive hypervolumes.

For all candidates, we assume quasi-circular orbits, and we carry out inference using multiple waveform models. Fuller details of these models and their development are given in A. G. Abac et al. (2026a). For each BBH candidate, we use the IMRPhenomXPHM_SpinTaylor (G. Pratten et al. 2021; M. Colleoni et al. 2025) and SEOBNRv5PHM (L. Pompili et al. 2023; A. Ramos-Buades et al. 2023) waveform models, each of which incorporates the effects of higher-order multipolar emission and spin precession. Depending on the inferred intrinsic parameters of these candidates, we use one to three additional waveform models. Many BBH candidates lie within the parameter-space coverage of the surrogate model NRSur7dq4 (V. Varma et al. 2019), which is built directly on numerical simulations of BBH coalescences, and in these cases, we use NRSur7dq4 for parameter estimation. For candidates with asymmetric masses or evidence of precession, we also use IMRPhenomXO4a (E. Hamilton et al. 2021; J. E. Thompson et al. 2024) because this waveform model includes a more complete physical description of precession effects (A. G. Abac et al. 2026a). For our BBH candidates, we report our posteriors by combining equal numbers of samples from each waveform model in order to mitigate uncertainties associated with our theoretical models (B. P. Abbott et al. 2016b). For the NSBH candidate GW230518_125908, we use IMRPhenomXPHM_SpinTaylor to produce our fiducial posteriors, while for GW230529_181500 (A. G. Abac et al. 2024b) we combine equal numbers of samples produced using IMRPhenomXPHM_SpinTaylor and SEOBNRv5PHM. Although these models incorporate key physical effects, they do not include the imprint of matter. Therefore, for both NSBH candidates, we use the IMRPhenomPv2_NRTidalv2 (T. Dietrich et al. 2019), IMRPhenomNSBH (J. E. Thompson et al. 2020), and SEOBNRv4_ROM_NRTidalv2_NSBH (A. Matas et al. 2020) models to constrain the tidal deformability of the NSs. The three tidal models do not include the effect of higher multipolar emission, and the latter two neglect the effects of spin precession on the waveform.

Our key results for O4a candidates with FAR $< 1\ \mathrm{yr}^{-1}$ are summarized in Table 3 and shown in Figures 2–4. Our default agnostic priors (A. G. Abac et al. 2026a) do not make strong assumptions about the nature of the underlying astrophysical population. In addition to the default prior, we also reweight the inferred posterior distribution of our BBH candidates using a population-informed prior (default model, Table 1 of A. G. Abac et al. 2026d), and show these population-informed measurements in Figure 2.

Highlights from the new candidates in GWTC-4.0 include the following:

1. GW230518_125908 is one of two high-significance candidates added to GWTC-4.0 whose sources have masses consistent with an NSBH binary. This candidate was previously reported as a probable online NSBH candidate, although no unambiguous multimessenger counterpart was identified in follow-up electromagnetic searches.[331] Its source properties are reported here for the first time. In addition to having more asymmetric masses than GW230529_181500, with mass ratio $q = 0.18^{+0.04}_{-0.03}$ and larger total mass $M = 9.61^{+0.76}_{-0.79}\ M_\odot$ , the spin of its primary is constrained to be relatively small $\chi_1 < 0.16$ (90% credible level), and we infer $\chi_{\rm eff} = -0.01^{+0.09}_{-0.11}$.
2. GW230529_181500 (A. G. Abac et al. 2024b) is the second high-significance candidate whose sources have masses consistent with an NSBH binary. Its component masses mean that the primary probably lies in the putative lower mass gap between 3 and 5 $M_\odot$ (C. D. Bailyn et al. 1998; F. Özel et al. 2010; W. M. Farr et al. 2011; L. Kreidberg et al. 2012). Relative to our previous results, we have corrected the implementation of calibration marginalization for this event (A. G. Abac et al. 2026a), corrected the normalization of the noise-weighted inner product used in the likelihood (A. G. Abac et al. 2026a; C. Talbot et al. 2025), and reweighted our priors to the reference cosmology used throughout GWTC-4.0. Our inferences of this candidate's source properties remain nearly identical to those previously presented (A. G. Abac et al. 2024b).
3. While GW230814_230901 (A. G. Abac et al. 2026c) is the highest-SNR candidate, with an SNR 42.1, its source properties are relatively common, with $m_1 = 33.6^{+2.8}_{-2.2}\ M_\odot$, $m_2 = 28.3^{+2.1}_{-3.0}\ M_\odot$ , and spins consistent with zero, with effective inspiral spin $\chi_{\rm eff} = -0.01^{+0.06}_{-0.08}$. Due to its high SNR, it is a promising signal for tests of relativity (A. G. Abac et al. 2026c).
4. The source of GW231028_153006 is a massive BBH, with total mass $M = 152^{+29}_{-14}\ M_\odot$ . It has a large effective inspiral spin $\chi_{\rm eff} = 0.44^{+0.16}_{-0.20}$ and unequal masses, with mass ratio $q = m_2/m_1 = 0.63^{+0.33}_{-0.35}$. Similar to GW231123_135430 (A. G. Abac et al. 2025b), this event displays significant systematic modeling uncertainties and multimodal posteriors.
5. GW231118_005626 is another candidate whose source has unequal masses and signatures of large spins. The binary has total mass $M = 30.9^{+5.3}_{-3.6}\ M_\odot$, $q = 0.55^{+0.37}_{-0.22}$, and large spins, with $\chi_{\rm eff} = 0.37^{+0.10}_{-0.11}$ and $\chi_1 = 0.65^{+0.28}_{-0.38}$.
6. The source of GW231123_135430 (A. G. Abac et al. 2025b) is, with high probability, the most massive BBH in our catalog of those with FAR $< 1\ \mathrm{yr}^{-1}$. We infer it to have total mass $M = 236^{+29}_{-48}\ M_\odot$ and large spins. It also displays significant systematic differences between inferences made with different waveform models.

In what follows, we give further detail about our inferences of masses, spins, matter effects, and locations of the sources of our high-purity candidates from O4a, highlighting additional sources that lie at the extremes of the parameter space (Sections 3.1–3.4). Some of our candidates display multiple modes in their inferred source parameters. This multimodality complicates the interpretation of sources where it is present (R. Abbott et al. 2023a), and it is not usually possible to isolate

[331] As reported in GCN Circulars (https://gcn.nasa.gov/circulars?query=S230518h) related to this event; see also G. S. H. Paek et al. (2025), M. Pillas et al. (2025), and the references therein.

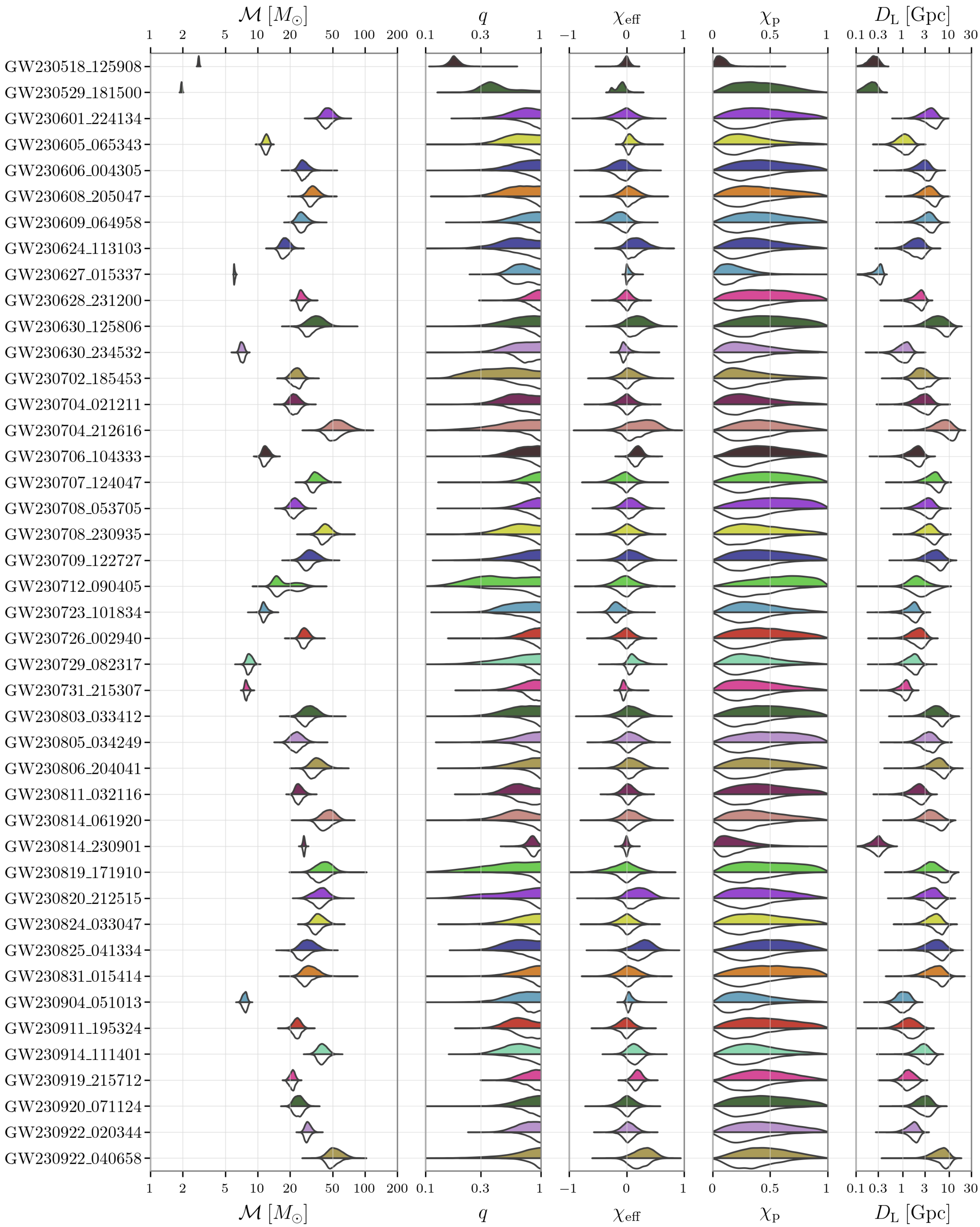

**Figure 2.** The marginal probability distributions for the source frame chirp mass $\mathcal{M}$, mass ratio $q$, effective inspiral spin $\chi_{\rm eff}$, effective precession spin $\chi_{\rm p}$, and luminosity distance $D_L$ for O4a candidates with FAR $< 1$ yr$^{-1}$. The colored upper halves of the plots shows the marginal posterior distributions using our default agnostic priors (A. G. Abac et al. 2026a), while the white lower halves show these marginal distributions after reweighting according to the inferred population model (A. G. Abac et al. 2026d) for each BBH. The two NSBH candidates have only the non-reweighted posterior distributions. The vertical thickness of each region is proportional to the marginal posterior probability at that value for each candidate.

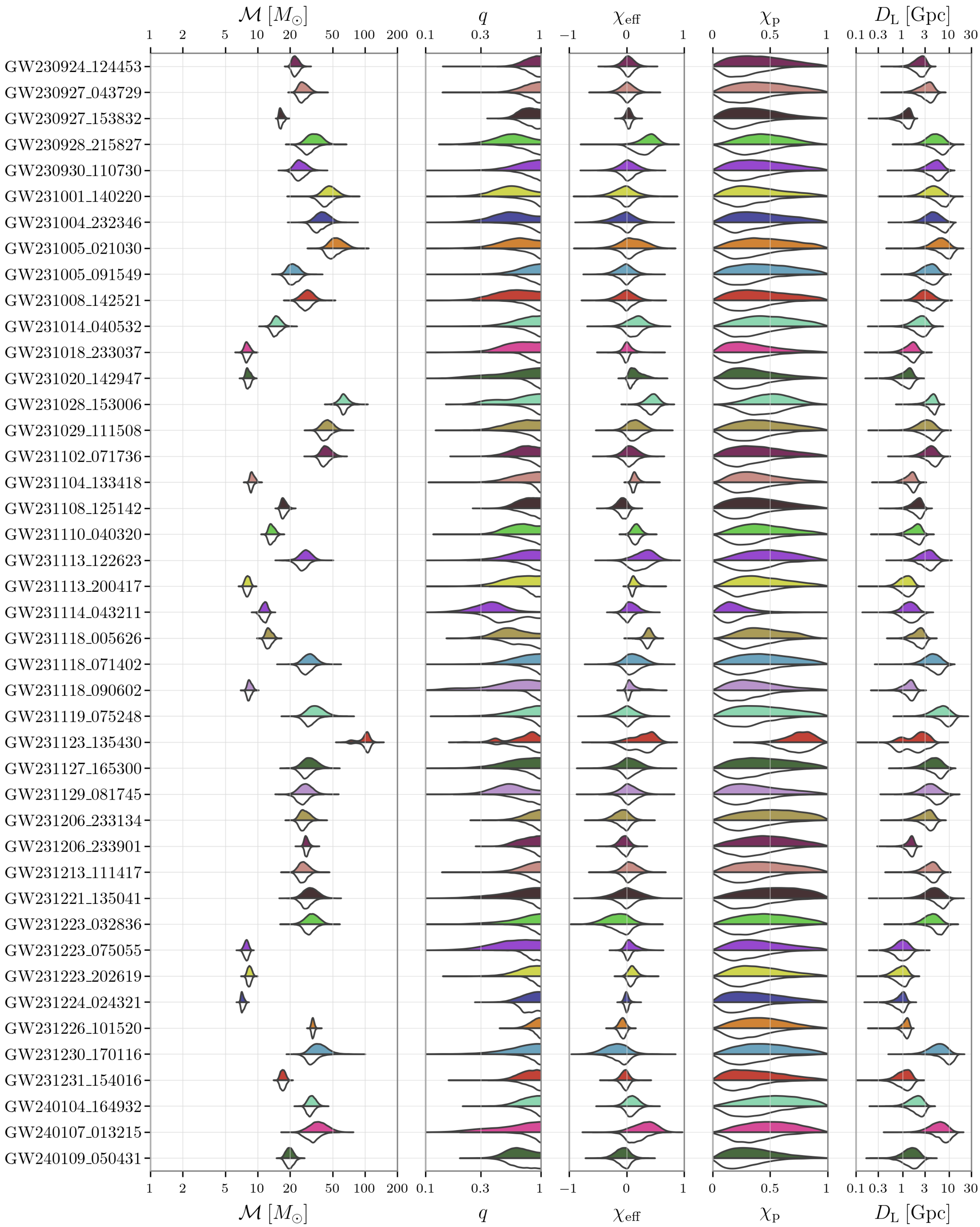

**Figure 2.** (Continued.)

**Table 3**
The Inferred Properties of GW Event Candidates from O4a with FAR $< 1$ yr$^{-1}$ and $p_{\rm astro} > 0.5$

| Candidate | $M$ ($M_\odot$) | $\mathcal{M}$ ($M_\odot$) | $m_1$ ($M_\odot$) | $m_2$ ($M_\odot$) | $\chi_{\rm eff}$ | $D_L$ (Gpc) | $z$ | $M_{\rm f}$ ($M_\odot$) | $\chi_{\rm f}$ | $\Delta\Omega$ (deg$^2$) | SNR |
|---|---|---|---|---|---|---|---|---|---|---|---|
| GW230518_125908 | $9.61^{+0.76}_{-0.79}$ | $2.80^{+0.06}_{-0.06}$ | $8.17^{+0.84}_{-0.92}$ | $1.45^{+0.13}_{-0.10}$ | $-0.01^{+0.09}_{-0.11}$ | $0.24^{+0.11}_{-0.10}$ | $0.05^{+0.02}_{-0.02}$ | $9.46^{+0.76}_{-0.80}$ | $0.38^{+0.03}_{-0.03}$ | 490 | $14.2^{+0.2}_{-0.4}$ |
| GW230529_181500 | $5.08^{+0.61}_{-0.60}$ | $1.94^{+0.04}_{-0.04}$ | $3.66^{+0.82}_{-1.21}$ | $1.42^{+0.60}_{-0.22}$ | $-0.10^{+0.12}_{-0.18}$ | $0.20^{+0.10}_{-0.10}$ | $0.04^{+0.02}_{-0.02}$ | $4.92^{+0.62}_{-0.63}$ | $0.58^{+0.08}_{-0.06}$ | 24,000 | $11.6^{+0.3}_{-0.4}$ |
| GW230601_224134 | $107^{+22}_{-15}$ | $45.0^{+10.0}_{-7.3}$ | $64^{+17}_{-13}$ | $44^{+14}_{-15}$ | $-0.03^{+0.27}_{-0.32}$ | $3.7^{+2.1}_{-1.8}$ | $0.60^{+0.28}_{-0.26}$ | $102^{+21}_{-14}$ | $0.67^{+0.12}_{-0.13}$ | 3300 | $12.3^{+0.2}_{-0.3}$ |
| GW230605_065343 | $28.6^{+4.0}_{-2.8}$ | $11.94^{+0.96}_{-0.94}$ | $17.2^{+6.5}_{-3.5}$ | $11.1^{+2.5}_{-2.7}$ | $0.06^{+0.16}_{-0.10}$ | $1.06^{+0.61}_{-0.48}$ | $0.21^{+0.10}_{-0.09}$ | $27.3^{+4.1}_{-2.7}$ | $0.69^{+0.05}_{-0.05}$ | 1000 | $10.5^{+0.3}_{-0.4}$ |
| GW230606_004305 | $63.4^{+13.4}_{-8.3}$ | $26.5^{+5.7}_{-3.6}$ | $37.6^{+13.4}_{-7.6}$ | $25.8^{+8.1}_{-8.3}$ | $-0.11^{+0.26}_{-0.31}$ | $2.7^{+1.5}_{-1.4}$ | $0.48^{+0.20}_{-0.22}$ | $60.7^{+12.8}_{-7.9}$ | $0.64^{+0.11}_{-0.14}$ | 1400 | $10.3^{+0.3}_{-0.4}$ |
| GW230608_205047 | $79^{+16}_{-11}$ | $32.8^{+7.3}_{-5.4}$ | $48^{+13}_{-11}$ | $31^{+11}_{-11}$ | $0.04^{+0.25}_{-0.26}$ | $3.5^{+2.2}_{-1.7}$ | $0.58^{+0.29}_{-0.26}$ | $75^{+15}_{-11}$ | $0.69^{+0.11}_{-0.14}$ | 2200 | $9.8^{+0.3}_{-0.5}$ |
| GW230609_064958 | $60.3^{+12.9}_{-8.0}$ | $25.5^{+5.8}_{-3.6}$ | $35.3^{+10.7}_{-6.9}$ | $25.2^{+7.5}_{-7.5}$ | $-0.14^{+0.22}_{-0.27}$ | $3.4^{+1.9}_{-1.7}$ | $0.56^{+0.25}_{-0.25}$ | $57.8^{+12.2}_{-7.7}$ | $0.64^{+0.09}_{-0.13}$ | 1700 | $9.8^{+0.3}_{-0.5}$ |
| GW230624_113103 | $43.8^{+11.1}_{-6.6}$ | $18.0^{+3.2}_{-2.4}$ | $27.2^{+13.4}_{-6.9}$ | $16.1^{+4.8}_{-4.5}$ | $0.17^{+0.29}_{-0.25}$ | $1.94^{+1.27}_{-0.97}$ | $0.35^{+0.19}_{-0.16}$ | $41.8^{+10.7}_{-6.3}$ | $0.72^{+0.12}_{-0.11}$ | 1300 | $9.7^{+0.4}_{-0.5}$ |
| GW230627_015337 | $14.19^{+0.77}_{-0.45}$ | $6.02^{+0.16}_{-0.07}$ | $8.4^{+1.7}_{-1.3}$ | $5.79^{+0.95}_{-0.92}$ | $0.02^{+0.08}_{-0.03}$ | $0.31^{+0.06}_{-0.13}$ | $0.07^{+0.01}_{-0.03}$ | $13.54^{+0.82}_{-0.45}$ | $0.68^{+0.02}_{-0.03}$ | 110 | $28.5^{+0.1}_{-0.1}$ |
| GW230628_231200 | $59.3^{+8.6}_{-4.9}$ | $25.5^{+3.8}_{-2.2}$ | $32.5^{+5.9}_{-3.9}$ | $27.1^{+4.9}_{-5.2}$ | $-0.01^{+0.16}_{-0.16}$ | $2.25^{+0.80}_{-1.07}$ | $0.40^{+0.12}_{-0.17}$ | $56.5^{+8.1}_{-4.6}$ | $0.69^{+0.08}_{-0.06}$ | 660 | $15.5^{+0.2}_{-0.3}$ |
| GW230630_125806 | $84^{+26}_{-19}$ | $35.0^{+11.4}_{-8.5}$ | $51^{+20}_{-14}$ | $33^{+15}_{-13}$ | $0.17^{+0.30}_{-0.31}$ | $5.4^{+5.1}_{-3.0}$ | $0.83^{+0.59}_{-0.41}$ | $80^{+25}_{-18}$ | $0.75^{+0.11}_{-0.16}$ | 4300 | $8.1^{+0.4}_{-0.5}$ |
| GW230630_234532 | $16.8^{+2.1}_{-1.4}$ | $7.06^{+0.57}_{-0.48}$ | $10.0^{+3.5}_{-1.8}$ | $6.6^{+1.5}_{-1.6}$ | $-0.04^{+0.16}_{-0.08}$ | $1.08^{+0.52}_{-0.48}$ | $0.21^{+0.09}_{-0.09}$ | $16.1^{+2.1}_{-1.4}$ | $0.66^{+0.04}_{-0.05}$ | 1500 | $9.4^{+0.3}_{-0.5}$ |
| GW230702_185453 | $60^{+15}_{-11}$ | $22.9^{+4.0}_{-3.5}$ | $40^{+20}_{-13}$ | $18.0^{+8.5}_{-6.5}$ | $0.05^{+0.29}_{-0.26}$ | $2.4^{+1.9}_{-1.1}$ | $0.43^{+0.27}_{-0.17}$ | $57^{+15}_{-11}$ | $0.64^{+0.13}_{-0.15}$ | 2500 | $9.5^{+0.3}_{-0.5}$ |
| GW230704_021211 | $52.7^{+10.8}_{-8.2}$ | $21.8^{+4.4}_{-3.2}$ | $32.4^{+11.9}_{-7.9}$ | $19.9^{+6.5}_{-6.0}$ | $-0.01^{+0.21}_{-0.24}$ | $2.7^{+1.8}_{-1.5}$ | $0.48^{+0.25}_{-0.23}$ | $50.5^{+10.6}_{-7.9}$ | $0.66^{+0.10}_{-0.13}$ | 1500 | $9.0^{+0.3}_{-0.5}$ |
| GW230704_212616 | $139^{+49}_{-31}$ | $55^{+23}_{-14}$ | $89^{+42}_{-27}$ | $49^{+30}_{-25}$ | $0.28^{+0.34}_{-0.43}$ | $7.2^{+6.1}_{-4.2}$ | $1.05^{+0.68}_{-0.54}$ | $132^{+47}_{-30}$ | $0.77^{+0.12}_{-0.19}$ | 7400 | $8.0^{+0.5}_{-0.6}$ |
| GW230706_104333 | $27.9^{+4.4}_{-3.0}$ | $11.8^{+1.7}_{-1.1}$ | $16.3^{+5.4}_{-3.0}$ | $11.5^{+2.6}_{-2.7}$ | $0.19^{+0.13}_{-0.14}$ | $1.93^{+0.92}_{-0.98}$ | $0.35^{+0.14}_{-0.16}$ | $26.5^{+4.3}_{-2.8}$ | $0.75^{+0.07}_{-0.06}$ | 1500 | $9.0^{+0.3}_{-0.5}$ |
| GW230707_124047 | $82^{+19}_{-12}$ | $35.1^{+8.6}_{-5.3}$ | $46.1^{+12.2}_{-8.2}$ | $36.4^{+10.6}_{-9.8}$ | $-0.05^{+0.25}_{-0.29}$ | $4.5^{+2.3}_{-2.3}$ | $0.71^{+0.29}_{-0.32}$ | $78^{+18}_{-11}$ | $0.68^{+0.09}_{-0.10}$ | 3200 | $10.6^{+0.2}_{-0.4}$ |
| GW230708_053705 | $51.9^{+10.2}_{-7.6}$ | $22.2^{+4.4}_{-3.3}$ | $29.1^{+8.2}_{-5.4}$ | $22.8^{+5.5}_{-5.4}$ | $0.07^{+0.23}_{-0.24}$ | $3.3^{+2.0}_{-1.7}$ | $0.55^{+0.27}_{-0.25}$ | $49.4^{+9.7}_{-7.2}$ | $0.72^{+0.09}_{-0.09}$ | 1900 | $8.3^{+0.4}_{-0.6}$ |
| GW230708_230935 | $103^{+21}_{-16}$ | $42.5^{+9.7}_{-7.8}$ | $64^{+20}_{-15}$ | $39^{+14}_{-15}$ | $0.01^{+0.27}_{-0.30}$ | $3.5^{+2.3}_{-1.6}$ | $0.58^{+0.29}_{-0.24}$ | $99^{+20}_{-15}$ | $0.67^{+0.12}_{-0.16}$ | 2600 | $9.2^{+0.3}_{-0.5}$ |
| GW230709_122727 | $74^{+20}_{-14}$ | $31.0^{+9.4}_{-6.6}$ | $45^{+16}_{-11}$ | $30^{+13}_{-13}$ | $0.08^{+0.30}_{-0.31}$ | $4.6^{+3.5}_{-2.4}$ | $0.73^{+0.43}_{-0.34}$ | $71^{+19}_{-13}$ | $0.71^{+0.12}_{-0.15}$ | 3600 | $8.5^{+0.3}_{-0.5}$ |
| GW230712_090405 | $46^{+21}_{-11}$ | $16.5^{+11.1}_{-3.2}$ | $32^{+17}_{-10}$ | $13.1^{+14.6}_{-5.9}$ | $-0.03^{+0.35}_{-0.31}$ | $2.0^{+2.6}_{-1.0}$ | $0.37^{+0.36}_{-0.17}$ | $45^{+20}_{-11}$ | $0.66^{+0.17}_{-0.22}$ | 1400 | $8.4^{+0.7}_{-0.9}$ |
| GW230723_101834 | $27.5^{+4.3}_{-2.6}$ | $11.40^{+1.64}_{-0.92}$ | $16.7^{+5.9}_{-3.4}$ | $10.6^{+2.8}_{-2.8}$ | $-0.18^{+0.20}_{-0.17}$ | $1.61^{+0.69}_{-0.89}$ | $0.30^{+0.11}_{-0.16}$ | $26.4^{+4.2}_{-2.6}$ | $0.61^{+0.09}_{-0.08}$ | 1100 | $9.7^{+0.3}_{-0.5}$ |
| GW230726_002940 | $63.6^{+10.3}_{-7.9}$ | $27.3^{+4.4}_{-3.4}$ | $35.6^{+9.0}_{-5.9}$ | $27.9^{+6.0}_{-6.0}$ | $-0.02^{+0.21}_{-0.23}$ | $2.0^{+1.2}_{-1.0}$ | $0.37^{+0.18}_{-0.17}$ | $60.6^{+9.7}_{-7.4}$ | $0.69^{+0.08}_{-0.08}$ | 28,000 | $10.2^{+0.2}_{-0.4}$ |
| GW230729_082317 | $20.3^{+5.0}_{-2.4}$ | $8.33^{+0.94}_{-0.75}$ | $12.3^{+7.6}_{-2.7}$ | $7.6^{+2.1}_{-2.6}$ | $0.12^{+0.22}_{-0.13}$ | $1.63^{+0.82}_{-0.76}$ | $0.31^{+0.13}_{-0.13}$ | $19.4^{+5.1}_{-2.4}$ | $0.71^{+0.06}_{-0.06}$ | 2200 | $8.2^{+0.3}_{-0.6}$ |
| GW230731_215307 | $18.3^{+1.6}_{-1.0}$ | $7.77^{+0.58}_{-0.36}$ | $10.4^{+2.9}_{-1.4}$ | $7.8^{+1.2}_{-1.6}$ | $-0.05^{+0.11}_{-0.06}$ | $1.11^{+0.35}_{-0.46}$ | $0.22^{+0.06}_{-0.08}$ | $17.4^{+1.6}_{-1.0}$ | $0.67^{+0.04}_{-0.03}$ | 710 | $11.9^{+0.2}_{-0.3}$ |
| GW230803_033412 | $74^{+22}_{-16}$ | $30.5^{+9.3}_{-6.7}$ | $45^{+19}_{-12}$ | $29^{+12}_{-10}$ | $0.06^{+0.29}_{-0.31}$ | $4.9^{+4.0}_{-2.6}$ | $0.77^{+0.47}_{-0.35}$ | $70^{+21}_{-15}$ | $0.71^{+0.12}_{-0.14}$ | 4100 | $7.7^{+0.4}_{-0.6}$ |
| GW230805_034249 | $54.9^{+13.8}_{-9.7}$ | $23.1^{+5.9}_{-4.2}$ | $32.1^{+12.5}_{-7.3}$ | $22.6^{+7.8}_{-7.5}$ | $0.06^{+0.27}_{-0.28}$ | $3.5^{+2.6}_{-1.7}$ | $0.58^{+0.33}_{-0.25}$ | $52.4^{+13.0}_{-9.2}$ | $0.71^{+0.11}_{-0.12}$ | 2400 | $9.0^{+0.3}_{-0.5}$ |
| GW230806_204041 | $85^{+24}_{-15}$ | $35.8^{+10.7}_{-6.9}$ | $51^{+18}_{-12}$ | $35^{+14}_{-13}$ | $0.08^{+0.28}_{-0.27}$ | $5.5^{+3.7}_{-2.9}$ | $0.84^{+0.44}_{-0.39}$ | $81^{+23}_{-15}$ | $0.71^{+0.11}_{-0.13}$ | 4600 | $8.5^{+0.3}_{-0.5}$ |
| GW230811_032116 | $57.9^{+8.7}_{-6.6}$ | $24.0^{+4.1}_{-2.4}$ | $35.6^{+8.7}_{-7.5}$ | $22.1^{+6.7}_{-5.1}$ | $0.02^{+0.18}_{-0.18}$ | $2.1^{+1.2}_{-1.1}$ | $0.38^{+0.17}_{-0.18}$ | $55.3^{+8.3}_{-6.4}$ | $0.69^{+0.09}_{-0.09}$ | 940 | $12.8^{+0.3}_{-0.4}$ |
| GW230814_061920 | $110^{+26}_{-20}$ | $45.5^{+11.9}_{-9.9}$ | $69^{+19}_{-17}$ | $42^{+17}_{-16}$ | $0.05^{+0.29}_{-0.28}$ | $4.0^{+3.4}_{-2.0}$ | $0.65^{+0.42}_{-0.28}$ | $105^{+24}_{-19}$ | $0.69^{+0.12}_{-0.14}$ | 5200 | $9.4^{+0.3}_{-0.5}$ |
| GW230814_230901 | $61.8^{+2.0}_{-2.1}$ | $26.75^{+0.89}_{-1.00}$ | $33.6^{+2.8}_{-2.2}$ | $28.3^{+2.1}_{-3.0}$ | $-0.01^{+0.06}_{-0.08}$ | $0.28^{+0.17}_{-0.13}$ | $0.06^{+0.03}_{-0.03}$ | $58.9^{+1.9}_{-1.9}$ | $0.68^{+0.02}_{-0.03}$ | 26,000 | $42.1^{+0.1}_{-0.1}$ |
| GW230819_171910 | $106^{+41}_{-22}$ | $42^{+14}_{-11}$ | $70^{+47}_{-21}$ | $35^{+20}_{-19}$ | $-0.04^{+0.36}_{-0.43}$ | $4.0^{+3.9}_{-2.2}$ | $0.65^{+0.49}_{-0.31}$ | $102^{+41}_{-22}$ | $0.64^{+0.19}_{-0.25}$ | 4800 | $8.9^{+0.4}_{-0.5}$ |
| GW230820_212515 | $96^{+22}_{-16}$ | $38.6^{+11.2}_{-9.8}$ | $62^{+23}_{-15}$ | $34^{+18}_{-18}$ | $0.21^{+0.31}_{-0.29}$ | $4.0^{+3.0}_{-2.1}$ | $0.65^{+0.38}_{-0.30}$ | $92^{+21}_{-16}$ | $0.74^{+0.12}_{-0.25}$ | 2100 | $8.4^{+0.4}_{-0.5}$ |
| GW230824_033047 | $88^{+19}_{-14}$ | $37.0^{+9.1}_{-6.3}$ | $53^{+16}_{-11}$ | $36^{+12}_{-13}$ | $0.00^{+0.23}_{-0.27}$ | $4.7^{+2.9}_{-2.3}$ | $0.74^{+0.36}_{-0.32}$ | $84^{+19}_{-13}$ | $0.68^{+0.10}_{-0.13}$ | 3700 | $10.0^{+0.2}_{-0.4}$ |
| GW230825_041334 | $71^{+22}_{-15}$ | $29.4^{+9.4}_{-6.3}$ | $43^{+17}_{-12}$ | $27.3^{+11.9}_{-8.9}$ | $0.28^{+0.25}_{-0.29}$ | $4.9^{+4.2}_{-2.9}$ | $0.77^{+0.50}_{-0.40}$ | $67^{+20}_{-15}$ | $0.79^{+0.08}_{-0.15}$ | 3400 | $8.1^{+0.5}_{-0.6}$ |
| GW230831_015414 | $73^{+26}_{-15}$ | $30.7^{+10.6}_{-6.5}$ | $42^{+18}_{-10}$ | $30^{+12}_{-10}$ | $0.03^{+0.30}_{-0.30}$ | $4.9^{+4.2}_{-2.9}$ | $0.77^{+0.50}_{-0.40}$ | $69^{+24}_{-14}$ | $0.71^{+0.12}_{-0.12}$ | 4000 | $8.1^{+0.3}_{-0.7}$ |
| GW230904_051013 | $17.9^{+2.4}_{-1.6}$ | $7.54^{+0.53}_{-0.60}$ | $10.6^{+4.1}_{-1.9}$ | $7.1^{+1.5}_{-1.9}$ | $0.05^{+0.15}_{-0.07}$ | $1.01^{+0.60}_{-0.42}$ | $0.20^{+0.10}_{-0.08}$ | $17.1^{+2.5}_{-1.6}$ | $0.69^{+0.04}_{-0.04}$ | 1800 | $10.2^{+0.3}_{-0.5}$ |
| GW230911_195324 | $55.3^{+7.8}_{-7.0}$ | $23.1^{+3.4}_{-3.0}$ | $33.6^{+8.1}_{-7.3}$ | $21.7^{+5.6}_{-5.7}$ | $-0.03^{+0.19}_{-0.23}$ | $1.37^{+1.16}_{-0.72}$ | $0.26^{+0.18}_{-0.13}$ | $53.0^{+7.4}_{-6.7}$ | $0.67^{+0.09}_{-0.11}$ | 27,000 | $10.6^{+0.3}_{-0.4}$ |
| GW230914_111401 | $95^{+15}_{-10}$ | $39.6^{+7.5}_{-6.2}$ | $59^{+12}_{-11}$ | $36^{+13}_{-12}$ | $0.12^{+0.19}_{-0.20}$ | $2.7^{+1.6}_{-1.2}$ | $0.47^{+0.23}_{-0.19}$ | $90.9^{+14.0}_{-9.7}$ | $0.71^{+0.09}_{-0.14}$ | 1900 | $16.2^{+0.2}_{-0.3}$ |
| GW230919_215712 | $49.0^{+4.2}_{-4.5}$ | $21.0^{+1.8}_{-1.9}$ | $27.3^{+5.5}_{-3.7}$ | $21.4^{+3.5}_{-4.3}$ | $0.18^{+0.12}_{-0.13}$ | $1.33^{+0.79}_{-0.50}$ | $0.26^{+0.13}_{-0.09}$ | $46.5^{+3.9}_{-4.3}$ | $0.75^{+0.06}_{-0.05}$ | 730 | $15.7^{+0.2}_{-0.3}$ |
| GW230920_071124 | $56.4^{+10.2}_{-8.0}$ | $23.9^{+4.6}_{-3.4}$ | $32.4^{+9.1}_{-6.2}$ | $23.8^{+6.6}_{-6.9}$ | $0.00^{+0.22}_{-0.23}$ | $2.9^{+1.8}_{-1.4}$ | $0.50^{+0.24}_{-0.21}$ | $53.9^{+9.6}_{-7.6}$ | $0.69^{+0.10}_{-0.10}$ | 2100 | $10.1^{+0.3}_{-0.4}$ |
| GW230922_020344 | $68.6^{+8.9}_{-6.5}$ | $29.2^{+3.6}_{-2.6}$ | $39.3^{+10.0}_{-6.3}$ | $29.2^{+5.6}_{-6.2}$ | $0.03^{+0.20}_{-0.21}$ | $1.64^{+0.73}_{-0.73}$ | $0.31^{+0.11}_{-0.12}$ | $65.4^{+8.4}_{-6.1}$ | $0.70^{+0.08}_{-0.08}$ | 330 | $11.8^{+0.3}_{-0.4}$ |
| GW230922_040658 | $125^{+37}_{-21}$ | $52^{+17}_{-11}$ | $76^{+29}_{-18}$ | $51^{+23}_{-24}$ | $0.31^{+0.25}_{-0.32}$ | $6.4^{+4.1}_{-3.5}$ | $0.96^{+0.47}_{-0.45}$ | $119^{+34}_{-21}$ | $0.79^{+0.08}_{-0.14}$ | 5100 | $11.4^{+0.2}_{-0.4}$ |
| GW230924_124453 | $51.8^{+7.0}_{-4.9}$ | $22.3^{+3.1}_{-2.1}$ | $28.8^{+5.9}_{-4.0}$ | $23.1^{+4.4}_{-4.4}$ | $0.02^{+0.18}_{-0.18}$ | $2.4^{+1.0}_{-1.0}$ | $0.42^{+0.15}_{-0.16}$ | $49.3^{+6.6}_{-4.6}$ | $0.70^{+0.07}_{-0.06}$ | 1300 | $12.9^{+0.2}_{-0.3}$ |
| GW230927_043729 | $61.8^{+12.6}_{-8.0}$ | $26.5^{+5.5}_{-3.5}$ | $34.9^{+8.7}_{-6.1}$ | $27.1^{+7.2}_{-6.5}$ | $0.01^{+0.20}_{-0.22}$ | $3.3^{+1.7}_{-1.7}$ | $0.55^{+0.23}_{-0.26}$ | $58.9^{+11.9}_{-7.6}$ | $0.69^{+0.08}_{-0.08}$ | 1500 | $10.5^{+0.2}_{-0.4}$ |
| GW230927_153832 | $38.3^{+3.3}_{-2.2}$ | $16.36^{+1.42}_{-0.82}$ | $21.9^{+3.7}_{-2.9}$ | $16.5^{+2.6}_{-2.5}$ | $0.03^{+0.08}_{-0.08}$ | $1.18^{+0.38}_{-0.53}$ | $0.23^{+0.06}_{-0.10}$ | $36.5^{+3.1}_{-2.1}$ | $0.69^{+0.04}_{-0.03}$ | 330 | $19.7^{+0.2}_{-0.2}$ |
| GW230928_215827 | $83^{+22}_{-18}$ | $33.6^{+9.9}_{-7.5}$ | $54^{+19}_{-15}$ | $29^{+14}_{-11}$ | $0.40^{+0.20}_{-0.28}$ | $5.0^{+3.9}_{-2.5}$ | $0.78^{+0.47}_{-0.34}$ | $79^{+21}_{-17}$ | $0.83^{+0.07}_{-0.15}$ | 3700 | $8.9^{+0.4}_{-0.6}$ |
| GW230930_110730 | $59^{+15}_{-10}$ | $24.9^{+6.3}_{-4.2}$ | $34.4^{+12.9}_{-7.8}$ | $24.4^{+8.3}_{-7.4}$ | $0.03^{+0.26}_{-0.27}$ | $4.9^{+3.2}_{-2.5}$ | $0.77^{+0.38}_{-0.34}$ | $56.3^{+14.2}_{-9.7}$ | $0.69^{+0.10}_{-0.12}$ | 3300 | $8.0^{+0.3}_{-0.5}$ |
| GW231001_140220 | $115^{+32}_{-22}$ | $46.4^{+14.3}_{-9.9}$ | $75^{+26}_{-22}$ | $40^{+18}_{-16}$ | $-0.04^{+0.32}_{-0.35}$ | $4.4^{+3.8}_{-2.4}$ | $0.71^{+0.47}_{-0.34}$ | $111^{+31}_{-21}$ | $0.64^{+0.15}_{-0.20}$ | 4300 | $9.6^{+0.3}_{-0.5}$ |

**Table 3**
(Continued)

| Candidate | $M$ ($M_\odot$) | $\mathcal{M}$ ($M_\odot$) | $m_1$ ($M_\odot$) | $m_2$ ($M_\odot$) | $\chi_{\rm eff}$ | $D_L$ (Gpc) | $z$ | $M_{\rm f}$ ($M_\odot$) | $\chi_{\rm f}$ | $\Delta\Omega$ (deg$^2$) | SNR |
|---|---|---|---|---|---|---|---|---|---|---|---|
| GW231004_232346 | $100^{+28}_{-19}$ | $40.1^{+11.7}_{-8.5}$ | $65^{+25}_{-19}$ | $35^{+16}_{-14}$ | $-0.05^{+0.30}_{-0.37}$ | $4.3^{+3.6}_{-2.2}$ | $0.69^{+0.44}_{-0.31}$ | $96^{+26}_{-19}$ | $0.64^{+0.14}_{-0.19}$ | 3500 | $8.2^{+0.3}_{-0.6}$ |
| GW231005_021030 | $132^{+37}_{-25}$ | $54^{+17}_{-12}$ | $83^{+31}_{-23}$ | $49^{+22}_{-20}$ | $0.10^{+0.35}_{-0.37}$ | $6.4^{+4.7}_{-3.3}$ | $0.95^{+0.54}_{-0.43}$ | $126^{+35}_{-24}$ | $0.72^{+0.14}_{-0.17}$ | 5600 | $9.4^{+0.3}_{-0.4}$ |
| GW231005_091549 | $50.2^{+11.3}_{-8.3}$ | $21.3^{+4.8}_{-3.5}$ | $28.8^{+10.1}_{-6.1}$ | $21.2^{+6.4}_{-6.1}$ | $-0.03^{+0.23}_{-0.26}$ | $3.8^{+2.6}_{-1.9}$ | $0.62^{+0.33}_{-0.27}$ | $48.0^{+10.8}_{-7.9}$ | $0.68^{+0.10}_{-0.11}$ | 2900 | $8.0^{+0.3}_{-0.6}$ |
| GW231008_142521 | $71^{+16}_{-13}$ | $28.9^{+6.8}_{-5.6}$ | $45^{+17}_{-12}$ | $25.5^{+10.7}_{-9.7}$ | $-0.01^{+0.26}_{-0.28}$ | $3.0^{+2.5}_{-1.3}$ | $0.51^{+0.33}_{-0.20}$ | $69^{+15}_{-13}$ | $0.66^{+0.13}_{-0.19}$ | 3000 | $8.9^{+0.4}_{-0.5}$ |
| GW231014_040532 | $35.5^{+7.0}_{-4.6}$ | $15.0^{+2.7}_{-1.9}$ | $20.6^{+7.8}_{-4.0}$ | $14.7^{+3.9}_{-4.3}$ | $0.17^{+0.26}_{-0.28}$ | $2.3^{+1.4}_{-1.2}$ | $0.42^{+0.20}_{-0.20}$ | $33.8^{+6.6}_{-4.3}$ | $0.75^{+0.10}_{-0.11}$ | 1800 | $8.7^{+0.4}_{-0.5}$ |
| GW231018_233037 | $19.1^{+3.3}_{-2.0}$ | $7.90^{+0.88}_{-0.68}$ | $11.6^{+4.9}_{-2.5}$ | $7.2^{+1.9}_{-2.0}$ | $0.01^{+0.17}_{-0.13}$ | $1.53^{+0.76}_{-0.73}$ | $0.29^{+0.12}_{-0.13}$ | $18.2^{+3.3}_{-2.0}$ | $0.67^{+0.05}_{-0.07}$ | 1600 | $8.2^{+0.3}_{-0.6}$ |
| GW231020_142947 | $19.8^{+6.0}_{-2.1}$ | $8.06^{+0.81}_{-0.53}$ | $12.1^{+9.1}_{-2.7}$ | $7.3^{+2.1}_{-2.9}$ | $0.15^{+0.25}_{-0.12}$ | $1.23^{+0.52}_{-0.63}$ | $0.24^{+0.09}_{-0.11}$ | $18.9^{+6.3}_{-2.0}$ | $0.72^{+0.06}_{-0.04}$ | 1500 | $10.5^{+0.3}_{-0.4}$ |
| GW231028_153006 | $152^{+29}_{-14}$ | $63^{+13}_{-10}$ | $95^{+33}_{-20}$ | $58^{+21}_{-25}$ | $0.44^{+0.16}_{-0.20}$ | $4.1^{+1.4}_{-1.9}$ | $0.67^{+0.18}_{-0.27}$ | $144^{+27}_{-13}$ | $0.84^{+0.05}_{-0.10}$ | 1300 | $21.0^{+0.2}_{-0.2}$ |
| GW231029_111508 | $106^{+22}_{-16}$ | $44.2^{+10.5}_{-8.0}$ | $65^{+17}_{-14}$ | $42^{+15}_{-15}$ | $0.14^{+0.24}_{-0.24}$ | $3.1^{+2.4}_{-1.7}$ | $0.53^{+0.32}_{-0.25}$ | $101^{+21}_{-15}$ | $0.72^{+0.10}_{-0.15}$ | 29,000 | $11.2^{+0.2}_{-0.3}$ |
| GW231102_071736 | $102^{+21}_{-13}$ | $43.4^{+9.3}_{-6.3}$ | $61^{+14}_{-12}$ | $43^{+13}_{-13}$ | $0.06^{+0.23}_{-0.22}$ | $3.8^{+2.1}_{-1.8}$ | $0.63^{+0.26}_{-0.26}$ | $98^{+19}_{-12}$ | $0.70^{+0.09}_{-0.10}$ | 3000 | $13.3^{+0.2}_{-0.3}$ |
| GW231104_133418 | $21.0^{+2.8}_{-1.7}$ | $8.84^{+0.87}_{-0.56}$ | $12.3^{+4.5}_{-2.1}$ | $8.6^{+1.8}_{-2.2}$ | $0.14^{+0.12}_{-0.07}$ | $1.47^{+0.54}_{-0.66}$ | $0.28^{+0.09}_{-0.11}$ | $20.0^{+2.8}_{-1.7}$ | $0.72^{+0.05}_{-0.04}$ | 1100 | $11.0^{+0.2}_{-0.4}$ |
| GW231108_125142 | $40.6^{+5.1}_{-3.3}$ | $17.3^{+2.1}_{-1.3}$ | $23.2^{+5.5}_{-3.6}$ | $17.4^{+3.2}_{-3.1}$ | $-0.08^{+0.13}_{-0.15}$ | $2.06^{+0.73}_{-0.91}$ | $0.37^{+0.11}_{-0.15}$ | $38.8^{+4.9}_{-3.2}$ | $0.66^{+0.06}_{-0.05}$ | 1000 | $12.4^{+0.2}_{-0.3}$ |
| GW231110_040320 | $32.2^{+4.3}_{-3.4}$ | $13.4^{+1.7}_{-1.2}$ | $19.5^{+5.8}_{-4.1}$ | $12.5^{+3.1}_{-2.9}$ | $0.17^{+0.13}_{-0.12}$ | $1.89^{+0.87}_{-0.91}$ | $0.35^{+0.13}_{-0.15}$ | $30.6^{+4.2}_{-3.3}$ | $0.74^{+0.06}_{-0.05}$ | 800 | $11.0^{+0.3}_{-0.4}$ |
| GW231113_122623 | $67^{+16}_{-13}$ | $27.9^{+6.3}_{-5.5}$ | $39.7^{+17.3}_{-9.6}$ | $26.4^{+9.0}_{-9.4}$ | $0.33^{+0.25}_{-0.33}$ | $3.4^{+2.4}_{-1.7}$ | $0.57^{+0.31}_{-0.25}$ | $63^{+15}_{-12}$ | $0.81^{+0.08}_{-0.16}$ | 2800 | $7.8^{+0.4}_{-0.7}$ |
| GW231113_200417 | $19.2^{+3.1}_{-1.9}$ | $8.01^{+0.68}_{-0.64}$ | $11.5^{+5.1}_{-2.3}$ | $7.4^{+1.8}_{-2.0}$ | $0.13^{+0.14}_{-0.08}$ | $1.16^{+0.64}_{-0.53}$ | $0.23^{+0.11}_{-0.10}$ | $18.3^{+3.2}_{-1.9}$ | $0.72^{+0.06}_{-0.04}$ | 1600 | $10.1^{+0.3}_{-0.5}$ |
| GW231114_043211 | $31.0^{+7.6}_{-4.8}$ | $11.6^{+1.1}_{-1.1}$ | $22.7^{+9.6}_{-6.4}$ | $8.2^{+2.6}_{-2.2}$ | $0.08^{+0.22}_{-0.16}$ | $1.38^{+0.87}_{-0.61}$ | $0.26^{+0.14}_{-0.11}$ | $30.0^{+7.8}_{-4.8}$ | $0.61^{+0.06}_{-0.06}$ | 1700 | $9.8^{+0.3}_{-0.5}$ |
| GW231118_005626 | $30.9^{+5.3}_{-3.6}$ | $12.6^{+1.6}_{-1.1}$ | $19.8^{+6.6}_{-4.9}$ | $10.9^{+3.3}_{-2.4}$ | $0.37^{+0.10}_{-0.11}$ | $2.19^{+0.93}_{-0.97}$ | $0.39^{+0.14}_{-0.16}$ | $29.3^{+5.3}_{-3.6}$ | $0.80^{+0.06}_{-0.05}$ | 1100 | $10.5^{+0.3}_{-0.5}$ |
| GW231118_071402 | $72^{+18}_{-14}$ | $30.4^{+7.9}_{-6.1}$ | $43^{+15}_{-10}$ | $29^{+11}_{-10}$ | $0.11^{+0.29}_{-0.28}$ | $4.3^{+3.5}_{-2.2}$ | $0.69^{+0.43}_{-0.32}$ | $69^{+17}_{-13}$ | $0.72^{+0.11}_{-0.14}$ | 3600 | $8.5^{+0.3}_{-0.5}$ |
| GW231118_090602 | $20.7^{+10.2}_{-2.3}$ | $8.37^{+0.76}_{-0.56}$ | $13.1^{+13.7}_{-3.3}$ | $7.3^{+2.1}_{-3.3}$ | $0.08^{+0.36}_{-0.09}$ | $1.35^{+0.54}_{-0.60}$ | $0.26^{+0.09}_{-0.11}$ | $19.7^{+10.6}_{-2.3}$ | $0.70^{+0.08}_{-0.04}$ | 1100 | $10.9^{+0.4}_{-0.4}$ |
| GW231119_075248 | $82^{+29}_{-18}$ | $34.4^{+12.4}_{-7.8}$ | $49^{+23}_{-13}$ | $33^{+15}_{-13}$ | $0.00^{+0.28}_{-0.30}$ | $6.7^{+5.5}_{-3.7}$ | $0.99^{+0.62}_{-0.48}$ | $79^{+27}_{-18}$ | $0.68^{+0.11}_{-0.16}$ | 5900 | $7.7^{+0.3}_{-0.5}$ |
| GW231123_135430 | $236^{+29}_{-48}$ | $101^{+13}_{-30}$ | $137^{+23}_{-18}$ | $101^{+22}_{-51}$ | $0.31^{+0.25}_{-0.41}$ | $2.2^{+2.0}_{-1.5}$ | $0.39^{+0.28}_{-0.25}$ | $222^{+27}_{-42}$ | $0.84^{+0.07}_{-0.19}$ | 1700 | $20.7^{+0.2}_{-0.3}$ |
| GW231127_165300 | $74^{+22}_{-15}$ | $30.5^{+9.5}_{-6.7}$ | $45^{+18}_{-12}$ | $29^{+12}_{-12}$ | $0.05^{+0.30}_{-0.32}$ | $4.5^{+3.6}_{-2.5}$ | $0.71^{+0.44}_{-0.35}$ | $71^{+21}_{-14}$ | $0.69^{+0.12}_{-0.17}$ | 4400 | $8.3^{+0.3}_{-0.5}$ |
| GW231129_081745 | $69^{+18}_{-14}$ | $27.7^{+7.7}_{-5.7}$ | $45^{+15}_{-13}$ | $23.8^{+10.0}_{-8.2}$ | $0.02^{+0.27}_{-0.26}$ | $3.8^{+3.6}_{-2.1}$ | $0.63^{+0.45}_{-0.31}$ | $66^{+17}_{-14}$ | $0.67^{+0.13}_{-0.16}$ | 3700 | $7.5^{+0.4}_{-0.7}$ |
| GW231206_233134 | $63.5^{+13.4}_{-8.8}$ | $27.2^{+5.9}_{-3.9}$ | $35.6^{+9.1}_{-6.3}$ | $28.1^{+7.5}_{-7.1}$ | $-0.09^{+0.22}_{-0.24}$ | $3.2^{+1.8}_{-1.8}$ | $0.54^{+0.25}_{-0.27}$ | $60.6^{+12.7}_{-8.4}$ | $0.67^{+0.09}_{-0.10}$ | 2400 | $11.0^{+0.3}_{-0.4}$ |
| GW231206_233901 | $66.0^{+5.3}_{-3.7}$ | $28.1^{+2.7}_{-1.7}$ | $37.6^{+6.6}_{-4.6}$ | $28.4^{+5.4}_{-6.0}$ | $-0.05^{+0.14}_{-0.15}$ | $1.49^{+0.34}_{-0.49}$ | $0.28^{+0.05}_{-0.08}$ | $63.1^{+5.0}_{-3.4}$ | $0.67^{+0.06}_{-0.07}$ | 350 | $21.0^{+0.1}_{-0.2}$ |
| GW231213_111417 | $62.5^{+14.3}_{-9.3}$ | $26.7^{+6.3}_{-4.2}$ | $35.5^{+10.3}_{-6.7}$ | $27.2^{+8.1}_{-7.5}$ | $0.06^{+0.24}_{-0.23}$ | $4.0^{+2.3}_{-2.0}$ | $0.65^{+0.30}_{-0.29}$ | $59.5^{+13.4}_{-8.8}$ | $0.71^{+0.09}_{-0.09}$ | 2600 | $9.7^{+0.2}_{-0.4}$ |
| GW231221_135041 | $76^{+21}_{-15}$ | $30.7^{+9.5}_{-6.8}$ | $47^{+21}_{-13}$ | $28^{+13}_{-13}$ | $0.01^{+0.36}_{-0.38}$ | $4.6^{+3.7}_{-2.5}$ | $0.73^{+0.45}_{-0.36}$ | $72^{+20}_{-15}$ | $0.69^{+0.15}_{-0.17}$ | 3700 | $7.8^{+0.4}_{-0.6}$ |
| GW231223_032836 | $76^{+18}_{-13}$ | $31.8^{+8.6}_{-7.0}$ | $46^{+16}_{-11}$ | $31^{+12}_{-14}$ | $-0.15^{+0.33}_{-0.37}$ | $4.2^{+3.1}_{-2.1}$ | $0.67^{+0.39}_{-0.30}$ | $73^{+17}_{-13}$ | $0.63^{+0.13}_{-0.17}$ | 4200 | $8.8^{+0.3}_{-0.5}$ |
| GW231223_075055 | $19.0^{+4.6}_{-2.0}$ | $7.79^{+0.58}_{-0.59}$ | $11.9^{+6.8}_{-2.8}$ | $6.8^{+2.0}_{-2.1}$ | $0.06^{+0.22}_{-0.12}$ | $0.97^{+0.58}_{-0.43}$ | $0.19^{+0.10}_{-0.08}$ | $18.1^{+4.8}_{-1.9}$ | $0.69^{+0.06}_{-0.05}$ | 1500 | $8.8^{+0.4}_{-0.6}$ |
| GW231223_202619 | $19.6^{+2.2}_{-1.5}$ | $8.36^{+0.66}_{-0.57}$ | $11.1^{+3.8}_{-1.6}$ | $8.3^{+1.4}_{-1.9}$ | $0.10^{+0.13}_{-0.11}$ | $0.89^{+0.49}_{-0.44}$ | $0.18^{+0.08}_{-0.08}$ | $18.7^{+2.2}_{-1.5}$ | $0.71^{+0.05}_{-0.05}$ | 27,000 | $9.8^{+0.2}_{-0.4}$ |
| GW231224_024321 | $16.66^{+1.29}_{-0.78}$ | $7.13^{+0.48}_{-0.29}$ | $9.3^{+2.2}_{-1.1}$ | $7.3^{+1.0}_{-1.3}$ | $-0.01^{+0.08}_{-0.06}$ | $0.95^{+0.29}_{-0.40}$ | $0.19^{+0.05}_{-0.07}$ | $15.89^{+1.26}_{-0.76}$ | $0.68^{+0.04}_{-0.03}$ | 390 | $12.9^{+0.2}_{-0.3}$ |
| GW231226_101520 | $74.8^{+4.1}_{-3.0}$ | $32.4^{+1.8}_{-1.4}$ | $40.1^{+4.4}_{-2.9}$ | $35.0^{+3.2}_{-4.9}$ | $-0.09^{+0.09}_{-0.10}$ | $1.18^{+0.24}_{-0.35}$ | $0.23^{+0.04}_{-0.06}$ | $71.4^{+3.8}_{-2.8}$ | $0.67^{+0.04}_{-0.04}$ | 190 | $33.7^{+0.1}_{-0.1}$ |
| GW231230_170116 | $89^{+49}_{-20}$ | $36.9^{+14.3}_{-8.4}$ | $54^{+50}_{-15}$ | $35^{+15}_{-14}$ | $-0.19^{+0.34}_{-0.35}$ | $5.7^{+4.8}_{-3.2}$ | $0.87^{+0.56}_{-0.43}$ | $85^{+47}_{-19}$ | $0.62^{+0.13}_{-0.20}$ | 6700 | $7.4^{+0.4}_{-0.7}$ |
| GW231231_154016 | $39.9^{+4.1}_{-3.4}$ | $17.1^{+1.6}_{-1.4}$ | $22.5^{+5.7}_{-3.3}$ | $17.2^{+2.9}_{-3.2}$ | $-0.03^{+0.12}_{-0.12}$ | $1.07^{+0.60}_{-0.53}$ | $0.21^{+0.10}_{-0.10}$ | $38.1^{+4.0}_{-3.2}$ | $0.67^{+0.06}_{-0.05}$ | 27,000 | $13.1^{+0.2}_{-0.3}$ |
| GW240104_164932 | $74.3^{+11.4}_{-8.5}$ | $31.8^{+5.1}_{-3.9}$ | $42.3^{+9.4}_{-6.7}$ | $32.1^{+7.5}_{-8.0}$ | $0.09^{+0.19}_{-0.19}$ | $1.91^{+1.09}_{-0.95}$ | $0.35^{+0.16}_{-0.16}$ | $70.6^{+10.7}_{-7.9}$ | $0.73^{+0.08}_{-0.08}$ | 28,000 | $14.1^{+0.2}_{-0.3}$ |
| GW240107_013215 | $91^{+29}_{-20}$ | $36.2^{+14.0}_{-9.5}$ | $59^{+27}_{-18}$ | $32^{+20}_{-16}$ | $0.30^{+0.30}_{-0.36}$ | $5.8^{+4.8}_{-3.2}$ | $0.87^{+0.56}_{-0.43}$ | $87^{+28}_{-19}$ | $0.79^{+0.10}_{-0.19}$ | 4800 | $8.5^{+0.4}_{-0.5}$ |
| GW240109_050431 | $47.1^{+6.7}_{-5.7}$ | $19.6^{+2.7}_{-2.2}$ | $28.8^{+7.5}_{-6.2}$ | $18.1^{+4.8}_{-4.1}$ | $-0.07^{+0.20}_{-0.23}$ | $1.51^{+1.01}_{-0.77}$ | $0.29^{+0.16}_{-0.13}$ | $45.1^{+6.5}_{-5.5}$ | $0.64^{+0.09}_{-0.09}$ | 27,000 | $10.4^{+0.2}_{-0.4}$ |

**Note.** For one-dimensional distributions, we provide the median and 90% symmetric credible intervals, while for the localization area $\Delta\Omega$ we provide the 90% credible area. The columns show the source-frame total mass $M$, source-frame chirp mass $\mathcal{M}$, source-frame component masses $m_i$, effective inspiral spin $\chi_{\rm eff}$, luminosity distance $D_L$, redshift $z$, source-frame remnant mass $M_{\rm f}$, remnant spin $\chi_{\rm f}$, localization area $\Delta\Omega$, and SNR.

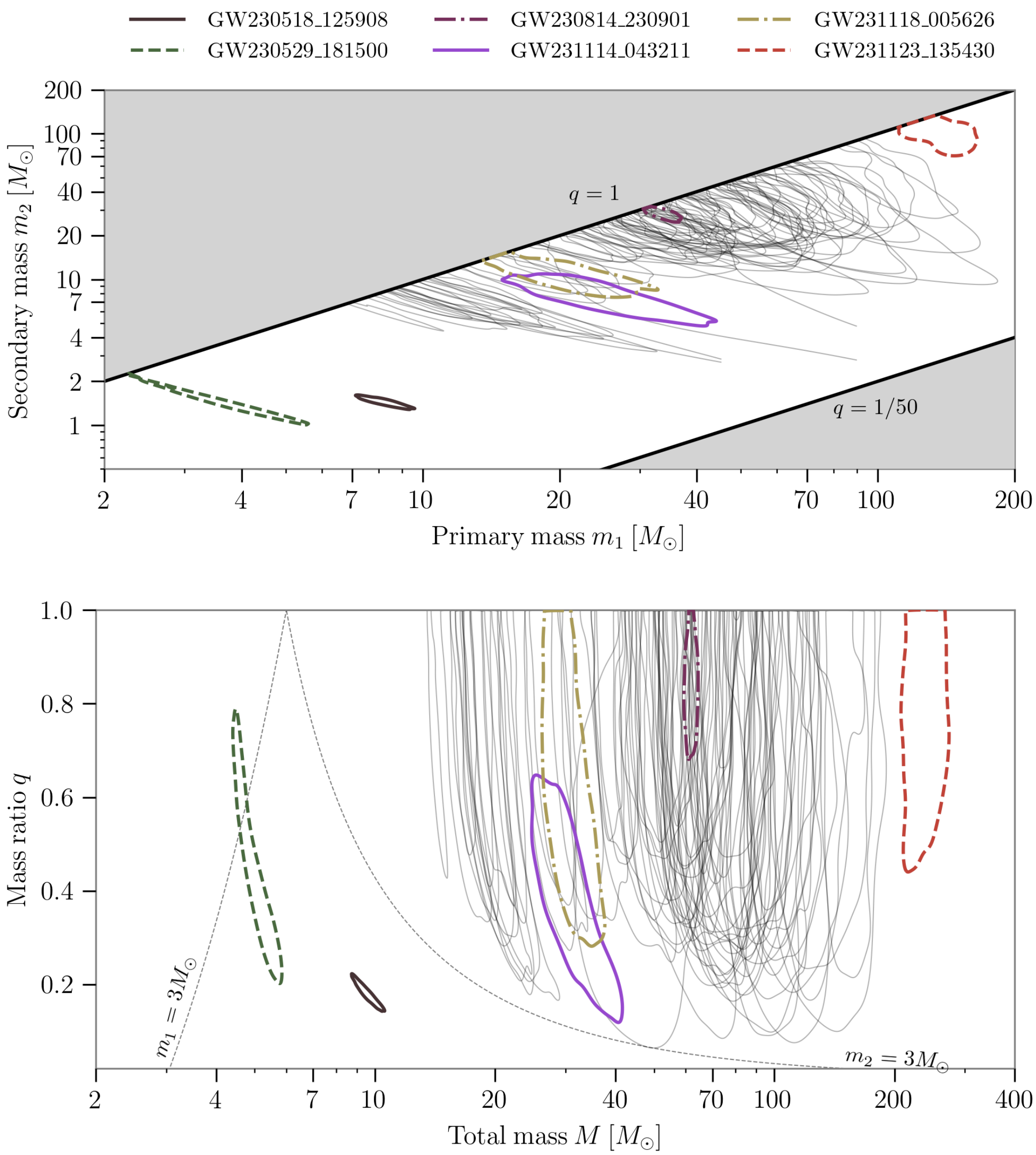

**Figure 3.** Credible-region contours for O4a candidates with FAR $< 1$ yr$^{-1}$. Top: credible-region contours for the inferred primary and secondary component masses $m_1$ and $m_2$. The upper shaded region denotes the area excluded by the convention $m_1 \geqslant m_2$. The lower shaded region denotes the most extreme mass ratio prior used by parameter-estimation analyses. Bottom: credible-region contours for the inferred total mass $M$ and mass ratio $q$. The dotted lines separate regions where the primary and secondary component masses are below 3 $M_\odot$. Each contour indicates the 90% credible region for a given candidate. We use colors to highlight candidates: GW230518_125908, which is an NSBH candidate; GW230529_181500, which is also an NSBH candidate with an inferred $\chi_{\rm eff} < 0$ at 92% credibility; GW231114_043211, which has the largest support for the most unequal inferred mass ratio of the unambiguous BBHs; GW231118_005626, which has inferred $\chi_{\rm eff} > 0$; GW230814_230901, which is the highest-SNR candidate observed in O4a; and GW231123_135430, which we infer to have the most massive source observed in O4a.

the probable reason for multiple modes. In Section 3.5, we discuss candidates that display multiple modes.

Systematic uncertainties in our modeling affect our parameter inference for some candidates. In Section 3.6, we discuss these cases in greater detail and present an analysis of the consistency between model-based and minimally modeled waveform reconstructions for a number of candidates.

### *3.1. Masses*

The masses of a compact binary source of GWs are often the most well-constrained parameters as they are the primary determinant of the phase evolution of the signal (A. G. Abac et al. 2025a, 2026a). The component masses are of particular interest since they indicate whether the compact objects are likely to be BHs or NSs; however, combinations of the two masses such as the chirp mass $\mathcal{M}$ or total mass $M$ are often more precisely measured than the individual masses (e.g., A. G. Abac et al. 2025a). For example, the chirp mass is the dominant parameter controlling the rate of binary inspiral, and so it is measured well in lower-mass systems where many cycles of inspiral can be observed (P. Kafka 1988; L. S. Finn & D. F. Chernoff 1993; C. Cutler et al. 1993; C. Cutler & É. E. Flanagan 1994). Meanwhile, the mass ratio $q \equiv m_2/m_1 \leqslant 1$ is generally less well measured (C. Cutler et al. 1993; C. Cutler & É. E. Flanagan 1994; E. Poisson & C. M. Will 1995).

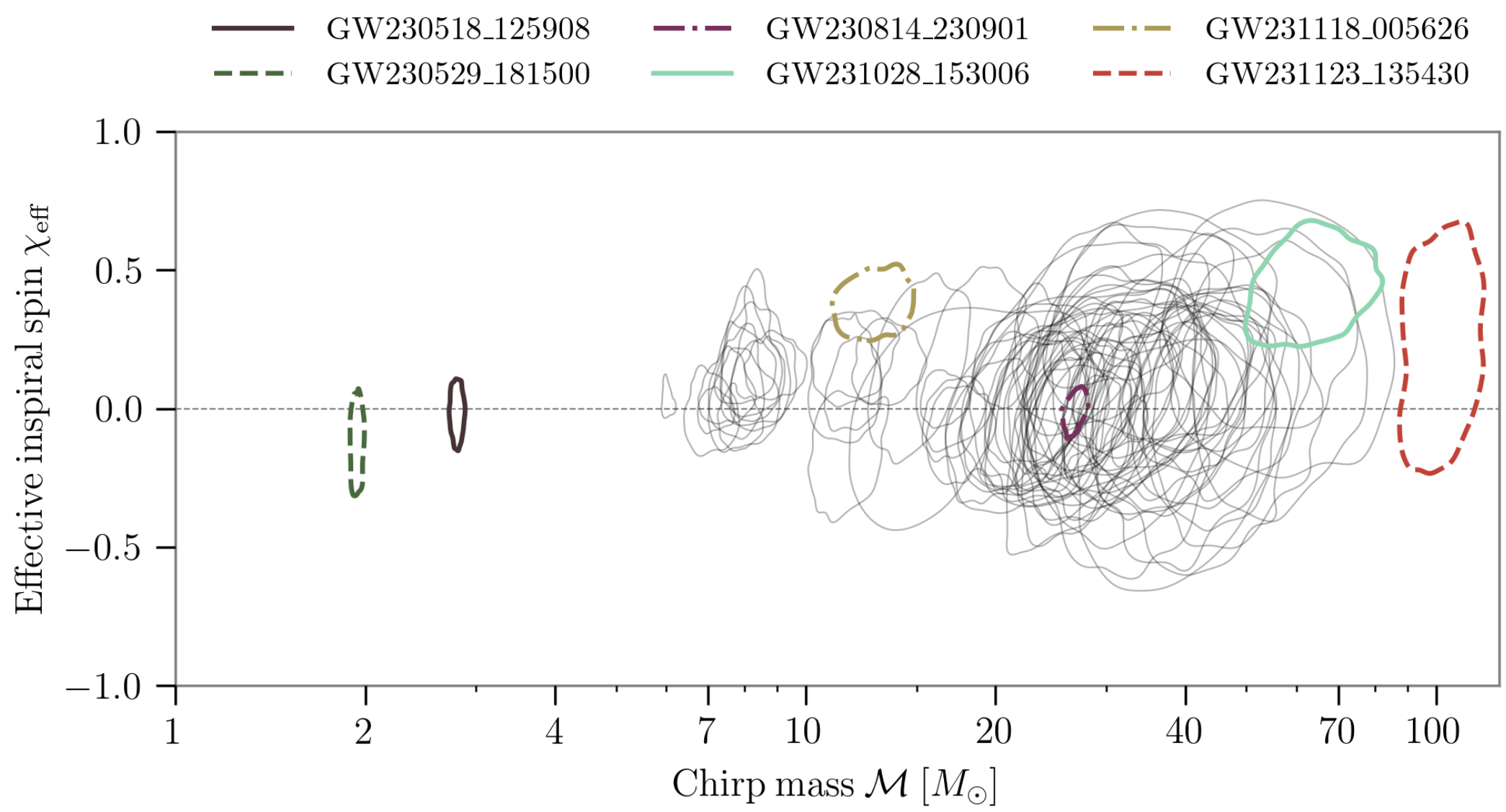

**Figure 4.** Credible-region contours in the chirp mass $\mathcal{M}$ and effective inspiral spin $\chi_{\rm eff}$ plane for O4a candidates with FAR $< 1$ yr$^{-1}$. Each contour indicates the 90% credible region for a given candidate. We use colors to highlight candidates GW230518_125908; GW230529_181500, which has an inferred $\chi_{\rm eff} < 0$ at 92% credibility; GW231028_153006 and GW231118_005626, which have inferred $\chi_{\rm eff} > 0$; GW230814_230901, which was the highest-SNR candidate observed in O4a; and GW231123_135430, which we infer to have the most massive source observed in O4a.

The detectors measure the redshifted masses $(1 + z)m_i$, where $z$ is the source redshift (A. Krolak & B. F. Schutz 1987). To recover the source-frame masses, we combine the measured redshifted masses with the inferred luminosity distance using an assumed cosmology (P. A. R. Ade et al. 2016). Due to the uncertainties in our estimation of the luminosity distance, source-frame mass parameters are generally less well constrained than their redshifted values. By default, we report source-frame mass values, using default agnostic priors as described above. Figure 2 shows the marginalized one-dimensional posteriors for the chirp mass $\mathcal{M}$ and mass ratio $q$ of each of the O4a candidates analyzed here. Figure 3 shows the marginalized two-dimensional posteriors for the individual component masses, $m_1$ and $m_2$, as well as for the total mass $M$ and mass ratio $q$. Similarly, Figure 4 shows the marginalized two-dimensional posteriors in $\mathcal{M}$ and effective inspiral spin $\chi_{\rm eff}$, described in Section 3.2. These representations of the inferred source masses of our newly added candidates display a large range, with total masses spanning nearly 2 orders of magnitude.

We use the inferred component masses to classify the probable nature of the binary components. For example, if one component has a mass above the maximum possible mass of an NS, we infer it to be a BH, even in the absence of other constraints on the presence of matter in the binary (see Section 3.3). The maximum possible nonspinning NS mass is unknown, although data from terrestrial experiments, electromagnetic observations of NSs, and GW detections have been used to bound this mass to $\sim$2.2–2.5 $M_\odot$ (e.g., M. C. Miller et al. 2019, 2021; T. E. Riley et al. 2019, 2021; T. Dietrich et al. 2020; P. Landry et al. 2020; I. Legred et al. 2021; Y. Lim et al. 2021; G. Raaijmakers et al. 2021; S. Huth et al. 2022; Y. -Z. Fan et al. 2024; N. Rutherford et al. 2024; B. Biswas & S. Rosswog 2025; L. Brandes & W. Weise 2025; J. Golomb et al. 2025).

The electromagnetic counterpart to GW170817 (B. P. Abbott et al. 2017a, 2017c) has been used together with models of the emission to place more stringent limits on the maximum mass of nonspinning NSs (e.g., B. Margalit & B. D. Metzger 2017; L. Rezzolla et al. 2018; M. Ruiz et al. 2018; B. P. Abbott et al. 2020c; A. Nathanail et al. 2021). On the other hand, spin increases the upper limit of possible NS masses. To account for the remaining uncertainty, we very conservatively select 3 $M_\odot$ as a robust upper limit on the maximum NS mass (C. E. Rhoades & R. Ruffini 1974; V. Kalogera & G. Baym 1996). As in GWTC-3.0 (R. Abbott et al. 2023a), we divide our candidates into two categories: *unambiguous* BBHs, where both components are BHs ($m_i > 3\,M_\odot$ at 99% probability), and *potential-NS binaries* where at least one component mass could have been an NS.

GWTC-4.0 also contains candidates with FAR $\geqslant 1$ yr$^{-1}$ for which at least one detection pipeline assigns a probability of astrophysical origin $p_{\rm astro} \geqslant 0.5$. Since we do not infer the source properties of these candidates, the classification of the sources of these candidates is provided through their multicomponent $p_{\rm astro}$ values. All such candidates with nonzero probability for containing an NS, as determined by $p_{\rm BNS} + P{\rm NSBH} > 0.001$, are given in Table 7 in Appendix A.

#### 3.1.1. Candidates with $m_2 > 3M_\odot$: Unambiguous BBHs

Our unambiguous BBH candidates span more than an order of magnitude in their inferred masses. Since the chirp mass is lower for asymmetric binaries at a fixed total mass, the systems with the largest and smallest chirp masses do not necessarily correspond to the most and least massive binaries. Nevertheless, we find that the same two transients, GW231123_135430 and GW230627_015337, lie at the extremes for both mass parameters. GW231123_135430 (A. G. Abac et al. 2025b) probably has the largest chirp mass of the candidates we analyze, with $\mathcal{M} = 101^{+13}_{-30}\,M_\odot$ , as well as the largest total mass with $M = 236^{+29}_{-48}\,M_\odot$. It is the most massive source in our catalog of those with FAR $< 1$ yr$^{-1}$. GW230627_015337 is the unambiguous BBH with the smallest chirp mass, $\mathcal{M} = 6.02^{+0.16}_{-0.07} M_\odot$, as well as the smallest total mass with $M = 14.19^{+0.77}_{-0.45} M_\odot$.

The individual components of our BBHs span masses between $5.79^{+0.95}_{-0.92} M_\odot$ and $137^{+23}_{-18} M_\odot$, with primary masses ranging from $8.4^{+1.7}_{-1.3} M_\odot$ for GW230627_015337 to $137^{+23}_{-18} M_\odot$ for GW231123_135430, and secondary masses ranging from $5.79^{+0.95}_{-0.92} M_\odot$ to $101^{+22}_{-51} M_\odot$ for GW230627_015337 and GW231123_135430, respectively. The BBH mass distribution inferred using all the candidates in GWTC-4.0 is discussed in depth in A. G. Abac et al. (2026d).

GW231123_135430 is probably more massive than the source of the less significant candidate GW190426_190642, which is inferred to have $M = 182.3^{+40.2}_{-35.7} M_\odot$ and an FAR > 1 yr$^{-1}$, although with $p_{\rm astro} \geqslant 0.5$ (R. Abbott et al. 2024). With a 94% probability of $m_1$ being greater than 120 $M_\odot$, GW231123_135430 has a component mass above the approximate upper boundary of the pair-instability supernova mass gap (W. A. Fowler & F. Hoyle 1964; Z. Barkat et al. 1967; C. L. Fryer et al. 2001; K. Belczynski et al. 2016; M. Spera & M. Mapelli 2017; S. Stevenson et al. 2019; S. E. Woosley & A. Heger 2021; A. K. Mehta et al. 2022). In addition to GW231123_135430, several transients have remnant mass $M_{\rm f} \geqslant 100$ $M_\odot$, satisfying a conventional threshold to be considered intermediate-mass BHs. The remnants of GW230704_212616, GW230922_040658, GW231005_021030, and GW231028_153006 all have > 90% probability of $M_{\rm f} \geqslant 100$ $M_\odot$.

At masses below 5 $M_\odot$, there is a putative lower mass gap, hypothesized based on X-ray binary observations (C. D. Bailyn et al. 1998; F. Özel et al. 2010; W. M. Farr et al. 2011; L. Kreidberg et al. 2012). We identify several unambiguous BBH transients whose secondary components could possibly fall into this mass gap: GW230627_015337, GW231020_142947, GW231118_090602, and GW231223_075055. The sources of GW231020_142947 and GW231118_090602 are inferred to have $m_2 \leqslant 5$ $M_\odot$ with 12% and 14% probability, respectively. Both GW230627_015337 and GW231223_075055 have similar and slightly smaller probabilities of having secondary source mass $m_2 \leqslant 5$ $M_\odot$, 8% and 9% respectively. No unambiguous BBH candidate had significant posterior support for a primary mass value in the lower mass gap.

The sources of several candidates have notable posterior support for unequal masses. GW231114_043211 has the most support of the unambiguous BBHs, with a mass ratio of $q = 0.36^{+0.27}_{-0.17}$. In addition, GW230702_185453, GW230712_090405, GW230928_215827, GW231001_140220, GW231004_232346, GW231118_005626, and GW231129_081745 all have mass ratios $\leqslant 0.85$ at the 90% credible level.

#### 3.1.2. Candidates with $m_2 < 3$ $M_\odot$

Two candidates are consistent with sources containing a secondary mass $m_2 < 3$ $M_\odot$, GW230518_125908 and GW230529_181500 (A. G. Abac et al. 2024b). We infer that these two candidates have negligible posterior support for $m_2 \geqslant 3$ $M_\odot$. No confident multimessenger counterparts were reported for either of these candidates or any other O4a candidate. Without an electromagnetic counterpart, we can only infer the presence of matter in a CBC via its impact on the GW signal. Matter can leave an imprint on the waveform during inspiral through tidal effects or modify the merger and postmerger dynamics, as discussed in Section 3.3.

The secondary of each system is consistent with the observed Galactic NS population (J. Antoniadis et al. 2016; J. Alsing et al. 2018; N. Farrow et al. 2019; K. El-Badry et al. 2024). Further, the mass of the primary of GW230518_125908 is well above 3 $M_\odot$, with $m_1 = 8.17^{+0.84}_{-0.92} M_\odot$ , leading us to conclude that this system is likely an NSBH. Further, the source of GW230518_125908 has clearly asymmetric masses, with a mass ratio of $q = 0.18^{+0.04}_{-0.03}$. The source of GW230529_181500 is most probably another NSBH, since its primary mass $m_1 = 3.66^{+0.82}_{-1.21} M_\odot$ is consistent with a low-mass BH (A. G. Abac et al. 2024b).

### 3.2. Spins

Compared to the masses, spins have a weaker impact on the GW emission and are more difficult to measure from observations (E. Poisson & C. M. Will 1995; E. Baird et al. 2013; S. Vitale et al. 2014; K. Chatziioannou et al. 2015; B. P. Abbott et al. 2016c; B. Farr et al. 2016; S. Vitale et al. 2017a; G. Pratten et al. 2020; J. García-Bellido et al. 2021). The component spins of compact binaries, $\chi_1$ and $\chi_2$, are typically poorly constrained since the leading-order spin contribution to the GW signal is determined by mass-weighted combinations of the components (T. Damour 2001; L. Blanchet 2014; M. Pürrer et al. 2016; K. K. Y. Ng et al. 2018; M. Zevin et al. 2020). Here, we focus on two specific mass-weighted spin parameters: the effective inspiral spin $\chi_{\rm eff}$ and the effective precession spin $\chi_{\rm p}$ (A. G. Abac et al. 2025a).

The effective inspiral spin $\chi_{\rm eff}$ (L. Santamaría et al. 2010; P. Ajith et al. 2011) is a mass-weighted combination of the components of the spin aligned with the Newtonian orbital angular momentum. It appears in the leading-order spin term due to spin–orbit coupling at 1.5 post-Newtonian order and is approximately conserved throughout the inspiral (É. Racine 2008). Positive and negative $\chi_{\rm eff}$ indicate that there is net spin aligned and antialigned, respectively, with the orbital angular momentum.

The effective precession spin $\chi_{\rm p}$ (P. Schmidt et al. 2015) measures the mass-weighted in-plane spin component that contributes to spin precession (T. A. Apostolatos et al. 1994; L. E. Kidder 1995). It is bounded between 0 and 1, with $\chi_{\rm p} = 0$ indicating no spin precession and $\chi_{\rm p} = 1$ indicating maximal precession. This parameter is typically weakly constrained, and posterior measurements of $\chi_{\rm p}$ are often dominated by the prior.

The spin orientations $\theta_i$ of a binary are of particular interest for the insight they provide to its evolutionary history (M. Fishbach et al. 2017; S. Stevenson et al. 2017; C. Talbot & E. Thrane 2017; S. Vitale et al. 2017b; D. Wysocki et al. 2019; M. Zevin et al. 2021). Compact binaries form via a myriad of channels but can be broadly classified as either dynamically assembled or formed via isolated binary evolution. Roughly speaking, in dynamically formed binaries, the spins are expected to be isotropically oriented, while binaries formed in isolation are expected to have spins more nearly aligned with the orbital axis. Nonzero $\chi_{\rm p}$ or negative $\chi_{\rm eff}$ are therefore more consistent with dynamically formed binaries than those formed in isolation. Further discussion of the connection between spin orientations and compact binary formation channels is given in A. G. Abac et al. (2026d).

Most of the candidates analyzed from O4a are consistent with having sources with $\chi_{\rm eff} = 0$, as seen in Figures 2 and 4. However, 16 candidates have sources with $\chi_{\rm eff} \geqslant 0$ with greater than 90% probability. Two sources with notably large $\chi_{\rm eff}$ values are that of GW231028_153006 with $\chi_{\rm eff} = 0.44^{+0.16}_{-0.20}$ and that of GW231118_005626 with $\chi_{\rm eff} = 0.37^{+0.10}_{-0.11}$. GW231123_135430 is inferred to have large component spins (A. G. Abac et al. 2025b) and has a 88% probability of $\chi_{\rm eff} > 0$. We find fewer candidates that probably have negative effective inspiral spins, with only three sources having $\chi_{\rm eff} < 0$ with greater than 90% probability. Of these, GW231226_101520 is remarkable. This candidate has the second largest SNR of those in GWTC-4.0, with an SNR of $33.7^{+0.1}_{-0.1}$. Its source has $\chi_{\rm eff} = -0.09^{+0.09}_{-0.10}$ and $\chi_{\rm eff} < 0$ with 93% probability. Another is the NSBH candidate GW230529_181500 (A. G. Abac et al. 2024b), whose source has $\chi_{\rm eff} < 0$ with 92% probability when analyzed with our fiducial set of waveforms (A. G. Abac et al. 2026a).

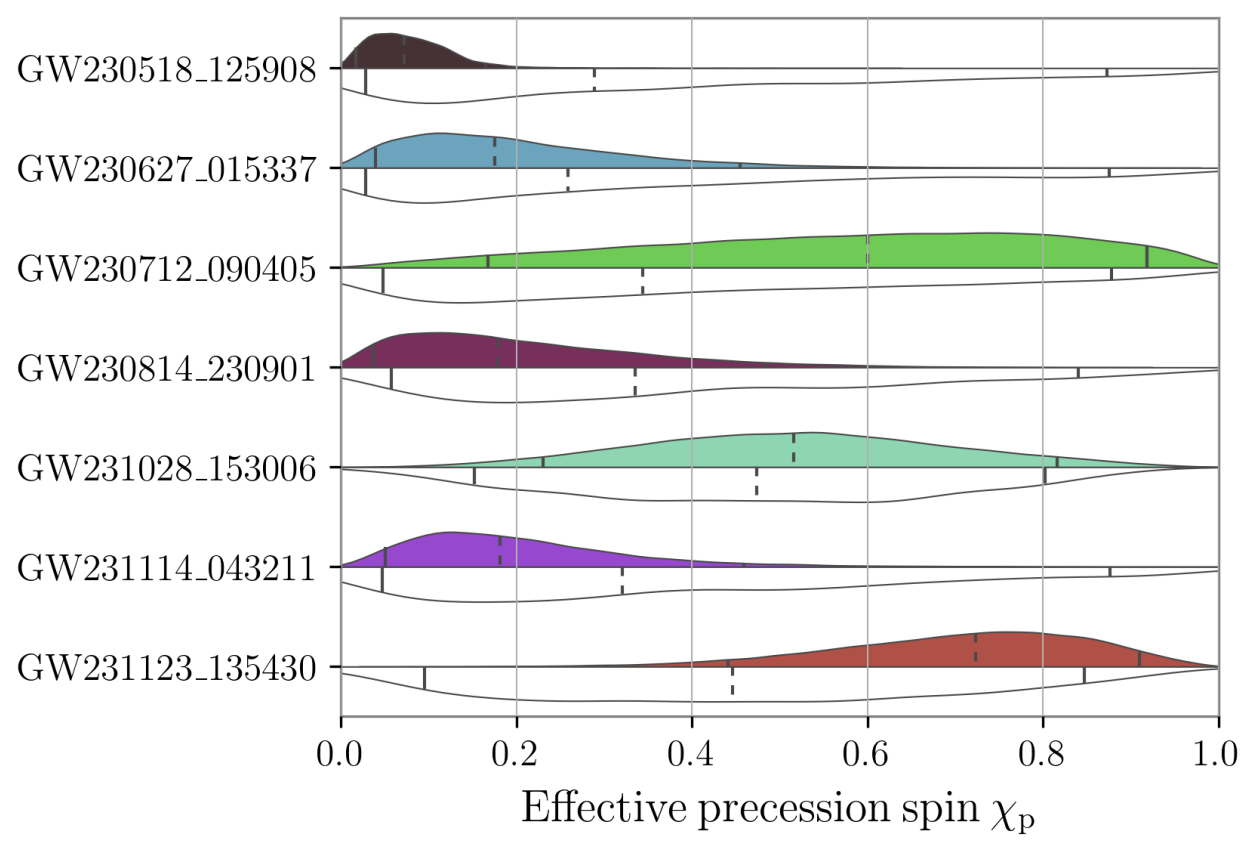

**Figure 5.** Posterior (upper, colored) and the effective prior (lower, white) probability distributions for the dimensionless effective precession spin $\chi_p$ for candidates GW230518_125908, GW230627_015337, GW230712_090405, GW230814_230901, GW231028_153006, GW231114_043211, and GW231123_135430. Vertical lines mark the median and symmetric 90% interval for the distributions. These candidates are the ones that show the greatest deviation between the posterior distribution and the effective prior over the $\chi_p$ parameter from the set of new candidates presented in this work.

Figure 5 shows the $\chi_p$ posterior probability distribution compared to the prior distribution after conditioning on the $\chi_{\rm eff}$ measurement (B. P. Abbott et al. 2019a), for a selection of candidates. These distributions would be the same if no information about the in-plane spin components had been extracted from the signal, and the selected candidates have the greatest difference between the two distributions. For most of the candidates, the $\chi_p$ posteriors are broad and uninformative.

Figure 6 shows the posterior distribution of the source component spin magnitudes $\chi_i$ and tilt angles $\theta_i$ inferred for a subset of the analyzed candidates. These are the candidates highlighted in Figure 5 (whose $\chi_p$ posteriors show the greatest deviation from their conditioned priors) or candidates that display systematic differences between inferences made with different waveform models (Section 3.6). We exclude exceptional candidates whose spin magnitudes and tilt angles have been reported elsewhere (A. G. Abac et al. 2025b, 2026c), and we additionally show the results for the high-SNR candidate GW231226_101520. In many other cases, the component spins of the sources are poorly measured, and our posteriors are similar to our priors. For those binaries where $\chi_{\rm eff}$ is constrained to be relatively small, the posteriors of the component spins may be concentrated in the equatorial plane even without positive evidence for precession, due to ruling out spins either relatively aligned or antialigned with the orbital angular momentum (B. P. Abbott et al. 2017d).

Of the sources analyzed from O4a, GW231123_135430 is inferred to have the highest primary spin magnitude, $\chi_1 = 0.90^{+0.08}_{-0.27}$ (A. G. Abac et al. 2025b). Two other systems that are inferred to have high primary spins are the sources of GW230928_215827 and GW231028_153006, for which $\chi_1 > 0.8$ with 44% and 47% probabilities, respectively.

The final spin $\chi_{\rm f}$ of the remnant BH following coalescence has contributions from the orbital angular momentum at merger and the spin angular momenta of the binary components. It is determined for our BBH candidates from the inferred component masses and spins, using fits to numerical relativity simulations (A. G. Abac et al. 2026a). The candidates with $\chi_{\rm f} > 0.75$ at 90% probability are GW231028_153006 and GW231118_005626.

### 3.3. Tidal Effects

Binary systems with at least one NS component display modifications to their gravitational waveform due to the presence of matter in the system. During the inspiral, tidal deformations of the NS constituents are imprinted on the waveform (É. É. Flanagan & T. Hinderer 2008; F. Pannarale et al. 2011; J. Vines et al. 2011). Tidal effects are quantified by the tidal deformability parameter of each component $\Lambda_i$, which are larger for stiffer NS equations of state and smaller for softer ones. Matter effects also modify the final merger and postmerger phases of the GW signal at relatively high frequencies (e.g., B. P. Abbott et al. 2017e, 2019d). For example, the GW emission from these phases may be truncated in NSBH systems if the NS is tidally disrupted before merger (K. Kyutoku et al. 2021). At the sensitive frequencies of current detectors, tidal effects are the dominant source of information about matter effects in GW signals.

For the new candidates detected in O4a, only two sources have component masses consistent with one of the two components being an NS, GW230518_125908 and GW230529_181500 (see Section 3.1.2). We have analyzed the GW signal from GW230518_125908 with three waveform models that include tidal effects, and in all cases, the data do not constrain $\Lambda_2$, as expected, given the signal's SNR and our mass inferences. Similarly, when analyzing GW230529_181500 with NSBH waveform models, the data do not constrain the tidal deformability $\Lambda_2$ of the NS (A. G. Abac et al. 2024b).

### 3.4. Localization

As GW detectors continue to improve their sensitivity, we are able to observe GW sources from a greater range of cosmic distances. The closest source observed in O4a is probably that of the NSBH candidate GW230529_181500, inferred to be at a luminosity distance $D_L = 0.20^{+0.10}_{-0.10}$ Gpc. The more massive NSBH source of GW230518_125908 was also inferred to be nearby, and probably closer than any BBH candidate ($D_L = 0.24^{+0.11}_{-0.10}$ Gpc).

During O4a, a number of relatively nearby BBH signals were also observed. The source that is probably the closest of these is also the source of the highest-SNR GW candidate detected through O4a, GW230814_230901 (A. G. Abac et al. 2026c), with $D_L = 0.28^{+0.17}_{-0.13}$ Gpc. However, GW230627_015337 ($D_L = 0.31^{+0.06}_{-0.13}$ Gpc), which has a large network SNR of 28.5 and is also the lightest BBH with $M = 14.19^{+0.77}_{-0.45}\,M_\odot$ may be closer; both events have considerable overlap between the credible ranges of their luminosity distances.

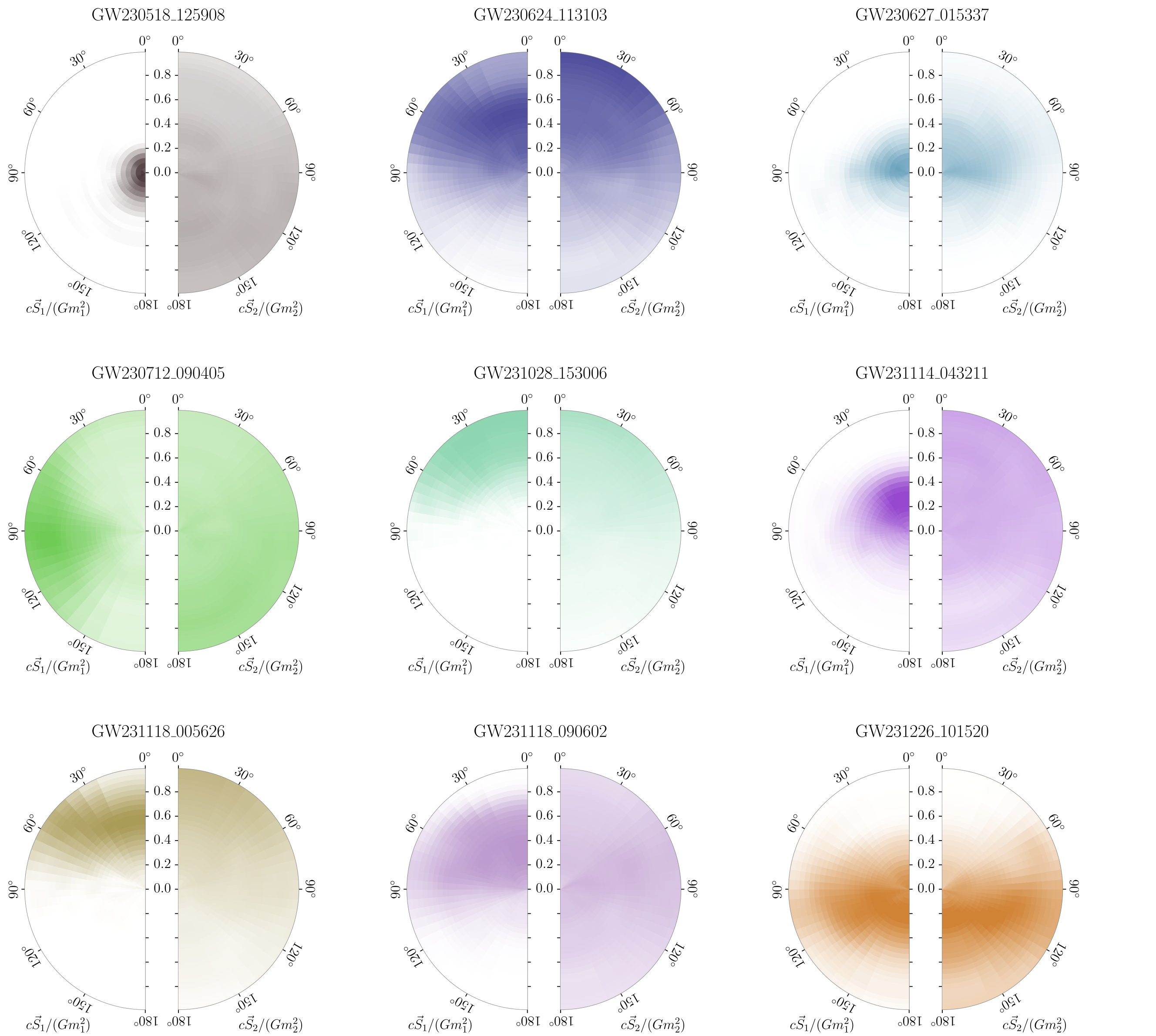

**Figure 6.** Posterior probability distributions for the dimensionless component spins $\chi_1 = c\boldsymbol{S}_1/(Gm_1^2)$ and $\chi_2 = c\boldsymbol{S}_2/(Gm_2^2)$ (with $\boldsymbol{S}_1$ and $\boldsymbol{S}_2$ the spin vectors of the components) relative to the orbital plane, marginalized over azimuthal angles, for candidates GW230518_125908, GW230624_113103, GW230627_015337, GW230712_090405, GW231028_153006, GW231114_043211, GW231118_005626, GW231118_090602, and GW231226_101520. In these plots, the histogram bins are constructed linearly in spin magnitude and the cosine of the tilt angles, such that they contain equal prior probability.

The farthest source of the candidates analyzed is probably for GW230704_212616, which is inferred to lie at $D_L = 7.2^{+6.1}_{-4.2}$ Gpc. As there are several candidates with sources at similar distances, GW230704_212616 has only a 22% probability of having the most distant source. The next most probable to have the most distant source is GW231119_075248 at $D_L = 6.7^{+5.5}_{-3.7}$ Gpc, with a 15% probability of being the most distant candidate. These distant candidates are at comparable, but probably larger, luminosity distances than the farthest sources detected in the third observing run (O3) with FAR $< 1$ yr$^{-1}$, for example, GW190805_211137 with $D_L = 6.1^{+3.7}_{-3.1}$ Gpc (R. Abbott et al. 2024).

The inferred sky location of each candidate depends largely on the number of observing GW detectors at the time of the detection (B. F. Schutz 1986; S. Fairhurst 2009, 2011; S. Nissanke et al. 2011, 2013; J. Veitch et al. 2012; K. Grover et al. 2014; M. M. Kasliwal & S. Nissanke 2014; L. P. Singer et al. 2014; C. P. L. Berry et al. 2015; B. P. Abbott et al. 2020a). During O4a, only the two LIGO detectors were operating with significant sensitivity to CBC signals. As a result, even the best localized new candidates in GWTC-4.0 have greater uncertainties in their sky location than many candidates detected previously with a network that included the Virgo detector (B. P. Abbott et al. 2019a, 2019d, 2021b, 2023a, 2024). While the NSBH candidate GW230529_181500 is inferred to be nearby, it was only detected in the Livingston detector, and as such, it is poorly localized in the sky (A. G. Abac et al. 2024b). Meanwhile, GW230518_125908 was observed with both LIGO detectors, and it is constrained to a sky area of 490 deg$^2$ (90% credible level). The best localized candidate in terms of sky area is the lowest-mass BBH candidate GW230627_015337, which is inferred to lie in a region of sky of 110 deg$^2$ (90% credible level).

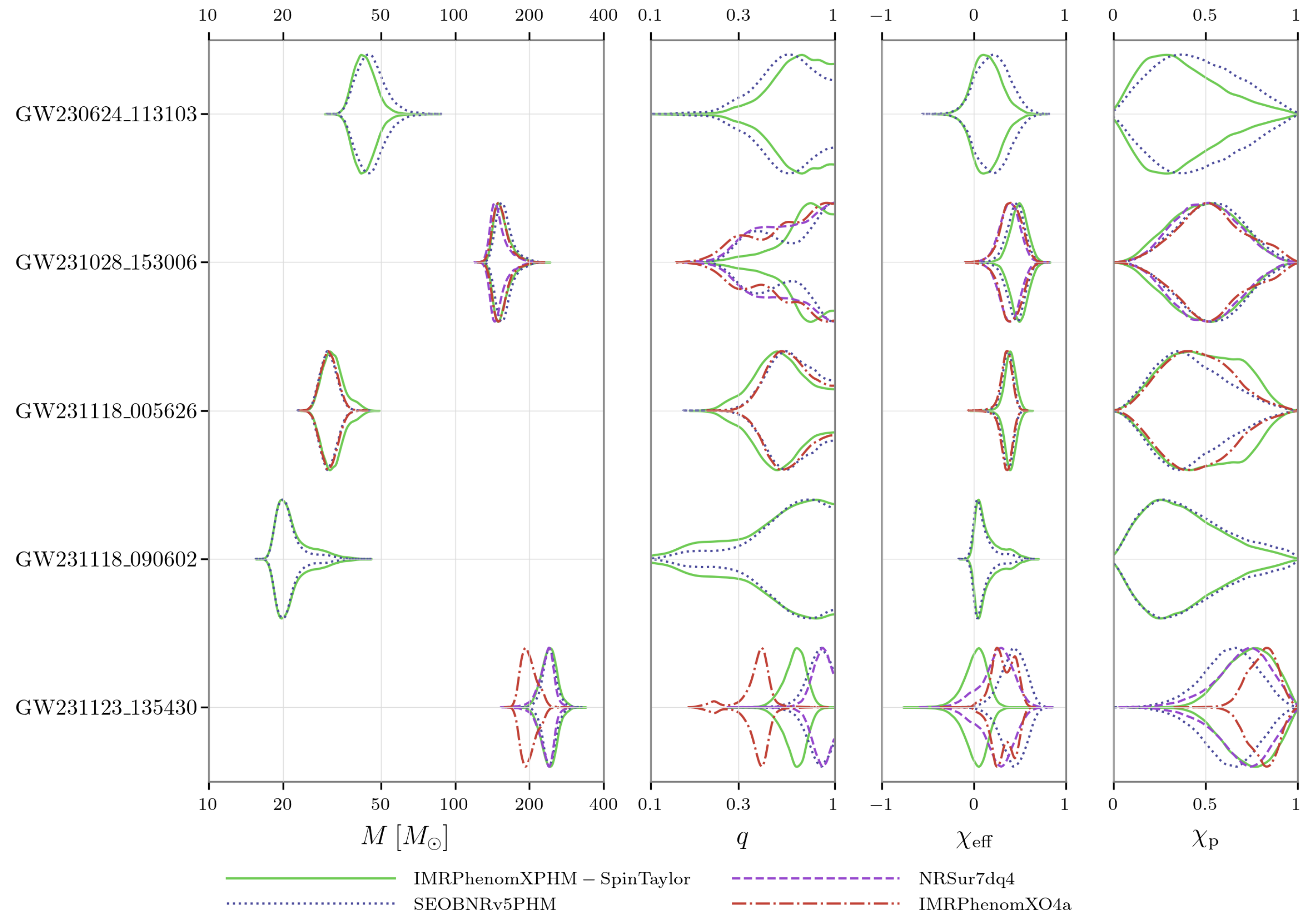

**Figure 7.** The marginal probability distributions for the (source-frame) total mass $M$, mass ratio $q$, effective inspiral spin $\chi_{\rm eff}$, and effective precession spin $\chi_{\rm p}$ for five O4a candidates that show significant waveform systematics.

The highest-SNR candidate GW230814_230901 was detected in only the Livingston detector. In common with other candidates identified in only a single observatory, it is only weakly localized by the antenna pattern of the detector. This is true also of the NSBH candidate GW230529_181500 despite it being inferred to be nearby.

The three-dimensional volume localization of each candidate depends on both its inferred distance and sky localization (L. P. Singer et al. 2016; W. Del Pozzo et al. 2018). Broadly speaking, nearby candidates have the best volume localization, provided they are observed in multiple detectors. The NSBH candidate GW230518_125908 is well localized as compared to the other candidates in O4a, to a volume 0.0019 Gpc$^3$ (90% credible level). The candidate with the best three-dimensional localization from O4a is GW230627_015337, which has a 90% credible volume of 0.00062 Gpc$^3$. The next best localized BBH is probably the second-highest-SNR candidate GW231226_101520, which has an SNR of 33.7 and is constrained to a comparatively much larger 90% credible volume of 0. 056 Gpc$^3$.

### 3.5. Multimodality

As with some of the candidates reported in GWTC-3.0 (R. Abbott et al. 2023a), a few of our candidates display multimodal posteriors. Multimodality can be an indication that the inferred parameters of a candidate lie in a region where our waveform models have a complex structure and where subtle changes in modeling may be important. This is especially true when models incorporate higher-order multipole moments (A. H. Nitz et al. 2021; H. S. Chia et al. 2022; H. Estellés et al. 2022b; A. K. Mehta et al. 2022) or precession (B. P. Abbott et al. 2019d, 2020d). Multiple modes can also arise from noise fluctuations in low-SNR signals (Y. Huang et al. 2018) due to the presence of glitches (J. Powell 2018; K. Chatziioannou et al. 2021; G. Ashton et al. 2022; S. Soni et al. 2025), or through the overlap of multiple signals (P. Relton & V. Raymond 2021), although this final possibility is unlikely at current sensitivities. As a result, multimodality is expected in at least some cases. In addition, if the inferences drawn using different waveform models display significant systematic differences for a given candidate, multimodality can be produced upon combining samples across each model for our final inferences.

Most mass posterior distributions are unimodal, as shown in Figure 2. Candidates that display multiple modes in the distributions of their inferred source masses are GW230712_090405, GW230723_101834, and GW231118_090602. The

**Table 4**
Median and 90% Symmetric Confidence Intervals (CIs) for Selected Source Properties as Inferred by Different Waveform Models

| Candidate | Model | $\mathcal{M}$ ($M_\odot$) | $q$ | $\chi_{\rm eff}$ | $\chi_{\rm p}$ | $D_L$ (Gpc) |
|---|---|---|---|---|---|---|
| GW230624_113103 | Mixed | $18.0^{+3.2}_{-2.4}$ | $0.59^{+0.35}_{-0.29}$ | $0.17^{+0.29}_{-0.25}$ | $0.39^{+0.42}_{-0.28}$ | $1.94^{+1.27}_{-0.97}$ |
| | IMRPhenomXPHM_SpinTaylor | $17.7^{+2.9}_{-2.2}$ | $0.63^{+0.32}_{-0.32}$ | $0.1^{+0.3}_{-0.2}$ | $0.34^{+0.43}_{-0.25}$ | $2.0^{+1.2}_{-1.0}$ |
| | SEOBNRv5PHM | $18.4^{+3.3}_{-2.5}$ | $0.55^{+0.37}_{-0.26}$ | $0.2^{+0.3}_{-0.3}$ | $0.44^{+0.40}_{-0.31}$ | $1.9^{+1.3}_{-0.9}$ |
| GW231028_153006 | Mixed | $63^{+13}_{-10}$ | $0.63^{+0.33}_{-0.35}$ | $0.44^{+0.16}_{-0.20}$ | $0.52^{+0.30}_{-0.29}$ | $4.1^{+1.4}_{-1.9}$ |
| | IMRPhenomXPHM_SpinTaylor | $64.3^{+13.4}_{-7.8}$ | $0.71^{+0.25}_{-0.36}$ | $0.5^{+0.1}_{-0.2}$ | $0.49^{+0.29}_{-0.30}$ | $4.5^{+1.2}_{-2.0}$ |
| | SEOBNRv5PHM | $63.5^{+13.4}_{-7.9}$ | $0.57^{+0.39}_{-0.29}$ | $0.5^{+0.1}_{-0.2}$ | $0.53^{+0.28}_{-0.29}$ | $4.0^{+1.3}_{-1.8}$ |
| | NRSur7dq4 | $59.8^{+14.0}_{-9.6}$ | $0.53^{+0.43}_{-0.26}$ | $0.4^{+0.2}_{-0.2}$ | $0.50^{+0.29}_{-0.26}$ | $4.1^{+1.4}_{-1.9}$ |
| | IMRPhenomXO4a | $62^{+12}_{-11}$ | $0.61^{+0.35}_{-0.35}$ | $0.4^{+0.2}_{-0.2}$ | $0.54^{+0.32}_{-0.29}$ | $4.0^{+1.6}_{-1.8}$ |
| GW231118_005626 | Mixed | $12.6^{+1.6}_{-1.1}$ | $0.55^{+0.37}_{-0.22}$ | $0.37^{+0.10}_{-0.11}$ | $0.44^{+0.34}_{-0.27}$ | $2.19^{+0.93}_{-0.97}$ |
| | IMRPhenomXPHM_SpinTaylor | $12.7^{+1.6}_{-1.1}$ | $0.50^{+0.39}_{-0.21}$ | $0.4^{+0.1}_{-0.1}$ | $0.48^{+0.31}_{-0.31}$ | $2.1^{+1.0}_{-1.0}$ |
| | SEOBNRv5PHM | $12.5^{+1.6}_{-1.1}$ | $0.58^{+0.35}_{-0.21}$ | $0.4^{+0.1}_{-0.1}$ | $0.40^{+0.36}_{-0.25}$ | $2.3^{+0.9}_{-1.0}$ |
| | IMRPhenomXO4a | $12.6^{+1.6}_{-1.1}$ | $0.56^{+0.35}_{-0.20}$ | $0.4^{+0.1}_{-0.1}$ | $0.45^{+0.31}_{-0.27}$ | $2.2^{+0.9}_{-0.9}$ |
| GW231118_090602 | Mixed | $8.37^{+0.76}_{-0.56}$ | $0.56^{+0.38}_{-0.41}$ | $0.08^{+0.36}_{-0.09}$ | $0.36^{+0.43}_{-0.27}$ | $1.35^{+0.54}_{-0.60}$ |
| | IMRPhenomXPHM_SpinTaylor | $8.37^{+0.76}_{-0.56}$ | $0.51^{+0.42}_{-0.38}$ | $0.1^{+0.4}_{-0.1}$ | $0.35^{+0.40}_{-0.25}$ | $1.4^{+0.5}_{-0.6}$ |
| | SEOBNRv5PHM | $8.38^{+0.77}_{-0.57}$ | $0.60^{+0.34}_{-0.33}$ | $0.07^{+0.21}_{-0.08}$ | $0.38^{+0.45}_{-0.28}$ | $1.3^{+0.6}_{-0.6}$ |
| GW231123_135430 | Mixed | $101^{+13}_{-30}$ | $0.74^{+0.22}_{-0.38}$ | $0.31^{+0.25}_{-0.41}$ | $0.72^{+0.19}_{-0.28}$ | $2.2^{+2.0}_{-1.5}$ |
| | IMRPhenomXPHM_SpinTaylor | $100^{+13}_{-15}$ | $0.61^{+0.13}_{-0.14}$ | $0.01^{+0.17}_{-0.24}$ | $0.75^{+0.18}_{-0.23}$ | $0.89^{+0.39}_{-0.34}$ |
| | SEOBNRv5PHM | $105^{+13}_{-11}$ | $0.82^{+0.15}_{-0.20}$ | $0.4^{+0.2}_{-0.2}$ | $0.64^{+0.22}_{-0.27}$ | $2.2^{+1.4}_{-1.0}$ |
| | NRSur7dq4 | $103^{+10}_{-14}$ | $0.85^{+0.13}_{-0.16}$ | $0.3^{+0.2}_{-0.4}$ | $0.72^{+0.20}_{-0.29}$ | $1.8^{+1.6}_{-1.0}$ |
| | IMRPhenomXO4a | $75^{+13}_{-10}$ | $0.39^{+0.07}_{-0.16}$ | $0.3^{+0.2}_{-0.2}$ | $0.82^{+0.11}_{-0.14}$ | $3.5^{+1.3}_{-1.4}$ |

**Note.** Values are given for a subset of the GW event candidates from O4a with FAR $< 1\ {\rm yr}^{-1}$ that show significant waveform systematics. The columns show chirp mass $\mathcal{M}$, mass ratio $q$, effective inspiral spin $\chi_{\rm eff}$, effective precession spin $\chi_{\rm p}$, and luminosity distance $D_L$ . The Mixed row gives our fiducial estimates, derived from combining samples from all available models equally (A. G. Abac et al. 2026a). For the exceptional event GW231123_135430, we include samples from an additional model IMRPhenomTPHM (H. Estellés et al. 2022a) in the Mixed samples (A. G. Abac et al. 2025b). We also evolve the spin quantities for all waveforms in GW231123_135430 to a large binary separation (A. G. Abac et al. 2026a).

sources of GW230712_090405 and GW230723_101834 are inferred to have bimodal chirp masses. Since the uncertainty in redshift tends to broaden our mass inferences in the source frame, the bimodalities for these two candidates are more easily seen in the redshifted chirp mass $(1 + z)\mathcal{M}$ posteriors. Additionally, GW231118_090602, a system with $\mathcal{M} = 8.37^{+0.76}_{-0.56}\ M_\odot$, shows multimodality in $m_2$.

Multimodality in posterior distributions for mass parameters often correlates with multiple modes in other parameters, especially the spin quantities, $\chi_{\rm eff}$ and $\chi_{\rm p}$. This is true of GW230723_101834, whose multiple modes are prominently seen in the two-dimensional marginalized distribution of the detector-frame chirp mass $(1 + z)\mathcal{M}$ and effective inspiral spin $\chi_{\rm eff}$. Similarly, for GW231118_090602, the mode at smaller $m_2$ values correlates with a mode of higher $\chi_{\rm eff}$ values.

The degree of multimodality can depend on the waveform model. For example, the bimodality in $(1 + z)\mathcal{M}$for GW230712_090405 is present for all models, but for IMRPhenomXO4a, the multimodality in this parameter is more prominent, with additional modes present. For GW231118_090602, $m_2$ is unimodal when analyzed using SEOBNRv5PHM and bimodal when using IMRPhenomXPHM_SpinTaylor. The two high-mass candidates GW231028_153006 and GW231123_135430 display both multimodal mass inferences for some waveform models and systematic differences between models that generate further multimodality upon combining samples. More detailed discussion of this multimodality for GW231123_135430 may be found in A. G. Abac et al. (2025b).

The candidate GW231001_140220 displayed bimodal posteriors for $\mathcal{M}$ in our preliminary analysis using the default lower-frequency cutoff (20 Hz) for our likelihood integration and the IMRPhenomXPHM_SpinTaylor waveform model. While our initial data-quality checks did not indicate a glitch coincident with this candidate, i.e., a glitch with support in time and frequency overlapping the candidate signal, further parameter-estimation analysis using data from individual detectors indicated possible non-Gaussian noise in the Livingston detector at the time of the detection. When analyzed with a lower-frequency cutoff of $f_{\rm low} = 40$ Hz in the likelihood integral for the Livingston data, we obtain unimodal posteriors while losing only a small amount of the total SNR for this candidate. We present our parameter inferences for this candidate using this narrower frequency range.

### *3.6. Waveform Systematics and Consistency*

The results presented in this Letter are the combined samples from a number of parameter-estimation analyses that use different waveform models (A. G. Abac et al. 2026a). The combined samples include results obtained with IMRPhenomXPHM_SpinTaylor and SEOBNRv5PHM for all candidates. Additionally, for candidates with source properties within the model's calibration region, we also include results obtained with NRSur7dq4, and where analyses were conducted using the IMRPhenomXO4a model for an event, these are also shown. For the high-mass candidate GW231123_135430, we additionally use the time-domain phenomenological model IMRPhenomTPHM (H. Estellés et al. 2022a), and samples are combined across the five models (A. G. Abac et al. 2025b). Here we focus on systematic

**Table 5**
The On-source and the Median Off-source Overlap (with 90% Confidence Intervals) Values and $p$-values for Three Minimally Modeled Waveform Reconstruction Methods on a Selection of Candidates

| Candidate | BayesWave | | | cWB-2G | | | cWB-BBH | | |
|---|---|---|---|---|---|---|---|---|---|
| | On-source | Off-source | $p$-value | On-source | Off-source | $p$-value | On-source | Off-source | $p$-value |
| GW230601_224134 | 0.85 | $0.89^{+0.06}_{-0.17}$ | 0.34 | 0.91 | $0.91^{+0.04}_{-0.07}$ | 0.43 | 0.93 | $0.91^{+0.04}_{-0.07}$ | 0.83 |
| GW230606_004305 | 0.91 | $0.78^{+0.11}_{-0.40}$ | 0.97 | 0.92 | $0.85^{+0.06}_{-0.11}$ | 0.99 | 0.94 | $0.86^{+0.06}_{-0.11}$ | 0.99 |
| GW230608_205047 | 0.80 | $0.83^{+0.09}_{-0.18}$ | 0.39 | 0.89 | $0.86^{+0.06}_{-0.12}$ | 0.73 | 0.91 | $0.86^{+0.06}_{-0.10}$ | 0.91 |
| GW230609_064958 | 0.75 | $0.71^{+0.16}_{-0.39}$ | 0.60 | 0.79 | $0.82^{+0.07}_{-0.14}$ | 0.28 | 0.92 | $0.84^{+0.07}_{-0.13}$ | 0.99 |
| GW230628_231200 | 0.93 | $0.89^{+0.05}_{-0.14}$ | 0.84 | 0.92 | $0.90^{+0.03}_{-0.05}$ | 0.79 | 0.92 | $0.91^{+0.03}_{-0.06}$ | 0.64 |
| GW230702_185453 | 0.60 | $0.64^{+0.22}_{-0.60}$ | 0.40 | 0.82 | $0.78^{+0.08}_{-0.14}$ | 0.73 | 0.88 | $0.81^{+0.08}_{-0.13}$ | 0.90 |
| GW230707_124047 | 0.88 | $0.85^{+0.08}_{-0.19}$ | 0.67 | 0.90 | $0.88^{+0.05}_{-0.09}$ | 0.69 | 0.91 | $0.89^{+0.05}_{-0.10}$ | 0.72 |
| GW230708_230935 | 0.80 | $0.78^{+0.14}_{-0.70}$ | 0.56 | 0.89 | $0.86^{+0.06}_{-0.12}$ | 0.80 | 0.93 | $0.87^{+0.06}_{-0.13}$ | 0.95 |
| GW230814_061920 | 0.77 | $0.82^{+0.11}_{-0.43}$ | 0.26 | 0.85 | $0.88^{+0.06}_{-0.10}$ | 0.31 | 0.87 | $0.89^{+0.05}_{-0.09}$ | 0.30 |
| GW230824_033047 | 0.89 | $0.84^{+0.08}_{-0.24}$ | 0.74 | 0.91 | $0.88^{+0.06}_{-0.09}$ | 0.71 | 0.96 | $0.89^{+0.05}_{-0.09}$ | 1.00 |
| GW230914_111401 | 0.95 | $0.93^{+0.03}_{-0.07}$ | 0.85 | 0.92 | $0.93^{+0.02}_{-0.04}$ | 0.29 | 0.95 | $0.94^{+0.03}_{-0.04}$ | 0.76 |
| GW230920_071124 | 0.83 | $0.76^{+0.14}_{-0.65}$ | 0.65 | 0.80 | $0.80^{+0.08}_{-0.14}$ | 0.48 | 0.84 | $0.83^{+0.07}_{-0.15}$ | 0.58 |
| GW230922_020344 | 0.79 | $0.86^{+0.08}_{-0.24}$ | 0.22 | 0.73 | $0.80^{+0.07}_{-0.19}$ | 0.16 | 0.84 | $0.84^{+0.05}_{-0.12}$ | 0.52 |
| GW230922_040658 | 0.87 | $0.89^{+0.06}_{-0.13}$ | 0.41 | 0.92 | $0.91^{+0.04}_{-0.08}$ | 0.75 | 0.94 | $0.92^{+0.03}_{-0.08}$ | 0.74 |
| GW230924_124453 | 0.84 | $0.81^{+0.09}_{-0.31}$ | 0.67 | 0.86 | $0.84^{+0.06}_{-0.10}$ | 0.71 | 0.87 | $0.86^{+0.05}_{-0.10}$ | 0.55 |
| GW230927_043729 | 0.83 | $0.79^{+0.11}_{-0.32}$ | 0.72 | 0.88 | $0.83^{+0.07}_{-0.12}$ | 0.89 | 0.90 | $0.85^{+0.06}_{-0.13}$ | 0.88 |
| GW231028_153006 | 0.99 | $0.97^{+0.02}_{-0.04}$ | 0.99 | 0.97 | $0.96^{+0.01}_{-0.02}$ | 0.74 | 0.97 | $0.97^{+0.01}_{-0.02}$ | 0.42 |
| GW231102_071736 | 0.91 | $0.92^{+0.04}_{-0.10}$ | 0.43 | 0.95 | $0.93^{+0.03}_{-0.05}$ | 0.88 | 0.96 | $0.94^{+0.03}_{-0.05}$ | 0.97 |
| GW231123_135430 | 0.97 | $0.96^{+0.02}_{-0.06}$ | 0.74 | 0.96 | $0.96^{+0.01}_{-0.03}$ | 0.57 | 0.98 | $0.96^{+0.01}_{-0.03}$ | 0.92 |
| GW231206_233134 | 0.89 | $0.87^{+0.07}_{-0.24}$ | 0.69 | 0.84 | $0.86^{+0.06}_{-0.10}$ | 0.36 | 0.88 | $0.88^{+0.05}_{-0.11}$ | 0.48 |
| GW231206_233901 | 0.96 | $0.95^{+0.02}_{-0.03}$ | 0.63 | 0.94 | $0.93^{+0.02}_{-0.03}$ | 0.76 | 0.96 | $0.95^{+0.02}_{-0.04}$ | 0.78 |
| GW231213_111417 | 0.84 | $0.74^{+0.15}_{-0.53}$ | 0.82 | 0.85 | $0.83^{+0.07}_{-0.12}$ | 0.66 | 0.90 | $0.85^{+0.06}_{-0.14}$ | 0.89 |
| GW231226_101520 | 0.99 | $0.98^{+0.01}_{-0.01}$ | 0.88 | 0.96 | $0.97^{+0.01}_{-0.01}$ | 0.19 | 0.97 | $0.97^{+0.01}_{-0.01}$ | 0.34 |

**Note.** The three minimally modeled waveform reconstruction methods used were BayesWave, cWB-2G (both designed for generic GW bursts), and cWB-BBH (specifically optimized for CBC signals with tailored frequency bands and time–frequency resolutions). The $p$-values are calculated by comparing the on-source value with the off-source distribution of overlaps.

differences among the two to four fiducial models used to analyze the bulk of our BBH candidates.

We find that for almost all the signals analyzed here, the differences between results obtained using different models are small compared to their statistical uncertainties. As for previous observations, differences are typically small and most noticeable for parameters like the spins (B. P. Abbott et al. 2016d, 2019a, 2019d, 2021b). For candidates that show significant differences between waveform models, potentially due to systematic uncertainties from waveform modeling, we plot a selection of the posteriors for all models in Figure 7. Additionally, in Table 4, we show the median and 90% credible intervals for these source parameters as inferred by each waveform model used in our analysis, along with our fiducial inferences from combining the samples across models (A. G. Abac et al. 2026a). The candidates we identified with significant systematics between waveform models are the following:

1. *GW230624_113103, a BBH candidate, whose source has a total mass $M = 43.8^{+11.1}_{-6.6}\,M_\odot$.* The inferred values of its mass ratio $q$ show visible differences between IMRPhenomXPHM_SpinTaylor and SEOBNRv5PHM results, with SEOB-NRv5PHM favoring more asymmetric masses. This results in systematic differences in the inferred component masses. These differences are correlated with our knowledge of the spin parameters of the source, and we favor larger values of both $\chi_{\mathrm{eff}}$ and $\chi_{\mathrm{p}}$ when analyzing this candidate with SEOBNRv5PHM as compared to IMRPhenomXPHM_SpinTaylor.
2. *GW231028_153006, which displays significant systematic variations in its source mass and spin inferences across waveform models.* For this candidate, no data-quality issues requiring mitigation were identified. We apply all four of our BBH waveform models in our inferences for GW231028_153006, and none of the models show close agreement with each other across component masses. Systematic differences in our inferences of $\chi_{\mathrm{eff}}$ are also visible, and a subdominant, second mode at high $\chi_{\mathrm{p}}$ values is visible in the samples drawn with IMRPhenomX04a. These systematic differences broaden our uncertainties in our overall combined results for this candidate, but it is clear that its source is massive, with $M = 152^{+29}_{-14}\,M_\odot$ and $m_1 = 95^{+33}_{-20}\,M_\odot$ . The source of this candidate probably has a large $\chi_{\mathrm{eff}}$ and has support for a large primary spin $\chi_1$.
3. *GW231118_005626, a BBH candidate, whose source has $M = 30.9^{+5.3}_{-3.6}\,M_\odot$ and credibly unequal masses $q = 0.55^{+37}_{-22}$.* The source has a large and well-measured effective inspiral spin, $\chi_{\mathrm{eff}} = 0.37^{+0.10}_{-0.11}$. For this candidate, the inferences drawn with SEOBNRv5PHM and IMRPhenomX04a are in good agreement but display some differences with those drawn using the IMRPhenomXPHM_SpinTaylor model. The total mass and chirp mass posteriors display a heavier tail toward larger values for IMRPhenomXPHM_SpinTaylor. More apparent are the differences in the $\chi_1$ and $\chi_p$ posteriors, which for IMRPhenomXPHM_SpinTaylor extend toward higher values than those obtained with the other two models.

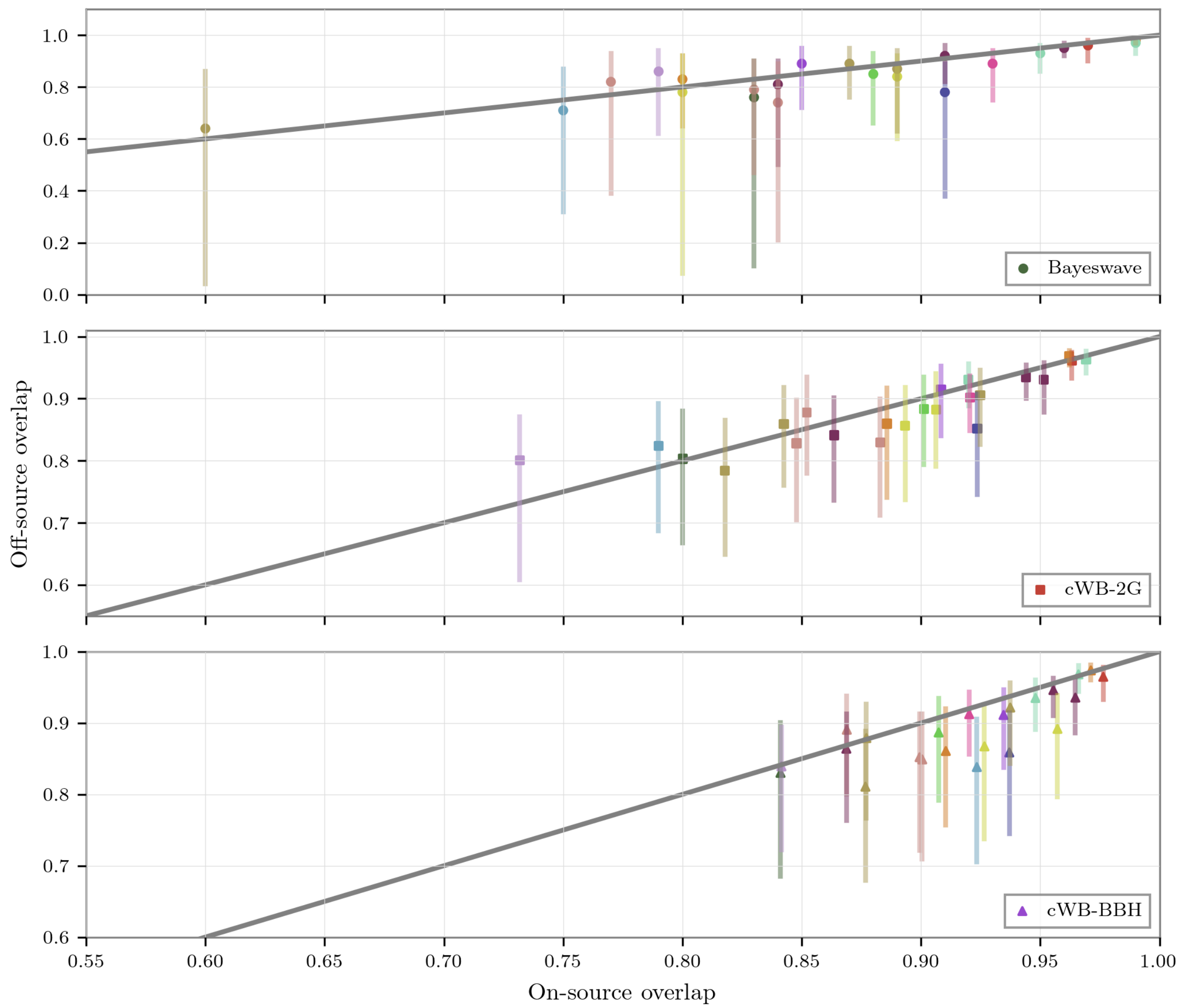

**Figure 8.** A comparison of the overlap between the on-source and off-source (with 90% confidence intervals) reconstructions for three different minimally modeled pipelines, BayesWave, cWB-2G, and cWB-BBH for candidates that had data from both interferometers and where the NRSur7dq4 waveform was used in parameter estimation, the network SNR was greater than 10, and the redshifted chirp mass $(1 + z)\mathcal{M} > 15 M_\odot$. The gray line denotes equal overlap between the on- and off-source reconstructions, indicating that there is no significant difference between the two.

4. *GW231118_090602, whose source is a relatively low-mass BBH with $M = 20.7^{+10.2}_{-2.3} M_\odot$ and mass ratio $q = 0.56^{+0.38}_{-0.41}$.* This candidate displays multimodal source mass and spin posteriors, as discussed in Section 3.5. This multimodality is present only in the posterior samples drawn using the IMRPhenomXPHM_SpinTaylor model, which includes a distinct high-likelihood mode at $q \sim 0.15$, which correlates with larger values of $\chi_{\rm eff} \sim 0.4$. This mode at asymmetric masses also correlates with systematically larger $\chi_p$ values than the broader mode at more equal masses, which has $0.3 \lesssim q \lesssim 1$. Meanwhile the samples drawn using SEOBNRv5PHM are unimodal in masses and spins and largely agree with the higher-$q$ mode from IMRPhenomXPHM_SpinTaylor, although they favor a larger median $\chi_p$ than the IMRPhenomXPHM_SpinTaylor results.
5. *GW231123_135430 displays significant systematic differences in its inferred source mass and spins, depending on the waveform model* (A. G. Abac et al. 2025b).

One possible source of systematic differences in our inferences is signal content that may be absent from our default CBC models. A method for testing for any missing signal content, or even to discover unexpected phenomena, is to estimate the overlap between the modeled reconstructions of our GW signals and the minimally modeled waveform reconstructions (A. G. Abac et al. 2026a). We test for missing signal content by selecting a subset of our O4a GW candidates and comparing their source inferences generated with the NRSur7dq4 waveform (V. Varma et al. 2019) to reconstructions made with minimal assumptions about the waveform morphology. We select 23 candidates using criteria to optimize performance of the minimally modeled reconstruction methods (A. G. Abac et al. 2026a).

To assess the statistical significance of the overlap between these reconstructions, we perform systematic injection studies (B. P. Abbott et al. 2019a; F. Salemi et al. 2019; S. Ghonge et al. 2020; R. Abbott et al. 2021b; N. K. Johnson-McDaniel et al. 2022).

We inject simulated signals with parameters drawn from the inferred posterior distributions[332] into nearby detector data not overlapping the candidate time. These off-source waveforms are then reconstructed using minimally modeled methods. By comparing the overlaps between the on-source reconstruction of the actual candidate and the distribution of off-source overlaps, we compute a $p$-value indicating the fraction of off-source overlaps greater than or equal to the on-source overlap.

We use three methods for reconstructing the signals with minimal assumptions about their morphology: BayesWave (N. J. Cornish & T. B. Littenberg 2015; T. B. Littenberg & N. J. Cornish 2015; N. J. Cornish et al. 2021; T. Gupta & N. J. Cornish 2024) and cWB-2G (S. Klimenko et al. 2016; M. Drago et al. 2021), which are designed for generic GW transients, and cWB-BBH (S. Klimenko 2022), which is optimized specifically for CBC signals. The results, summarized in Table 5 and shown in Figure 8, show no significant deviations between the on-source and off-source reconstructions across all three methods. However, as noted in A. G. Abac et al. (2026a), all three pipelines show some level of bias in their respective tests to assess the $p$-values.

Taken together, these studies show that while a few of the new candidates added to GWTC-4.0 show noticeable systematic uncertainties in our inferences of their source properties, there is as yet no strong evidence for missing signal content in our models.

## 4. Conclusion

We present GWTC-4.0, which contains 218 CBC events with $p_{\rm astro} \geqslant 0.5$ and which we find unlikely to be of instrumental origin after further event validation. Analyzing data from O4a, this version of the catalog adds 128 GW candidates consistent with BBHs and NSBHs to the cumulative catalog of events, more than doubling the census of CBCs passing these criteria from the first three observing runs (R. Abbott et al. 2023a). For high-significance candidates with FARs $< 1\ {\rm yr}^{-1}$, we also estimate their source properties. These include GW230518_125908, a new NSBH signal observed during the engineering run preceding O4a with detailed analysis presented for the first time in this catalog; GW230529_181500, which most likely originates from an NSBH binary with a primary mass $\leqslant 5 M_\odot$ (A. G. Abac et al. 2024b); GW230814_061920, the highest-SNR event observed so far (A. G. Abac et al. 2026c); and GW231123_135430, the most massive BBH in the catalog (A. G. Abac et al. 2025b) with an FAR $< 1\ {\rm yr}^{-1}$.

Additional results related to candidates in the catalog are interpreted in other papers of the The LIGO-Virgo-KAGRA Gravitational Wave Transient Catalog (GWTC): Release 4 focus issue. This includes inferring the mass and spin distributions of the CBCs we have observed (A. G. Abac et al. 2026d), testing general relativity in the strong-field regime (A. G. Abac et al. 2026f, 2026g, 2026h), providing independent measures of local cosmology (A. G. Abac et al. 2026e), and searching for gravitationally lensed counterparts to our candidates (A. G. Abac et al. 2026i).

The data products associated with the results described here, as well as a complete list of all candidates with FARs $\leqslant 2\ {\rm day}^{-1}$ and the underlying strain data, are publicly available through GWOSC and are described in detail in A. G. Abac et al. (2026b). Past releases of the public strain data have led to additional GW candidates (R. Magee et al. 2019; A. H. Nitz et al. 2019, 2020a, 2021, 2023; T. Venumadhav et al. 2019, 2020; B. Zackay et al. 2019, 2021; S. Olsen et al. 2022; P. Kumar & T. Dent 2024; A. E. Koloniari et al. 2025; T. Mishra et al. 2025), and the data products have enabled myriad studies probing the nature of individual detections, population properties, and other astrophysical inferences.

On 2024 January 16, the LIGO interferometers paused operations to begin a commissioning break to further improve sensitivity. The second part of the fourth observing run (O4b) began on 2024 April 10 during which the LIGO detectors were joined by the Virgo detector (F. Acernese et al. 2015). Then, on 2025 January 28 the third part of the fourth observing run (O4c) started. On 2025 June 11, during this observing run, the KAGRA detector (T. Akutsu et al. 2019) also rejoined the network. Increasing the number of detectors will greatly improve the ability of the network to infer source localizations and also provide a higher rate of detections; analysis of this data will be presented in a future version of the GWTC releases.

In the coming years, the LVK network will undergo additional upgrades to further improve its sensitivity and uncover hitherto hidden parts of the gravitational universe (B. P. Abbott et al. 2020a). Future GW transients could include novel CBC sources, such as subsolar-mass compact binaries (B. P. Abbott et al. 2018, 2019e, 2022a, 2023c; A. H. Nitz & Y. F. Wang 2021a, 2021b) or other exotica but also new classes of GW transients such as supernovae (R. Abbott et al. 2021c), cosmic strings (R. Abbott et al. 2021d), and bursts of unknown origin (R. Abbott et al. 2021g; A. G. Abac et al. 2025c). We additionally anticipate long-lived GW signals from rapidly rotating NSs (R. Abbott et al. 2022b; A. G. Abac et al. 2025c) and the stochastic background (R. Abbott et al. 2021e, 2021f) of the Universe. As the detectors improve in sensitivity, we therefore expect to deepen our understanding of the Universe.

## Acknowledgments

This material is based upon work supported by NSF's LIGO Laboratory, which is a major facility fully funded by the National Science Foundation. The authors also gratefully acknowledge the support of the Science and Technology Facilities Council (STFC) of the United Kingdom, the Max–Planck–Society (MPS), and the State of Niedersachsen/Germany for support of the construction of Advanced LIGO and construction and operation of the GEO 600 detector. Additional support for Advanced LIGO was provided by the Australian Research Council. The authors gratefully acknowledge the Italian Istituto Nazionale di Fisica Nucleare (INFN), the French Centre National de la Recherche Scientifique (CNRS), and the Netherlands Organization for Scientific

[332] These results were generated prior to a correction to the likelihood used for inference (A. G. Abac et al. 2026a, Section 5.10), using the posterior distributions published in version 1 of this work's data release (LIGO Scientific Collaboration et al. 2025b). We recomputed overlaps and $p$-values for a subset of events using the corrected posteriors and found no significant changes; the differences in the posteriors are small enough that they do not affect the conclusions of this analysis.

Research (NWO) for the construction and operation of the Virgo detector and the creation and support of the EGO consortium. The authors also gratefully acknowledge research support from these agencies, as well as by the Council of Scientific and Industrial Research of India; the Department of Science and Technology, India; the Science & Engineering Research Board (SERB), India; the Ministry of Human Resource Development, India; the Spanish Agencia Estatal de Investigación (AEI); the Spanish Ministerio de Ciencia, Innovación y Universidades; the European Union NextGenerationEU/PRTR (PRTR-C17.I1); the ICSC—CentroNazionale di Ricerca in High Performance Computing, Big Data and Quantum Computing, funded by the European Union NextGenerationEU; the Comunitat Autonòma de les Illes Balears through the Conselleria d'Educació i Universitats; the Conselleria d'Innovació, Universitats, Ciència i Societat Digital de la Generalitat Valen-ciana and the CERCA Programme Generalitat de Catalunya, Spain; the Polish National Agency for Academic Exchange; the National Science Centre of Poland and the European Union—European Regional Development Fund; the Foundation for Polish Science (FNP); the Polish Ministry of Science and Higher Education; the Swiss National Science Foundation (SNSF); the Russian Science Foundation; the European Commission; the European Social Funds (ESF); the European Regional Development Funds (ERDF); the Royal Society; the Scottish Funding Council; the Scottish Universities Physics Alliance; the Hungarian Scientific Research Fund (OTKA); the French Lyon Institute of Origins (LIO); the Belgian Fonds de la Recherche Scientifique (FRS-FNRS), Actions de Recherche Concertées (ARC) and Fonds Wetenschappelijk Onderzoek—Vlaanderen (FWO), Belgium; the Paris Île-de-France Region; the National Research, Development and Innovation Office of Hungary (NKFIH); the National Research Foundation of Korea; the Natural Sciences and Engineering Research Council of Canada (NSERC); the Canadian Foundation for Innovation (CFI); the Brazilian Ministry of Science, Technology, and Innovations; the International Center for Theoretical Physics South American Institute for Fundamental Research (ICTP-SAIFR); the Research Grants Council of Hong Kong; the National Natural Science Foundation of China (NSFC); the Israel Science Foundation (ISF); the US–Israel Binational Science Fund (BSF); the Leverhulme Trust; the Research Corporation; the National Science and Technology Council (NSTC), Taiwan; the United States Department of Energy; and the Kavli Foundation. The authors gratefully acknowledge the support of the NSF, STFC, INFN, and CNRS for provision of computational resources.

This work was supported by MEXT; the JSPS Leading-edge Research Infrastructure Program, JSPS Grant-in-Aid for Specially Promoted Research 26000005, JSPS Grant-in-Aid for Scientific Research on Innovative Areas 2402: 24103006, 24103005, and 2905: JP17H06358, JP17H06361, and JP17H06364, JSPS Core-to-Core Program A. Advanced Research Networks, JSPS Grants-in-Aid for Scientific Research (S) 17H06133 and 20H05639, JSPS Grant-in-Aid for Transformative Research Areas (A) 20A203: JP20H05854; the joint research program of the Institute for Cosmic Ray Research, University of Tokyo; the National Research Foundation (NRF); the Computing Infrastructure Project of the Global Science experimental Data hub Center (GSDC) at KISTI; the Korea Astronomy and Space Science Institute (KASI); the Ministry of Science and ICT (MSIT) in Korea; Academia Sinica (AS), the AS Grid Center (ASGC), and the National Science and Technology Council (NSTC) in Taiwan under grants including the Science Vanguard Research Program; the Advanced Technology Center (ATC) of NAOJ; and the Mechanical Engineering Center of KEK.

Additional acknowledgments for support of individual authors may be found in the following document: https://dcc.ligo.org/LIGO-M2300033/public. For the purpose of open access, the authors have applied a Creative Commons Attribution (CC BY) license to any Author Accepted Manuscript version arising. We request that citations to this Letter use "A. G. Abac et al. (LIGO–Virgo–KAGRA Collaboration), …" or similar phrasing, depending on journal convention.

*Facility*: LIGO.

*Software:* Calibration of the LIGO strain data was performed with a GSTLAL-based calibration software pipeline (A. D. Viets et al. 2018). Data-quality products and event-validation results were computed using the DMT (J. Zweizig 2006), DQR (LIGO Scientific Collaboration & Virgo Collaboration 2018), DQSEGDB (R. P. Fisher et al. 2021), GWDETCHAR (D. Macleod et al. 2021b), HVETO (J. R. Smith et al. 2011), iDQ (R. Essick et al. 2020), OMICRON (F. Robinet et al. 2020) and PYTHONVIRGOTOOLS (Virgo Collaboration 2021) software packages and contributing software tools. Analyses in this catalog relied upon the LALSUITE software library (LIGO–Virgo–KAGRA Collaboration 2018; K. Wette 2020). The detection of the signals and subsequent significance evaluations in this catalog were performed with the GSTLAL-based inspiral software pipeline (C. Messick et al. 2017; S. Sachdev et al. 2019; C. Hanna et al. 2020; K. Cannon et al. 2021), with the MBTA pipeline (T. Adams et al. 2016; F. Aubin et al. 2021), and with the PYCBC (S. A. Usman et al. 2016; A. H. Nitz et al. 2017; G. S. Davies et al. 2020) and the CWB (S. Klimenko & G. Mitselmakher 2004; S. Klimenko et al. 2011, 2016) packages. Estimates of the noise spectra and glitch models were obtained using BAYESWAVE (N. J. Cornish & T. B. Littenberg 2015; T. B. Littenberg et al. 2016; N. J. Cornish et al. 2021; T. Gupta & N. J. Cornish 2024). Source-parameter estimation was performed with the BILBY library (G. Ashton et al. 2019; I. M. Romero-Shaw et al. 2020) using the DYNESTY nested sampling package (J. S. Speagle 2020). PESUMMARY was used to postprocess and collate parameter-estimation results (C. Hoy & V. Raymond 2021). The various stages of the parameter-estimation analysis were managed with the ASIMOV library (D. Williams et al. 2023). Plots were prepared with MATPLOTLIB (J. D. Hunter 2007), SEABORN (M. Waskom 2021) and GWPY (D. M. Macleod et al. 2021a). NUMPY (C. R. Harris et al. 2020) and SCIPY (P. Virtanen et al. 2020) were used in the preparation of the manuscript.

## Data Availability

All strain data analyzed as part of GWTC-4.0 are publicly available through GWOSC. The details of this data release and information about the digital version of the GWTC are described in detail in A. G. Abac et al. (2026b). We also provide data releases of the search pipeline results and initial source localization (LIGO–Virgo–KAGRA Collaboration 2025b), parameter-estimation samples (LIGO–Virgo–KAGRA Collaboration 2025c), glitch modeling (LIGO–Virgo–KAGRA Collaboration 2025d), and data-quality products (LIGO–Virgo–KAGRA Collaboration 2025e) Finally, we also note that the search sensitivity estimates are also publicly available (LIGO–Virgo–KAGRA Collaboration 2025f, 2025g).

## Appendix A
## Additional Search Results

We present the individual-detector SNRs for all candidates with FAR $\leqslant 1\ \mathrm{yr}^{-1}$ in Table 6, extending the information provided in Table 1 and offering additional context on the search pipeline responses to the signals observed in data from individual detectors.

In Table 7, we provide the calculated probabilities that a candidate comes from a BBH ($p_{\mathrm{BBH}}$), an NSBH ($p_{\mathrm{NSBH}}$), or a BNS ($p_{\mathrm{BNS}}$) for all new candidates in GWTC-4.0 with the maximum $p_{\mathrm{astro}} > 0.5$ and minimum FARs $> 1\ \mathrm{yr}^{-1}$ across pipelines. We only show systems where $p_{\mathrm{NSBH}} + p_{\mathrm{BNS}} > 0.001$; the remaining marginal candidates are consistent with BBHs. GstLAL estimates the relative probabilities of different astrophysical source types using the component masses of the matching template and SNR of the signal (A. Ray et al. 2023); MBTA uses the chirp mass and mass ratio of the identifying template (N. Andres et al. 2022), and PyCBC only uses the chirp mass, along with an estimate of the source luminosity distance (T. Dal Canton et al. 2021; V. Villa-Ortega et al. 2022). Neglecting mass ratio information can systematically impact the source categorization (V. Villa--Ortega et al. 2022). The full details of the $p_{\mathrm{astro}}$ calculations are described in A. G. Abac et al. (2026a).

**Table 6**
Individual-detector SNRs for All Candidates with FAR $\leqslant 1\ \mathrm{yr}^{-1}$

| Candidate | cWB-BBH | | GstLAL | | MBTA | | PyCBC | |
|---|---|---|---|---|---|---|---|---|
| | H | L | H | L | H | L | H | L |
| GW230518_125908 | … | … | 10.0 | 9.3 | 10.6 | 9.4 | 10.2 | 9.0 |
| GW230529_181500 | … | … | … | 11.8 | … | 11.4 | … | 11.7 |
| GW230601_224134 | 9.3 | 9.6 | 8.6 | 8.2 | 8.4 | 9.1 | 8.7 | 8.5 |
| GW230605_065343 | *4.4* | *6.1* | 6.8 | 8.3 | 6.7 | 8.8 | 7.2 | 8.8 |
| GW230606_004305 | 9.1 | 6.3 | 9.4 | 5.4 | *9.3* | *5.7* | 9.4 | 5.2 |
| GW230608_205047 | 8.1 | 5.8 | 8.2 | 6.2 | 8.3 | 5.9 | … | … |
| GW230609_064958 | 6.9 | 8.1 | 5.6 | 8.2 | *5.7* | *8.7* | 5.7 | 7.7 |
| GW230624_113103 | 8.5 | 7.6 | 7.8 | 6.3 | 8.1 | 6.4 | 8.1 | 6.1 |
| GW230627_015337 | 21.4 | 17.7 | 22.1 | 17.8 | 21.7 | 18.3 | 22.0 | 18.4 |
| GW230628_231200 | 12.3 | 10.8 | 12.2 | 9.3 | 12.5 | 9.8 | 12.8 | 9.4 |
| GW230630_070659 | … | … | 4.4 | 8.7 | … | … | … | … |
| GW230630_125806 | 6.1 | 6.7 | 5.4 | 6.0 | *5.0* | *6.5* | 5.2 | 6.2 |
| GW230630_234532 | … | … | 7.3 | 6.4 | 7.4 | 6.7 | 7.2 | 6.6 |
| GW230702_185453 | 7.1 | 7.3 | 6.2 | 7.6 | 5.9 | 7.9 | 5.6 | 7.3 |
| GW230704_021211 | … | … | 5.0 | 7.9 | *5.0* | *7.8* | 5.1 | 7.6 |
| GW230704_212616 | *5.6* | *6.2* | *4.5* | *7.0* | 4.6 | 7.3 | … | … |
| GW230706_104333 | … | … | 5.7 | 7.3 | … | … | *5.9* | *6.6* |
| GW230707_124047 | 8.4 | 8.4 | 5.7 | 8.4 | 5.7 | 8.6 | 6.2 | 8.4 |
| GW230708_053705 | … | … | *6.7* | *5.4* | *7.1* | *5.3* | 7.1 | 5.4 |
| GW230708_230935 | *8.1* | *5.9* | 7.6 | 5.9 | 7.6 | 6.0 | 7.4 | 5.8 |
| GW230709_122727 | 7.2 | 7.2 | 6.9 | 7.1 | *7.3* | *7.0* | 6.8 | 7.3 |
| GW230712_090405 | 5.8 | 7.5 | *4.0* | *7.2* | … | … | *4.5* | *6.9* |
| GW230723_101834 | … | … | 7.4 | 6.6 | 7.4 | 6.7 | 7.6 | 6.6 |
| GW230726_002940 | … | … | … | 10.5 | … | … | … | *10.0* |
| GW230729_082317 | … | … | 7.0 | 6.4 | … | … | *7.0* | *6.3* |
| GW230731_215307 | … | … | 8.5 | 8.8 | 8.0 | 8.9 | 8.0 | 8.8 |
| GW230803_033412 | *7.1* | *6.0* | *5.8* | *5.5* | *6.6* | *5.6* | 6.1 | 5.5 |
| GW230805_034249 | *6.3* | *7.2* | 6.1 | 7.0 | *6.3* | *7.0* | 6.1 | 7.1 |
| GW230806_204041 | 7.0 | 6.3 | 7.3 | 5.4 | 7.2 | 6.0 | 6.9 | 5.9 |
| GW230811_032116 | 9.0 | 10.2 | 6.7 | 11.0 | 6.8 | 11.4 | 6.4 | 10.6 |
| GW230814_061920 | 8.8 | 7.0 | 8.1 | 6.2 | 7.9 | 6.2 | 7.1 | 6.4 |
| GW230814_230901 | … | … | … | 42.3 | … | … | … | 43.0 |
| GW230819_171910 | 7.3 | 6.7 | 7.1 | 5.6 | … | … | *6.9* | *5.7* |
| GW230820_212515 | *5.9* | *5.2* | 7.5 | 5.1 | 7.6 | 5.3 | 7.4 | 5.1 |
| GW230824_033047 | 7.8 | 7.8 | 6.8 | 8.0 | 6.9 | 8.1 | 6.6 | 8.4 |
| GW230825_041334 | *6.3* | *6.2* | 7.0 | 5.1 | *6.8* | *5.1* | 6.9 | 5.2 |
| GW230831_015414 | *5.6* | *5.4* | 6.9 | 5.2 | *6.7* | *5.4* | 6.5 | 5.4 |
| GW230904_051013 | … | … | 6.8 | 8.0 | 6.1 | 8.5 | 6.2 | 8.1 |
| GW230911_195324 | … | … | 10.7 | … | … | … | 11.1 | … |
| GW230914_111401 | 10.7 | 13.4 | 10.2 | 12.2 | 10.5 | 12.9 | 10.3 | 12.2 |
| GW230919_215712 | 12.6 | 11.2 | 11.9 | 11.1 | 11.4 | 11.4 | 11.5 | 11.8 |
| GW230920_071124 | 8.0 | 7.7 | 7.1 | 7.2 | 7.2 | 7.2 | 6.9 | 6.6 |
| GW230922_020344 | 8.9 | 9.9 | 6.6 | 10.4 | 6.6 | 10.3 | 6.5 | 10.0 |
| GW230922_040658 | 8.8 | 8.8 | 7.6 | 8.7 | 7.6 | 8.7 | 8.1 | 8.3 |
| GW230924_124453 | 8.6 | 10.3 | 9.9 | 8.8 | 9.7 | 9.1 | 9.7 | 8.8 |

**Table 6**
(Continued)

| Candidate | cWB-BBH | | GSTLAL | | MBTA | | PyCBC | |
|---|---|---|---|---|---|---|---|---|
| | H | L | H | L | H | L | H | L |
| GW230927_043729 | 9.1 | 7.9 | 8.9 | 7.0 | 8.8 | 6.8 | 8.7 | 6.9 |
| GW230927_153832 | 13.5 | 15.1 | 11.8 | 16.0 | 12.1 | 16.2 | 11.6 | 15.8 |
| GW230928_215827 | 7.8 | 7.1 | 6.7 | 6.7 | *6.9* | *6.3* | 6.7 | 6.8 |
| GW230930_110730 | *6.4* | *6.2* | 5.9 | 6.1 | *6.0* | *6.1* | 6.1 | 5.7 |
| GW231001_140220 | 8.6 | 7.6 | 7.2 | 7.4 | 7.6 | 7.4 | 7.2 | 6.7 |
| GW231004_232346 | 5.2 | 7.3 | *4.0* | *6.8* | … | … | … | *7.0* |
| GW231005_021030 | 7.1 | 7.6 | 6.2 | 6.9 | 6.6 | 7.2 | 6.6 | 7.3 |
| GW231005_091549 | *8.2* | *7.4* | 6.3 | 6.2 | *6.1* | *6.1* | *6.2* | *5.9* |
| GW231008_142521 | … | … | 6.7 | 6.5 | *6.2* | *6.6* | 6.1 | 6.2 |
| GW231014_040532 | *5.5* | *6.6* | 5.9 | 6.8 | *6.1* | *6.2* | *6.2* | *6.1* |
| GW231018_233037 | … | … | *5.4* | *6.8* | 5.3 | 7.4 | *5.0* | *7.0* |
| GW231020_142947 | … | … | 9.9 | 6.5 | 10.2 | 6.3 | 9.9 | 6.5 |
| GW231028_153006 | 14.1 | 17.4 | 12.0 | 17.2 | 12.5 | 18.0 | 11.8 | 18.4 |
| GW231029_111508 | … | … | … | 10.8 | … | … | … | … |
| GW231102_071736 | 11.0 | 11.1 | 9.8 | 9.7 | 10.3 | 10.6 | 8.8 | 10.1 |
| GW231104_133418 | … | … | 7.7 | 8.3 | 7.8 | 8.4 | 8.0 | 8.7 |
| GW231108_125142 | 8.1 | 9.7 | 8.1 | 9.7 | 8.4 | 9.3 | 8.0 | 9.3 |
| GW231110_040320 | … | … | 6.7 | 9.2 | 6.8 | 9.3 | 6.7 | 8.9 |
| GW231113_122623 | … | … | 5.6 | 6.2 | *5.4* | *6.6* | 5.6 | 6.5 |
| GW231113_200417 | … | … | 8.1 | 6.3 | 8.0 | 6.2 | 8.3 | 6.5 |
| GW231114_043211 | … | … | 7.4 | 6.7 | 7.4 | 6.5 | 7.6 | 6.0 |
| GW231118_005626 | … | … | 7.5 | 7.2 | 7.7 | 7.4 | 7.4 | 7.4 |
| GW231118_071402 | 6.0 | 7.0 | 7.1 | 5.9 | 6.8 | 6.2 | 6.9 | 6.1 |
| GW231118_090602 | … | … | 7.8 | 7.4 | 7.8 | 7.8 | 7.7 | 7.6 |
| GW231119_075248 | *5.4* | *5.7* | 5.4 | 6.0 | *5.3* | *6.0* | 5.2 | 6.5 |
| GW231123_135430 | 13.3 | 17.2 | 13.1 | 15.2 | 10.9 | 15.5 | 12.2 | 15.8 |
| GW231127_165300 | 7.2 | 6.9 | 8.0 | 5.6 | 7.9 | 5.4 | 7.9 | 5.4 |
| GW231129_081745 | 7.0 | 6.3 | 6.7 | 5.1 | *6.6* | *5.2* | *6.8* | *5.2* |
| GW231206_233134 | 9.1 | 9.0 | 7.0 | 9.6 | 6.6 | 9.7 | 6.9 | 9.2 |
| GW231206_233901 | 12.2 | 18.2 | 12.1 | 16.9 | 11.8 | 17.8 | 11.7 | 17.5 |
| GW231213_111417 | 6.6 | 7.5 | 6.3 | 8.0 | 6.5 | 8.1 | 6.3 | 7.9 |
| GW231221_135041 | 7.8 | 6.2 | *7.1* | *4.5* | *6.4* | *4.9* | *7.1* | *4.4* |
| GW231223_032836 | 7.2 | 7.3 | 6.5 | 6.8 | *6.5* | *6.4* | 6.5 | 6.2 |
| GW231223_075055 | … | … | *7.0* | *6.2* | *7.0* | *6.2* | 7.2 | 6.1 |
| GW231223_202619 | … | … | *10.0* | … | … | … | 10.0 | … |
| GW231224_024321 | … | … | 8.1 | 10.2 | 8.6 | 11.0 | 8.4 | 10.3 |
| GW231226_101520 | 26.9 | 21.9 | 26.2 | 22.0 | 26.4 | 20.8 | 26.3 | 20.3 |
| GW231230_170116 | 6.3 | 5.3 | *6.4* | *4.7* | … | … | … | … |
| GW231231_154016 | … | … | 13.4 | … | … | … | 13.4 | … |
| GW240104_164932 | … | … | 14.8 | … | … | … | 12.2 | … |
| GW240107_013215 | 7.8 | 5.3 | 7.1 | 5.6 | 8.0 | 5.3 | 7.4 | 5.3 |
| GW240109_050431 | … | … | 10.4 | … | … | … | 10.0 | … |

**Note.** LIGO Hanford and LIGO Livingston are denoted by H and L, respectively. Entries in italics indicate candidates that were recovered with an FAR $\geqslant 1\ \mathrm{yr}^{-1}$ by a given analysis. Dashes (–) indicate that a candidate was not found by an analysis.

**Table 7**
Multicomponent $p_{astro}$ for All Candidates with FAR > 1 yr$^{-1}$, $p_{astro}$ > 0.5, and $p_{BNS} + p_{NSBH} > 0.001$

| Candidate | cWB-BBH | GstLAL | | | | MBTA | | | | PyCBC | | | |
|---|---|---|---|---|---|---|---|---|---|---|---|---|---|
| | $p_{astro}$ | $p_{BBH}$ | $p_{NSBH}$ | $p_{BNS}$ | $p_{astro}$ | $p_{BBH}$ | $p_{NSBH}$ | $p_{BNS}$ | $p_{astro}$ | $p_{BBH}$ | $p_{NSBH}$ | $p_{BNS}$ | $p_{astro}$ |
| GW230729_082317 | ⋯ | 0.95 | <0.01 | <0.01 | 0.95 | ⋯ | ⋯ | ⋯ | ⋯ | *0.73* | *0.04* | *<0.01* | *0.77* |
| GW230831_134621 | ⋯ | *0.11* | *<0.01* | *<0.01* | *0.11* | *0.34* | *<0.01* | *<0.01* | *0.34* | *0.69* | *0.12* | *<0.01* | *0.80* |
| GW230902_172430 | ⋯ | *0.02* | *<0.01* | *<0.01* | *0.02* | *0.07* | *<0.01* | *<0.01* | *0.07* | *0.51* | *0.08* | *<0.01* | *0.59* |
| GW230904_152545 | ⋯ | – | ⋯ | ⋯ | ⋯ | *0.03* | *0.15* | *0.02* | *0.19* | *<0.01* | *0.72* | *0.02* | *0.74* |
| GW230920_064709 | ⋯ | *<0.01* | *<0.01* | *<0.01* | *<0.01* | *0.03* | *<0.01* | *<0.01* | *0.03* | *0.16* | *0.65* | *<0.01* | *0.81* |
| GW231013_135504 | ⋯ | ⋯ | ⋯ | ⋯ | ⋯ | ⋯ | ⋯ | ⋯ | ⋯ | *0.10* | *0.44* | *<0.01* | *0.54* |
| GW231120_022103 | ⋯ | *0.53* | *<0.01* | *<0.01* | *0.53* | *0.45* | *<0.01* | *<0.01* | *0.45* | *0.63* | *0.27* | *<0.01* | *0.90* |
| GW240105_151143 | ⋯ | ⋯ | ⋯ | ⋯ | ⋯ | ⋯ | ⋯ | ⋯ | ⋯ | *0.34* | *0.37* | *<0.01* | *0.70* |

**Note.** These candidates do not meet the criterion for source property estimation. Entries in italics indicate candidates that were recovered with an FAR > 1 yr$^{-1}$ by a given analysis. Dashes (–) indicate that a candidate was not found by an analysis. The BBH, BNS, and NSBH categories are defined by the masses of the search template that recovered the candidate and are not necessarily indicative of true astrophysical population.

## Appendix B
## Glitch Mitigation

When a glitch is identified around the time of a candidate, we carry out further procedures to mitigate its impact on our inferences described in Section 4 of A. G. Abac et al. (2026a). For the candidates identified in O4a, we either model and coherently subtract the glitch with the BayesWave algorithm (N. J. Cornish & T. B. Littenberg 2015; T. B. Littenberg & N. J. Cornish 2015; N. J. Cornish et al. 2021; S. Hourihane et al. 2022; A. G. Abac et al. 2026a), or we integrate the parameter-estimation likelihood in a narrower frequency band to exclude the effect of the glitch by increasing the low-frequency cutoff $f_{low}$. For cases where we applied BayesWave, Table 8 shows input parameters into the algorithm: the reference trigger time of the CBC candidate as determined by GW searches and the bands in time and frequency space where the glitch is a priori identified to have power. For cases where we narrow the frequency band, we give the $f_{low}$ value used to set the lower bound; the upper bound $f_{high}$ is the Nyquist frequency multiplied by a roll-off factor as described in A. G. Abac et al. (2026a).

**Table 8**
List of O4a Candidates with FAR < 1 yr$^{-1}$, for Which Glitch Mitigation Was Performed

| Candidate | GPS Time (s) | Detector | Time Window (s) | Frequency Range (Hz) | Flow (Hz) |
|---|---|---|---|---|---|
| GW230601_224134 | ⋯ | H | ⋯ | ⋯ | 20.93 |
| GW230601_224134 | ⋯ | L | ⋯ | ⋯ | 20.93 |
| GW230606_004305 | 1370047403.79 | H | [1.11, 1.31] | [8.0, 512] | ⋯ |
| GW230702_185453 | ⋯ | H | ⋯ | ⋯ | 20 |
| GW230707_124047 | 1372768865.35 | H | [−0.15, 0.1] | [15.0, 30.0] | ⋯ |
| GW230708_053705 | 1372829843.12 | H | [−2.02, 0.98] | [10.0, 50.0] | ⋯ |
| GW230709_122727 | ⋯ | H | ⋯ | ⋯ | 50 |
| GW230729_082317 | ⋯ | H | ⋯ | ⋯ | 50 |
| GW230731_215307 | ⋯ | H | ⋯ | ⋯ | 40 |
| GW230803_033412 | ⋯ | H | ⋯ | ⋯ | 30 |
| GW230806_204041 | 1375389659.94 | H | [1.8, 2.0] | [25.0, 65.0] | ⋯ |
| GW230819_171910 | 1376500768.45 | L | [−3.2, −2.8] | [8.0, 512] | ⋯ |
| GW230814_061920 | ⋯ | H | ⋯ | ⋯ | 21.48 |
| GW230814_061920 | ⋯ | L | ⋯ | ⋯ | 21.48 |
| GW230814_230901 | ⋯ | L | ⋯ | ⋯ | 24 |
| GW230824_033047 | ⋯ | H | ⋯ | ⋯ | 22.18 |
| GW230824_033047 | ⋯ | L | ⋯ | ⋯ | 22.18 |
| GW230831_015414 | ⋯ | L | ⋯ | ⋯ | 22 |
| GW230911_195324 | ⋯ | H | ⋯ | ⋯ | 28.38 |
| GW230920_071124 | ⋯ | H | ⋯ | ⋯ | 40 |
| GW231001_140220 | ⋯ | L | ⋯ | ⋯ | 40 |
| GW231014_040532 | ⋯ | H | ⋯ | ⋯ | 50 |
| GW231018_233037 | ⋯ | H | ⋯ | ⋯ | 30 |
| GW231020_142947 | ⋯ | H | ⋯ | ⋯ | 45 |
| GW231102_071736 | ⋯ | H | ⋯ | ⋯ | 20.13 |
| GW231102_071736 | ⋯ | L | ⋯ | ⋯ | 20.13 |
| GW231113_122623 | 1383913601.88 | L | [0.01, 0.22] | [70.0, 120.0] | ⋯ |
| GW231114_043211 | 1383971549.25 | H | [−0.95, −0.6] | [10.0, 30.0] | ⋯ |
| GW231118_005626 | ⋯ | H | ⋯ | ⋯ | 30 |
| GW231118_071402 | ⋯ | H | ⋯ | ⋯ | 50 |

**Table 8**
(Continued)

| Candidate | GPS Time (s) | Detector | Time Window (s) | Frequency Range (Hz) | Flow (Hz) |
|---|---|---|---|---|---|
| GW231118_090602 | 1384333580.01 | H | [−4.81, −4.51] | [15.0, 50.0] | ⋯ |
| GW231123_135430 | 1384782888.63 | H | [−1.7, −1.1] | [15.0, 30.0] | ⋯ |
| GW231127_165300 | ⋯ | H | ⋯ | ⋯ | 50 |
| GW231129_081745 | 1385281083.64 | L | [1.4, 1.8] | [10.0, 170.0] | ⋯ |
| GW231129_081745 | ⋯ | H | ⋯ | ⋯ | 60 |
| GW231206_233134 | ⋯ | H | ⋯ | ⋯ | 40 |
| GW231206_233134 | ⋯ | L | ⋯ | ⋯ | 30 |
| GW231221_135041 | 1387201859.32 | H | [0.3, 0.4] | [200.0, 450.0] | ⋯ |
| GW231223_032836 | 1387337334.05 | H | [−0.55, −0.25] | [10.0, 25.0] | ⋯ |
| GW231223_075055 | ⋯ | H | ⋯ | ⋯ | 40 |
| GW231223_075055 | ⋯ | L | ⋯ | ⋯ | 30 |
| GW231223_202619 | ⋯ | H | ⋯ | ⋯ | 40 |
| GW231224_024321 | ⋯ | H | ⋯ | ⋯ | 40 |
| GW240107_013215 | ⋯ | H | ⋯ | ⋯ | 40 |

**Note.** For each candidate, we show the GPS time and the interferometer(s) where glitch subtraction was applied (H and L indicate LIGO Hanford and LIGO Livingston, respectively). For candidates where glitch subtraction was performed using BayesWave, we provide the time and frequency windows used for subtraction. For candidates where the low-frequency cutoff, $f_{\rm low}$, was changed (from the standard 20 Hz) to excise contaminated data, we quote the cutoff used.

## Appendix C
## Spectrograms of Selected Events

Here we provide time–frequency spectrograms (J. C. Brown 1991; S. Chatterji et al. 2004) for several candidates of particular interest. GW230630_070659 is discussed in Section 2.1.2 and is presented in Table 1 with other candidates identified by our searches with both $p_{\rm astro} \geqslant 0.5$ and FAR $< 1\ {\rm yr}^{-1}$. GW230824_135331 and GW240105_151143 are discussed in Sections 2.1.3 and 2.1.2, respectively, and further information on them is in Table 2, together with other candidates with $p_{\rm astro} \geqslant 0.5$ but which do not meet the threshold for detailed investigation of their source properties or event validation.

The top panels of Figure 9 show the time–frequency spectrograms for GW230630_070659, which has been determined to be likely of instrumental origin. This candidate was found only by the GstLAL matched-filter search pipeline. The event-validation procedures discussed in Section 2.1.2 identified excess power from scattered light (D. J. Ottaway et al. 2012; S. Soni et al. 2025) in both detectors at the time of the candidate.

The middle panels of Figure 9 show the time–frequency spectrograms of GW230824_135331. This candidate was recovered only by the minimally modeled cWB-BBH search pipeline, with $p_{\rm astro} = 0.58$. It was not recovered, even as a subthreshold candidate with FAR $< 2\ {\rm day}^{-1}$ and $p_{\rm astro} < 0.5$, by any matched-filter pipelines; it is the only candidate with $p_{\rm astro} \geqslant 0.5$ in this catalog with that distinction. The cWB-BBH search pipeline has the potential to detect BBH candidates impacted by physical effects neglected in the matched-filter searches, such as precession and eccentricity (e.g., A. G. Abac et al. 2024a; T. Mishra et al. 2025), in addition to non-CBC transients. However, the computation of $p_{\rm astro}$ assumes a prior CBC source population (A. G. Abac et al. 2026a). These assumptions may not be valid if a source arises from a population that our matched-filter CBC searches are not sensitive to (R. Abbott et al. 2023a). Our event-validation procedures were not applied to GW230824_135331, since it has an FAR $> 1\ {\rm yr}^{-1}$. However, the appearance of this candidate in the time–frequency spectrograms is not clearly a genuine CBC signal. The spectrograms also show the presence of excess power in LHO at the time of the event.

The bottom panels of Figure 9 present the time–frequency spectrograms of GW240105_151143, a high-SNR (>25) candidate detected by only the PyCBC matched-filter search pipeline. There are several regions with significant excess power in LHO that are not consistent with a CBC signal and could contribute to its identification by only one search pipeline with a large SNR. Our event-validation procedures were not applied to GW230824_135331, since it has an FAR $> 1\ {\rm yr}^{-1}$.

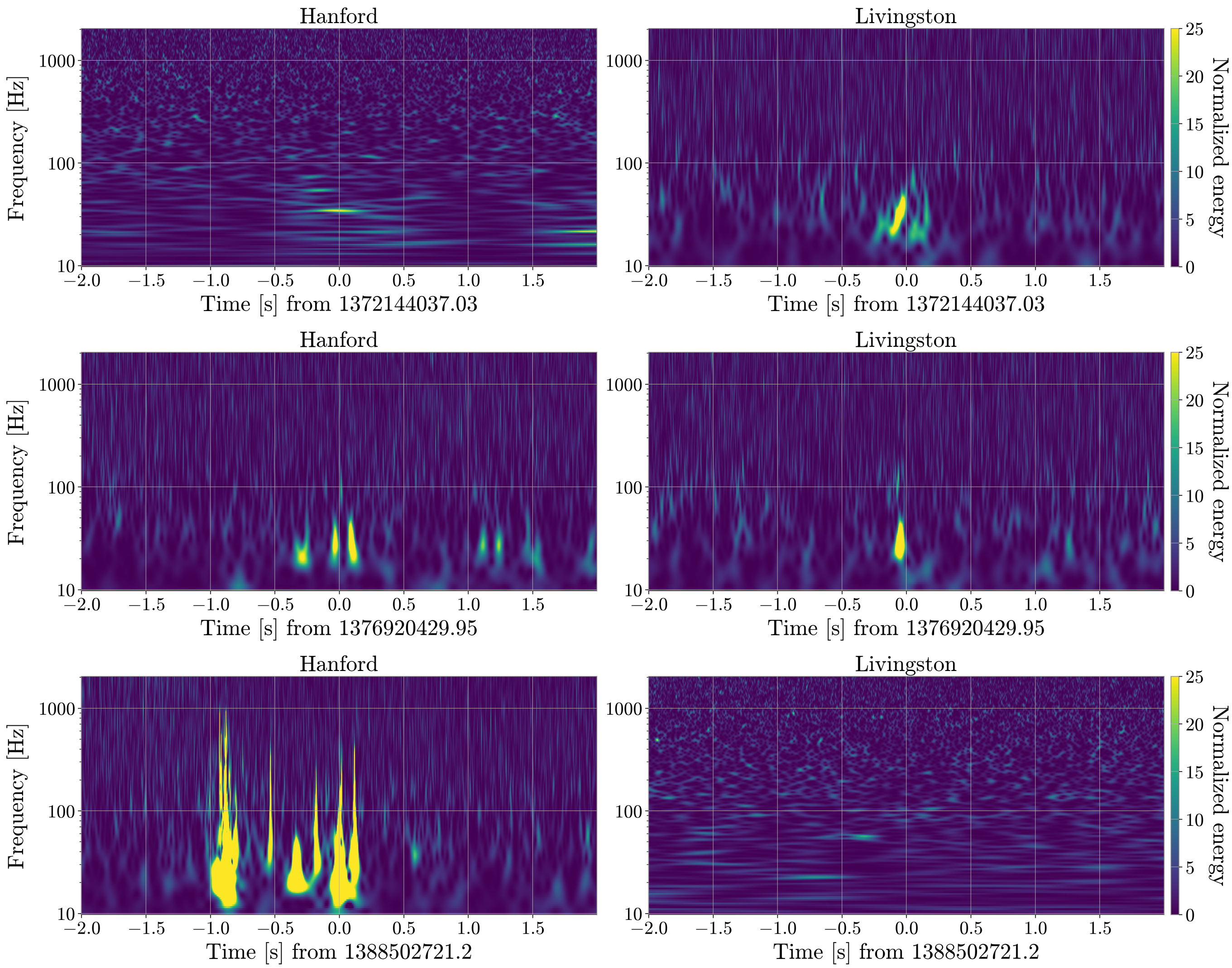

**Figure 9.** Spectrograms for three candidates of interest. Top panels: spectrograms for GW230630_070659, a candidate identified by GSTLAL with FAR $< 1$ yr$^{-1}$ but which event validation indicates is likely of instrumental origin. We do not carry out parameter estimation on this candidate. Middle panels: spectrograms for GW230824_135331, a cWB-BBH candidate that is not identified as a candidate or subthreshold candidate by any matched-filter search. Excess power is visible in LHO around the time of the candidate. Bottom panels: spectrograms for GW240105_151143, a candidate identified by only PyCBC. Significant excess power is present in LHO at the time of the candidate.

## ORCID iDs

A. G. Abac https://orcid.org/0000-0003-4786-2698
K. Ackley https://orcid.org/0000-0002-8648-0767
C. Adamcewicz https://orcid.org/0000-0001-5525-6255
S. Adhicary https://orcid.org/0009-0004-2101-5428
N. Adhikari https://orcid.org/0000-0002-4559-8427
R. X. Adhikari https://orcid.org/0000-0002-5731-5076
S. Afroz https://orcid.org/0009-0004-4459-2981
D. Agarwal https://orcid.org/0000-0002-8735-5554
M. Agathos https://orcid.org/0000-0002-9072-1121
O. D. Aguiar https://orcid.org/0000-0002-2139-4390
L. Aiello https://orcid.org/0000-0003-2771-8816
A. Ain https://orcid.org/0000-0003-4534-4619
P. Ajith https://orcid.org/0000-0001-7519-2439
S. Akcay https://orcid.org/0000-0003-2216-421X
T. Akutsu https://orcid.org/0000-0003-0733-7530
S. Albanesi https://orcid.org/0000-0001-7345-4415
A. Allocca https://orcid.org/0000-0002-5288-1351
P. A. Altin https://orcid.org/0000-0001-8193-5825
S. Alvarez-Lopez https://orcid.org/0009-0003-8040-4936
A. Amato https://orcid.org/0000-0001-9557-651X
F. Amicucci https://orcid.org/0009-0005-2139-4197
S. B. Anderson https://orcid.org/0000-0003-2219-9383
W. G. Anderson https://orcid.org/0000-0003-0482-5942
M. Andia https://orcid.org/0000-0003-3675-9126
M. Andrés-Carcasona https://orcid.org/0000-0002-8738-1672
T. Andrić https://orcid.org/0000-0002-9277-9773
S. Ansoldi https://orcid.org/0000-0002-5613-7693
J. M. Antelis https://orcid.org/0000-0003-3377-0813
S. Antier https://orcid.org/0000-0002-7686-3334
S. K. Apple https://orcid.org/0009-0007-4490-5804
K. Arai https://orcid.org/0000-0001-8916-8915
A. Araya https://orcid.org/0000-0002-6884-2875
M. C. Araya https://orcid.org/0000-0002-6018-6447
M. Arca Sedda https://orcid.org/0000-0002-3987-0519
J. S. Areeda https://orcid.org/0000-0003-0266-7936
F. Armato https://orcid.org/0000-0002-8856-8877

S. Armstrong https://orcid.org/6512-3515-4685-5112
N. Arnaud https://orcid.org/0000-0001-6589-8673
M. Arogeti https://orcid.org/0000-0001-5124-3350
S. M. Aronson https://orcid.org/0000-0001-7080-8177
K. G. Arun https://orcid.org/0000-0002-6960-8538
G. Ashton https://orcid.org/0000-0001-7288-2231
Y. Aso https://orcid.org/0000-0002-1902-6695
P. Astone https://orcid.org/0000-0003-4981-4120
F. Attadio https://orcid.org/0009-0008-8916-1658
F. Aubin https://orcid.org/0000-0003-1613-3142
K. AultONeal https://orcid.org/0000-0002-6645-4473
G. Avallone https://orcid.org/0000-0001-5482-0299
E. A. Avila https://orcid.org/0009-0008-9329-4525
S. Babak https://orcid.org/0000-0001-7469-4250
S. Bae https://orcid.org/0000-0003-2429-3357
S. Bagnasco https://orcid.org/0000-0001-6062-6505
L. Baiotti https://orcid.org/0000-0003-0458-4288
R. Bajpai https://orcid.org/0000-0003-0495-5720
T. Baker https://orcid.org/0000-0001-5470-7616
G. Baldi https://orcid.org/0000-0001-8963-3362
N. Baldicchi https://orcid.org/0009-0009-8888-291X
S. Banagiri https://orcid.org/0000-0001-7852-7484
B. Banerjee https://orcid.org/0000-0002-8008-2485
D. Bankar https://orcid.org/0000-0002-6068-2993
P. Baral https://orcid.org/0000-0001-6308-211X
M. Baratti https://orcid.org/0009-0003-5744-8025
P. Barneo https://orcid.org/0000-0002-8883-7280
F. Barone https://orcid.org/0000-0002-8069-8490
B. Barr https://orcid.org/0000-0002-5232-2736
L. Barsotti https://orcid.org/0000-0001-9819-2562
M. Barsuglia https://orcid.org/0000-0002-1180-4050
D. Barta https://orcid.org/0000-0001-6841-550X
M. A. Barton https://orcid.org/0000-0002-9948-306X
A. Basalaev https://orcid.org/0000-0001-5623-2853
R. Bassiri https://orcid.org/0000-0001-8171-6833
A. Basti https://orcid.org/0000-0003-2895-9638
M. Bawaj https://orcid.org/0000-0003-3611-3042
J. C. Bayley https://orcid.org/0000-0003-2306-4106
A. C. Baylor https://orcid.org/0000-0003-0918-0864
F. Beirnaert https://orcid.org/0000-0002-4003-7233
M. Bejger https://orcid.org/0000-0002-4991-8213
D. Belardinelli https://orcid.org/0000-0001-9332-5733
A. S. Bell https://orcid.org/0000-0003-1523-0821
L. Bellizzi https://orcid.org/0000-0002-2071-0400
W. Benoit https://orcid.org/0000-0003-4750-9413
I. Bentara https://orcid.org/0009-0000-5074-839X
J. D. Bentley https://orcid.org/0000-0002-4736-7403
S. Bera https://orcid.org/0000-0003-0907-6098
F. Bergamin https://orcid.org/0000-0002-1113-9644
B. K. Berger https://orcid.org/0000-0002-4845-8737
S. Bernuzzi https://orcid.org/0000-0002-2334-0935
M. Beroiz https://orcid.org/0000-0001-6486-9897
C. P. L. Berry https://orcid.org/0000-0003-3870-7215
D. Bersanetti https://orcid.org/0000-0002-7377-415X
J. Betzwieser https://orcid.org/0000-0003-1533-9229
D. Beveridge https://orcid.org/0000-0002-1481-1993
G. Bevilacqua https://orcid.org/0000-0002-7298-6185
N. Bevins https://orcid.org/0000-0002-4312-4287
S. A. Bhat https://orcid.org/0000-0002-6783-1840
D. Bhattacharjee https://orcid.org/0000-0001-6623-9506
S. Bhaumik https://orcid.org/0000-0001-8492-2202
V. Biancalana https://orcid.org/0000-0002-1642-5391
G. Billingsley https://orcid.org/0000-0002-4141-2744
A. Binetti https://orcid.org/0000-0001-6449-5493
S. Bini https://orcid.org/0000-0002-0267-3562
O. Birnholtz https://orcid.org/0000-0002-7562-9263
S. Biscoveanu https://orcid.org/0000-0001-7616-7366
M. Bitossi https://orcid.org/0000-0002-9862-4668
M.-A. Bizouard https://orcid.org/0000-0002-4618-1674
J. K. Blackburn https://orcid.org/0000-0002-3838-2986
N. Bode https://orcid.org/0000-0002-7101-9396
G. Boileau https://orcid.org/0000-0002-3576-6968
M. Boldrini https://orcid.org/0000-0001-9861-821X
G. N. Bolingbroke https://orcid.org/0000-0002-7350-5291
L. D. Bonavena https://orcid.org/0000-0002-2630-6724
R. Bondarescu https://orcid.org/0000-0003-0330-2736
F. Bondu https://orcid.org/0000-0001-6487-5197
E. Bonilla https://orcid.org/0000-0002-6284-9769
M. S. Bonilla https://orcid.org/0000-0003-4502-528X
R. Bonnand https://orcid.org/0000-0001-5013-5913
V. Boschi https://orcid.org/0000-0001-8665-2293
Y. Bothra https://orcid.org/0000-0002-9380-6390
P. R. Brady https://orcid.org/0000-0002-4611-9387
M. Branchesi https://orcid.org/0000-0003-1643-0526
T. Briant https://orcid.org/0000-0002-6013-1729
E. Brockmueller https://orcid.org/0000-0002-1489-942X
A. F. Brooks https://orcid.org/0000-0003-4295-792X
M. L. Brozzetti https://orcid.org/0000-0002-5260-4979
R. Bruntz https://orcid.org/0000-0002-0840-8567
F. Bucci https://orcid.org/0000-0003-1726-3838
O. Bulashenko https://orcid.org/0000-0003-1720-4061
A. Buonanno https://orcid.org/0000-0002-5433-1409
R. Buscicchio https://orcid.org/0000-0002-7387-6754
C. Buy https://orcid.org/0000-0003-2872-8186
G. S. Cabourn Davies https://orcid.org/0000-0002-4289-3439
R. Cabrita https://orcid.org/0000-0003-0133-1306
V. Cáceres-Barbosa https://orcid.org/0000-0001-9834-4781
L. Cadonati https://orcid.org/0000-0002-9846-166X
G. Cagnoli https://orcid.org/0000-0002-7086-6550
C. Cahillane https://orcid.org/0000-0002-3888-314X
J. Calderón Bustillo https://orcid.org/0000-0002-4171-5884
S. R. Callos https://orcid.org/0000-0003-0639-9342
G. Caneva Santoro https://orcid.org/0000-0002-2935-1600
K. C. Cannon https://orcid.org/0000-0003-4068-6572
E. Capocasa https://orcid.org/0000-0003-3762-6958
E. Capote https://orcid.org/0009-0007-0246-713X
G. Capurri https://orcid.org/0000-0003-0889-1015
J. B. Carlin https://orcid.org/0000-0001-5694-0809
M. Carpinelli https://orcid.org/0000-0002-8205-930X
J. J. Carter https://orcid.org/0000-0001-8845-0900
G. Carullo https://orcid.org/0000-0001-9090-1862
J. Casanueva Diaz https://orcid.org/0000-0002-2948-5238
C. Casentini https://orcid.org/0000-0001-8100-0579
M. Cavaglià https://orcid.org/0000-0002-3835-6729
R. Cavalieri https://orcid.org/0000-0001-6064-0569
G. Cella https://orcid.org/0000-0002-0752-0338
P. Cerdá-Durán https://orcid.org/0000-0003-4293-340X
E. Cesarini https://orcid.org/0000-0001-9127-3167
A. Chakraborty https://orcid.org/0009-0004-4937-4633
P. Chakraborty https://orcid.org/0000-0002-0994-7394

S. Chalathadka Subrahmanya https://orcid.org/0000-0002-9207-4669
J. C. L. Chan https://orcid.org/0000-0002-3377-4737
S. Chao https://orcid.org/0000-0003-3853-3593
P. Charlton https://orcid.org/0000-0002-4263-2706
E. Chassande-Mottin https://orcid.org/0000-0003-3768-9908
C. Chatterjee https://orcid.org/0000-0001-8700-3455
Debarati Chatterjee https://orcid.org/0000-0002-0995-2329
Deep Chatterjee https://orcid.org/0000-0003-0038-5468
S. Chaty https://orcid.org/0000-0002-5769-8601
K. Chatziioannou https://orcid.org/0000-0002-5833-413X
A. Chen https://orcid.org/0000-0001-9174-7780
D. Chen https://orcid.org/0000-0003-1433-0716
H. Y. Chen https://orcid.org/0000-0001-5403-3762
Yitian Chen https://orcid.org/0000-0002-8664-9702
P. Chessa https://orcid.org/0000-0001-9092-3965
H. T. Cheung https://orcid.org/0000-0003-3905-0665
F. Chiadini https://orcid.org/0000-0002-9339-8622
A. Chincarini https://orcid.org/0000-0003-4094-9942
M. L. Chiofalo https://orcid.org/0000-0002-6992-5963
A. Chiummo https://orcid.org/0000-0003-2165-2967
S. Choudhary https://orcid.org/0000-0003-0949-7298
N. Christensen https://orcid.org/0000-0002-6870-4202
S. S. Y. Chua https://orcid.org/0000-0001-8026-7597
G. Ciani https://orcid.org/0000-0003-4258-9338
P. Ciecielag https://orcid.org/0000-0002-5871-4730
M. Cieślar https://orcid.org/0000-0001-8912-5587
M. Cifaldi https://orcid.org/0009-0007-1566-7093
J. A. Clark https://orcid.org/0000-0003-3243-1393
T. A. Clarke https://orcid.org/0000-0002-6714-5429
E. Codazzo https://orcid.org/0000-0001-7170-8733
P.-F. Cohadon https://orcid.org/0000-0003-3452-9415
S. Colace https://orcid.org/0009-0007-9429-1847
M. Colleoni https://orcid.org/0000-0002-7214-9088
S. Colloms https://orcid.org/0009-0009-9828-3646
A. Colombo https://orcid.org/0000-0002-7439-4773
L. Conti https://orcid.org/0000-0003-2731-2656
T. R. Corbitt https://orcid.org/0000-0002-5520-8541
I. Cordero-Carrión https://orcid.org/0000-0002-1985-1361
S. Corezzi https://orcid.org/0000-0002-3437-5949
N. J. Cornish https://orcid.org/0000-0002-7435-0869
A. Corsi https://orcid.org/0000-0001-8104-3536
M. W. Coughlin https://orcid.org/0000-0002-8262-2924
P. Couvares https://orcid.org/0000-0002-2823-3127
R. Coyne https://orcid.org/0000-0002-5243-5917
J. D. E. Creighton https://orcid.org/0000-0003-3600-2406
P. Cremonese https://orcid.org/0000-0001-6472-8509
J. R. Cudell https://orcid.org/0000-0002-2003-4238
T. J. Cullen https://orcid.org/0000-0001-8075-4088
A. Cumming https://orcid.org/0000-0003-4096-7542
E. Cuoco https://orcid.org/0000-0002-6528-3449
M. Cusinato https://orcid.org/0000-0003-4075-4539
L. V. Da Conceição https://orcid.org/0000-0002-5042-443X
T. Dal Canton https://orcid.org/0000-0001-5078-9044
S. Dal Pra https://orcid.org/0000-0002-1057-2307
G. Dálya https://orcid.org/0000-0003-3258-5763
B. D'Angelo https://orcid.org/0000-0001-9143-8427
S. Danilishin https://orcid.org/0000-0001-7758-7493
S. D'Antonio https://orcid.org/0000-0003-0898-6030
L. P. Dartez https://orcid.org/0000-0002-2216-0465
V. Dattilo https://orcid.org/0000-0002-8816-8566
D. Davis https://orcid.org/0000-0001-5620-6751
M. C. Davis https://orcid.org/0000-0001-7663-0808
P. Davis https://orcid.org/0009-0004-5008-5660
E. J. Daw https://orcid.org/0000-0002-3780-5430
M. Dax https://orcid.org/0000-0001-8798-0627
J. De Bolle https://orcid.org/0000-0002-5179-1725
J. Degallaix https://orcid.org/0000-0002-1019-6911
U. Deka https://orcid.org/0000-0002-5942-4487
M. De Laurentis https://orcid.org/0000-0002-3815-4078
F. De Lillo https://orcid.org/0000-0003-4977-0789
S. Della Torre https://orcid.org/0000-0002-7669-0859
W. Del Pozzo https://orcid.org/0000-0003-3978-2030
F. De Marco https://orcid.org/0000-0002-5411-9424
F. De Matteis https://orcid.org/0000-0001-7860-9754
T. Dent https://orcid.org/0000-0003-1354-7809
A. Depasse https://orcid.org/0000-0003-1014-8394
R. De Pietri https://orcid.org/0000-0003-1556-8304
R. De Rosa https://orcid.org/0000-0002-4004-947X
C. De Rossi https://orcid.org/0000-0002-5825-472X
M. Desai https://orcid.org/0009-0003-4448-3681
R. DeSalvo https://orcid.org/0000-0002-4818-0296
A. Dhani https://orcid.org/0000-0001-9930-9101
M. C. Díaz https://orcid.org/0000-0002-7555-8856
M. Di Cesare https://orcid.org/0009-0003-0411-6043
T. Dietrich https://orcid.org/0000-0003-2374-307X
C. Di Fronzo https://orcid.org/0000-0002-2693-6769
M. Di Giovanni https://orcid.org/0000-0003-4049-8336
T. Di Girolamo https://orcid.org/0000-0003-2339-4471
J. Ding https://orcid.org/0000-0003-1693-3828
S. Di Pace https://orcid.org/0000-0001-6759-5676
I. Di Palma https://orcid.org/0000-0003-1544-8943
F. Di Renzo https://orcid.org/0000-0002-5447-3810
Divyajyoti https://orcid.org/0000-0002-2787-1012
A. Dmitriev https://orcid.org/0000-0002-0314-956X
Z. Doctor https://orcid.org/0000-0002-2077-4914
N. Doerksen https://orcid.org/0009-0002-3776-5026
L. D'Onofrio https://orcid.org/0000-0001-9546-5959
K. L. Dooley https://orcid.org/0000-0002-1636-0233
S. Doravari https://orcid.org/0000-0001-8750-8330
M. Drago https://orcid.org/0000-0002-3738-2431
J. C. Driggers https://orcid.org/0000-0002-6134-7628
L. Dunn https://orcid.org/0000-0002-1769-6097
P.-A. Duverne https://orcid.org/0000-0002-3906-0997
D. D'Urso https://orcid.org/0000-0002-8215-4542
P. Dutta Roy https://orcid.org/0000-0001-8874-4888
H. Duval https://orcid.org/0000-0002-2475-1728
M. Ebersold https://orcid.org/0000-0003-4631-1771
T. Eckhardt https://orcid.org/0000-0002-1224-4681
G. Eddolls https://orcid.org/0000-0002-5895-4523
A. Effler https://orcid.org/0000-0001-8242-3944
J. Eichholz https://orcid.org/0000-0002-2643-163X
M. Emma https://orcid.org/0000-0001-7943-0262
R. Enficiaud https://orcid.org/0000-0003-3908-1912
L. Errico https://orcid.org/0000-0003-2112-0653
M. Esposito https://orcid.org/0009-0009-8482-9417
R. C. Essick https://orcid.org/0000-0001-8196-9267
H. Estellés https://orcid.org/0000-0001-6143-5532
M. Evans https://orcid.org/0000-0001-8459-4499
J. M. Ezquiaga https://orcid.org/0000-0002-7213-3211
F. Fabrizi https://orcid.org/0000-0002-3809-065X
V. Fafone https://orcid.org/0000-0003-1314-1622

S. Fairhurst https://orcid.org/0000-0001-8480-1961
A. M. Farah https://orcid.org/0000-0002-6121-0285
B. Farr https://orcid.org/0000-0002-2916-9200
W. M. Farr https://orcid.org/0000-0003-1540-8562
G. Favaro https://orcid.org/0000-0002-0351-6833
M. Favata https://orcid.org/0000-0001-8270-9512
M. Fays https://orcid.org/0000-0002-4390-9746
M. Fazio https://orcid.org/0000-0002-9057-9663
R. Felicetti https://orcid.org/0009-0005-6263-5604
E. Fenyvesi https://orcid.org/0000-0003-2777-3719
T. Fernandes https://orcid.org/0009-0006-6820-2065
S. Ferraiuolo https://orcid.org/0009-0005-5582-2989
F. Fidecaro https://orcid.org/0000-0002-6189-3311
P. Figura https://orcid.org/0000-0002-8925-0393
A. Fiori https://orcid.org/0000-0003-3174-0688
I. Fiori https://orcid.org/0000-0002-0210-516X
M. Fishbach https://orcid.org/0000-0002-1980-5293
R. Fittipaldi https://orcid.org/0000-0003-2096-7983
V. Fiumara https://orcid.org/0000-0003-3644-217X
S. M. Fleischer https://orcid.org/0000-0001-7884-9993
J. A. Font https://orcid.org/0000-0001-6650-2634
B. Fornal https://orcid.org/0000-0003-3271-2080
F. Frasconi https://orcid.org/0000-0003-4204-6587
Z. Frei https://orcid.org/0000-0002-0181-8491
A. Freise https://orcid.org/0000-0001-6586-9901
O. Freitas https://orcid.org/0000-0002-2898-1256
R. Frey https://orcid.org/0000-0003-0341-2636
G. G. Fronzé https://orcid.org/0000-0003-0966-4279
M. Fuentes-Garcia https://orcid.org/0000-0003-3390-8712
B. Gadre https://orcid.org/0000-0002-1534-9761
J. R. Gair https://orcid.org/0000-0002-1671-3668
S. Galaudage https://orcid.org/0000-0002-1819-0215
A. Gamboa https://orcid.org/0000-0001-8391-5596
D. Ganapathy https://orcid.org/0000-0003-3028-4174
A. Ganguly https://orcid.org/0000-0001-7394-0755
B. Garaventa https://orcid.org/0000-0003-2490-404X
J. García-Bellido https://orcid.org/0000-0002-9370-8360
C. García-Quirós https://orcid.org/0000-0002-8059-2477
J. W. Gardner https://orcid.org/0000-0002-8592-1452
J. Gargiulo https://orcid.org/0000-0002-3507-6924
X. Garrido https://orcid.org/0000-0002-7088-5831
A. Garron https://orcid.org/0000-0002-1601-797X
F. Garufi https://orcid.org/0000-0003-1391-6168
C. Gasbarra https://orcid.org/0000-0001-8335-9614
F. Gautier https://orcid.org/0000-0001-8006-9590
V. Gayathri https://orcid.org/0000-0002-7167-9888
G. Gemme https://orcid.org/0000-0002-1127-7406
A. Gennai https://orcid.org/0000-0003-0149-2089
V. Gennari https://orcid.org/0000-0002-0190-9262
R. George https://orcid.org/0000-0002-7797-7683
O. Gerberding https://orcid.org/0000-0001-7740-2698
L. Gergely https://orcid.org/0000-0003-3146-6201
Archisman Ghosh https://orcid.org/0000-0003-0423-3533
Shaon Ghosh https://orcid.org/0000-0001-9901-6253
Suprovo Ghosh https://orcid.org/0000-0002-1656-9870
Tathagata Ghosh https://orcid.org/0000-0001-9848-9905
J. A. Giaime https://orcid.org/0000-0002-3531-817X
C. Gier https://orcid.org/0000-0003-0897-7943
S. Gkaitatzis https://orcid.org/0000-0001-9420-7499
J. Glanzer https://orcid.org/0009-0000-0808-0795
F. Glotin https://orcid.org/0000-0003-2637-1187
P. Godwin https://orcid.org/0000-0002-7489-4751
A. S. Goettel https://orcid.org/0000-0002-6215-4641
E. Goetz https://orcid.org/0000-0003-2666-721X
S. Gomez Lopez https://orcid.org/0000-0002-9557-4706
B. Goncharov https://orcid.org/0000-0003-3189-5807
G. González https://orcid.org/0000-0003-0199-3158
P. Goodarzi https://orcid.org/0009-0008-1093-6706
A. W. Goodwin-Jones https://orcid.org/0000-0002-0395-0680
R. Gouaty https://orcid.org/0000-0001-5372-7084
A. Grado https://orcid.org/0000-0002-0501-8256
V. Graham https://orcid.org/0000-0003-3633-0135
A. E. Granados https://orcid.org/0000-0003-2099-9096
M. Granata https://orcid.org/0000-0003-3275-1186
V. Granata https://orcid.org/0000-0003-2246-6963
R. Gray https://orcid.org/0000-0002-5556-9873
A. C. Green https://orcid.org/0000-0002-6287-8746
S. R. Green https://orcid.org/0000-0002-6987-6313
H. L. Griggs https://orcid.org/0000-0001-5018-7908
C. Grimaud https://orcid.org/0000-0001-7736-7730
H. Grote https://orcid.org/0000-0002-0797-3943
S. Grunewald https://orcid.org/0000-0003-4641-2791
D. Guerra https://orcid.org/0000-0003-0029-5390
D. Guetta https://orcid.org/0000-0002-7349-1109
G. M. Guidi https://orcid.org/0000-0002-3061-9870
F. Gulminelli https://orcid.org/0000-0003-4354-2849
H. Guo https://orcid.org/0000-0002-3777-3117
W. Guo https://orcid.org/0000-0002-4320-4420
Y. Guo https://orcid.org/0000-0002-6959-9870
Anuradha Gupta https://orcid.org/0000-0002-5441-9013
I. Gupta https://orcid.org/0000-0001-6932-8715
V. Gupta https://orcid.org/0000-0002-7672-0480
F. Guzman https://orcid.org/0000-0001-9136-929X
M. Haberland https://orcid.org/0000-0001-9816-5660
E. D. Hall https://orcid.org/0000-0001-9018-666X
R. Hamburg https://orcid.org/0000-0003-0761-6388
E. Z. Hamilton https://orcid.org/0000-0003-0098-9114
G. Hammond https://orcid.org/0000-0002-1414-3622
C. Hanna https://orcid.org/0000-0002-0965-7493
O. A. Hannuksela https://orcid.org/0000-0002-3887-7137
A. G. Hanselman https://orcid.org/0000-0002-8304-0109
S. Harikumar https://orcid.org/0000-0002-2653-7282
T. Harmark https://orcid.org/0000-0002-2795-7035
J. Harms https://orcid.org/0000-0002-7332-9806
G. M. Harry https://orcid.org/0000-0002-8905-7622
I. W. Harry https://orcid.org/0000-0002-5304-9372
C. J. Haster https://orcid.org/0000-0001-8040-9807
K. Haughian https://orcid.org/0000-0002-1223-7342
A. Heffernan https://orcid.org/0000-0003-3355-9671
J. Heinze https://orcid.org/0000-0001-8692-2724
H. Heitmann https://orcid.org/0000-0003-0625-5461
F. Hellman https://orcid.org/0000-0002-9135-6330
A. F. Helmling-Cornell https://orcid.org/0000-0002-7709-8638
G. Hemming https://orcid.org/0000-0001-5268-4465
O. Henderson-Sapir https://orcid.org/0000-0002-1613-9985
M. Hendry https://orcid.org/0000-0001-8322-5405
M. H. Hennig https://orcid.org/0000-0003-1531-8460
C. Henshaw https://orcid.org/0000-0002-4206-3128
M. Heurs https://orcid.org/0000-0002-5577-2273
A. L. Hewitt https://orcid.org/0000-0002-1255-3492
Y. Himemoto https://orcid.org/0000-0002-6856-3809

I. J. Hollows https://orcid.org/0000-0002-3404-6459
D. E. Holz https://orcid.org/0000-0002-0175-5064
J. Hough https://orcid.org/0000-0003-3242-3123
S. Hourihane https://orcid.org/0000-0002-9152-0719
E. J. Howell https://orcid.org/0000-0001-7891-2817
C. G. Hoy https://orcid.org/0000-0002-8843-6719
H.-F. Hsieh https://orcid.org/0000-0002-8947-723X
W.-F. Hsu https://orcid.org/0000-0001-5234-3804
Q. Hu https://orcid.org/0000-0002-3033-6491
H. Y. Huang https://orcid.org/0000-0002-1665-2383
Y. Huang https://orcid.org/0000-0002-2952-8429
V. Hui https://orcid.org/0000-0002-0233-2346
S. Husa https://orcid.org/0000-0002-0445-1971
L. Iampieri https://orcid.org/0009-0004-1161-2990
G. A. Iandolo https://orcid.org/0000-0003-1155-4327
G. Iannone https://orcid.org/0000-0001-8347-7549
G. Iorio https://orcid.org/0000-0003-0293-503X
P. Iosif https://orcid.org/0000-0003-1621-7709
J. Irwin https://orcid.org/0000-0002-2364-2191
M. Isi https://orcid.org/0000-0001-8830-8672
K. S. Isleif https://orcid.org/0000-0001-7032-9440
Y. Itoh https://orcid.org/0000-0003-2694-8935
B. R. Iyer https://orcid.org/0000-0002-4141-5179
P.-E. Jacquet https://orcid.org/0000-0001-9552-0057
S. P. Jadhav https://orcid.org/0000-0003-0554-0084
A. L. James https://orcid.org/0000-0001-9165-0807
A. Jan https://orcid.org/0000-0003-2050-7231
K. Jani https://orcid.org/0000-0003-1007-8912
J. Janquart https://orcid.org/0000-0003-2888-7152
S. Jaraba https://orcid.org/0000-0002-4759-143X
P. Jaranowski https://orcid.org/0000-0001-8085-3414
R. Jaume https://orcid.org/0000-0001-8691-3166
J. Jiang https://orcid.org/0000-0002-0154-3854
H.-B. Jin https://orcid.org/0000-0002-6217-2428
S. J. Jin https://orcid.org/0000-0003-3697-3501
N. K. Johnson-McDaniel https://orcid.org/0000-0001-5357-9480
M. C. Johnston https://orcid.org/0000-0002-0663-9193
D. H. Jones https://orcid.org/0000-0003-3987-068X
P. Joshi https://orcid.org/0000-0002-4148-4932
L. Ju https://orcid.org/0000-0002-7951-4295
K. Jung https://orcid.org/0000-0003-4789-8893
J. Junker https://orcid.org/0000-0002-3051-4374
H. B. Kabagoz https://orcid.org/0000-0002-0900-8557
T. Kajita https://orcid.org/0000-0003-1207-6638
V. Kalogera https://orcid.org/0000-0001-9236-5469
M. Kalomenopoulos https://orcid.org/0000-0001-6677-949X
M. Kamiizumi https://orcid.org/0000-0001-7216-1784
N. Kanda https://orcid.org/0000-0001-6291-0227
S. Kandhasamy https://orcid.org/0000-0002-4825-6764
G. Kang https://orcid.org/0000-0002-6072-8189
S. J. Kapadia https://orcid.org/0000-0001-5318-1253
D. P. Kapasi https://orcid.org/0000-0001-8189-4920
M. Kasprzack https://orcid.org/0000-0003-4618-5939
R. Kaushik https://orcid.org/0000-0003-4888-5154
D. Keitel https://orcid.org/0000-0002-2824-626X
L. J. Kemperman https://orcid.org/0009-0009-5254-8397
J. Kennington https://orcid.org/0000-0002-6899-3833
R. Kesharwani https://orcid.org/0009-0002-2528-5738
J. S. Key https://orcid.org/0000-0003-0123-7600
F. Y. Khalili https://orcid.org/0000-0001-7068-2332
F. Khan https://orcid.org/0000-0001-6176-853X
W. Kiendrebeogo https://orcid.org/0000-0002-9108-5059
N. Kijbunchoo https://orcid.org/0000-0002-2874-1228
K. Kim https://orcid.org/0000-0003-1653-3795
M. H. Kim https://orcid.org/0009-0009-9894-3640
S. Kim https://orcid.org/0000-0003-1437-4647
Y.-M. Kim https://orcid.org/0000-0001-8720-6113
C. Kimball https://orcid.org/0000-0001-9879-6884
J. S. Kissel https://orcid.org/0000-0002-1702-9577
A. M. Knee https://orcid.org/0000-0003-0703-947X
N. Knust https://orcid.org/0000-0002-5984-5353
S. M. Koehlenbeck https://orcid.org/0000-0002-3842-9051
K. Kohri https://orcid.org/0000-0003-3764-8612
K. Kokeyama https://orcid.org/0000-0002-2896-1992
S. Koley https://orcid.org/0000-0002-5793-6665
P. Kolitsidou https://orcid.org/0000-0002-6719-8686
A. E. Koloniari https://orcid.org/0000-0002-0546-5638
K. Komori https://orcid.org/0000-0002-4092-9602
A. K. H. Kong https://orcid.org/0000-0002-5105-344X
A. Kontos https://orcid.org/0000-0002-1347-0680
M. Korobko https://orcid.org/0000-0002-3839-3909
A. Koushik https://orcid.org/0000-0002-7638-4544
N. Kouvatsos https://orcid.org/0000-0002-5497-3401
N. V. Krishnendu https://orcid.org/0000-0002-3483-7517
A. Królak https://orcid.org/0000-0003-4514-7690
J. Kubisz https://orcid.org/0000-0001-7258-8673
S. Kulkarni https://orcid.org/0000-0001-8057-0203
A. Kulur Ramamohan https://orcid.org/0000-0003-3681-1887
Praveen Kumar https://orcid.org/0000-0002-2288-4252
Prayush Kumar https://orcid.org/0000-0001-5523-4603
J. Kume https://orcid.org/0000-0003-3126-5100
K. Kuns https://orcid.org/0000-0003-0630-3902
S. Kuroyanagi https://orcid.org/0000-0001-6538-1447
S. Kuwahara https://orcid.org/0009-0009-2249-8798
K. Kwak https://orcid.org/0000-0002-2304-7798
S. Kwon https://orcid.org/0009-0006-3770-7044
D. Laghi https://orcid.org/0000-0001-7462-3794
M. Lalleman https://orcid.org/0000-0002-2254-010X
R. N. Lang https://orcid.org/0000-0002-4804-5537
R. Langgin https://orcid.org/0000-0002-5116-6217
B. Lantz https://orcid.org/0000-0002-7404-4845
I. L. Rosa https://orcid.org/0000-0003-0107-1540
A. Lartaux-Vollard https://orcid.org/0000-0003-1714-365X
P. D. Lasky https://orcid.org/0000-0003-3763-1386
J. Lawrence https://orcid.org/0000-0003-1222-0433
M. Laxen https://orcid.org/0000-0001-7515-9639
C. Lazarte https://orcid.org/0000-0002-6964-9321
A. Lazzarini https://orcid.org/0000-0002-5993-8808
P. Leaci https://orcid.org/0000-0002-3997-5046
Y. K. Lecoeuche https://orcid.org/0000-0002-9186-7034
H. M. Lee https://orcid.org/0000-0003-4412-7161
H. W. Lee https://orcid.org/0000-0002-1998-3209
K. Lee https://orcid.org/0000-0003-0470-3718
R.-K. Lee https://orcid.org/0000-0002-7171-7274
Sungho Lee https://orcid.org/0000-0001-6034-2238
M. Le Jean https://orcid.org/0009-0003-8047-3958
A. Lemaître https://orcid.org/0000-0002-6865-9245
M. Lenti https://orcid.org/0000-0002-2765-3955
M. Leonardi https://orcid.org/0000-0002-7641-0060
N. Leroy https://orcid.org/0000-0002-2321-1017

M. Lethuillier https://orcid.org/0000-0001-6185-2045
K. L. Li https://orcid.org/0000-0001-8229-2024
X. Li https://orcid.org/0000-0002-3780-7735
E. T. Lin https://orcid.org/0000-0002-0030-8051
L. C.-C. Lin https://orcid.org/0000-0003-4083-9567
Y.-C. Lin https://orcid.org/0000-0003-4939-1404
A. Liu https://orcid.org/0000-0003-1081-8722
G. C. Liu https://orcid.org/0000-0001-5663-3016
Jian Liu https://orcid.org/0000-0001-6726-3268
J. Llobera-Querol https://orcid.org/0000-0003-3322-6850
R. K. L. Lo https://orcid.org/0000-0003-1561-6716
A. Longo https://orcid.org/0000-0003-4254-8579
D. Lopez https://orcid.org/0000-0003-3342-9906
A. Lorenzo-Medina https://orcid.org/0009-0006-0860-5700
G. Losurdo https://orcid.org/0000-0003-0452-746X
T. P. Lott, IV https://orcid.org/0009-0002-2864-162X
J. D. Lough https://orcid.org/0000-0002-5160-0239
C. O. Lousto https://orcid.org/0000-0002-6400-9640
N. Lu https://orcid.org/0000-0002-8861-9902
L. Lucchesi https://orcid.org/0000-0002-5916-8014
D. Lumaca https://orcid.org/0000-0002-3628-1591
A. P. Lundgren https://orcid.org/0000-0002-0363-4469
A. W. Lussier https://orcid.org/0000-0002-4507-1123
R. Macas https://orcid.org/0000-0002-6096-8297
D. M. Macleod https://orcid.org/0000-0002-1395-8694
I. A. O. MacMillan https://orcid.org/0000-0002-6927-1031
A. Macquet https://orcid.org/0000-0001-5955-6415
S. Maenaut https://orcid.org/0000-0003-1464-2605
R. M. Magee https://orcid.org/0000-0001-9769-531X
E. Maggio https://orcid.org/0000-0002-1960-8185
M. Magnozzi https://orcid.org/0000-0003-4512-8430
D. Malakar https://orcid.org/0000-0003-4234-4023
U. Mali https://orcid.org/0009-0003-1285-2788
L. Mallick https://orcid.org/0000-0001-8624-9162
A.-K. Malz https://orcid.org/0009-0004-7196-4170
M. Mancarella https://orcid.org/0000-0002-0675-508X
V. Mandic https://orcid.org/0000-0001-6333-8621
V. Mangano https://orcid.org/0000-0001-7902-8505
G. L. Mansell https://orcid.org/0000-0003-4736-6678
M. Manske https://orcid.org/0000-0002-7778-1189
M. Mantovani https://orcid.org/0000-0002-4424-5726
M. Mapelli https://orcid.org/0000-0001-8799-2548
C. Marinelli https://orcid.org/0000-0002-3596-4307
F. Marion https://orcid.org/0000-0002-8184-1017
S. Marsat https://orcid.org/0000-0001-9449-1071
F. Martelli https://orcid.org/0000-0003-3761-8616
I. W. Martin https://orcid.org/0000-0001-7300-9151
R. M. Martin https://orcid.org/0000-0001-9664-2216
V. Martinez https://orcid.org/0000-0001-5852-2301
J. C. Martins https://orcid.org/0000-0002-6099-4831
M. Masso-Reid https://orcid.org/0000-0001-6177-8105
S. Mastrogiovanni https://orcid.org/0000-0003-1606-4183
T. Matcovich https://orcid.org/0009-0004-1209-008X
M. Matiushechkina https://orcid.org/0000-0002-9957-8720
N. Mavalvala https://orcid.org/0000-0003-0219-9706
D. E. McClelland https://orcid.org/0000-0001-6210-5842
L. McCuller https://orcid.org/0000-0003-0851-0593
G. I. McGhee https://orcid.org/0000-0001-5038-2658
J. McIver https://orcid.org/0000-0003-0316-1355
A. McLeod https://orcid.org/0000-0001-5424-8368
I. McMahon https://orcid.org/0000-0002-4529-1505
R. McTeague https://orcid.org/0009-0004-3329-6079
D. Meacher https://orcid.org/0000-0001-5882-0368
M. Melching https://orcid.org/0009-0001-4899-9955
C. S. Menoni https://orcid.org/0000-0001-9185-2572
R. A. Mercer https://orcid.org/0000-0001-8372-3914
J. R. Mérou https://orcid.org/0000-0002-5776-6643
C. Messick https://orcid.org/0000-0002-8230-3309
M. Meyer-Conde https://orcid.org/0000-0003-2230-6310
F. Meylahn https://orcid.org/0000-0002-9556-142X
A. Miani https://orcid.org/0000-0001-7737-3129
C. Michel https://orcid.org/0000-0003-0606-725X
Y. Michimura https://orcid.org/0000-0002-2218-4002
H. Middleton https://orcid.org/0000-0001-5532-3622
D. P. Mihaylov https://orcid.org/0000-0002-8820-407X
S. J. Miller https://orcid.org/0000-0001-5670-7046
M. Millhouse https://orcid.org/0000-0002-8659-5898
E. Milotti https://orcid.org/0000-0001-7348-9765
V. Milotti https://orcid.org/0000-0003-4732-1226
Ll. M. Mir https://orcid.org/0000-0002-4276-715X
L. Mirasola https://orcid.org/0009-0004-0174-1377
M. Miravet-Tenés https://orcid.org/0000-0002-8766-1156
C.-A. Miritescu https://orcid.org/0000-0002-7716-0569
C. Mishra https://orcid.org/0000-0002-8115-8728
T. Mishra https://orcid.org/0000-0002-7881-1677
S. Mitra https://orcid.org/0000-0002-0800-4626
V. P. Mitrofanov https://orcid.org/0000-0002-6983-4981
O. Miyakawa https://orcid.org/0000-0002-9085-7600
S. Miyoki https://orcid.org/0000-0002-1213-8416
G. Mo https://orcid.org/0000-0001-6331-112X
L. Mobilia https://orcid.org/0009-0000-3022-2358
S. R. Mohite https://orcid.org/0000-0003-1356-7156
M. Molina-Ruiz https://orcid.org/0000-0003-4892-3042
A. More https://orcid.org/0000-0001-7714-7076
S. More https://orcid.org/0000-0002-2986-2371
C. Moreno https://orcid.org/0000-0002-0496-032X
E. A. Moreno https://orcid.org/0000-0001-5666-3637
S. Morisaki https://orcid.org/0000-0002-8445-6747
Y. Moriwaki https://orcid.org/0000-0002-4497-6908
G. Morras https://orcid.org/0000-0002-9977-8546
A. Moscatello https://orcid.org/0000-0001-5480-7406
M. Mould https://orcid.org/0000-0001-5460-2910
P. Mourier https://orcid.org/0000-0001-8078-6901
B. Mours https://orcid.org/0000-0002-6444-6402
C. M. Mow-Lowry https://orcid.org/0000-0002-0351-4555
L. Muccillo https://orcid.org/0009-0000-6237-0590
F. Muciaccia https://orcid.org/0000-0003-0850-2649
D. Mukherjee https://orcid.org/0000-0001-7335-9418
Suvodip Mukherjee https://orcid.org/0000-0002-3373-5236
N. Mukund https://orcid.org/0000-0002-8666-9156
P. G. Murray https://orcid.org/0000-0002-8218-2404
D. Nabari https://orcid.org/0009-0006-8500-7624
N. Nagarajan https://orcid.org/0000-0003-3695-0078
K. Nakamura https://orcid.org/0000-0001-6148-4289
H. Nakano https://orcid.org/0000-0001-7665-0796
D. Nanadoumgar-Lacroze https://orcid.org/0009-0009-7255-8111
P. Narayan https://orcid.org/0009-0009-0599-532X
I. Nardecchia https://orcid.org/0000-0001-5558-2595
L. Naticchioni https://orcid.org/0000-0003-2918-0730
R. K. Nayak https://orcid.org/0000-0002-6814-7792
A. Nelson https://orcid.org/0000-0002-5909-4692

A. Neunzert https://orcid.org/0000-0003-0323-0111
L. Nguyen Quynh https://orcid.org/0000-0002-1828-3702
A. B. Nielsen https://orcid.org/0000-0001-8694-4026
A. Nishizawa https://orcid.org/0000-0003-3562-0990
W. Niu https://orcid.org/0000-0003-1470-532X
J. Novak https://orcid.org/0000-0002-6029-4712
R. Nowicki https://orcid.org/0009-0008-6626-0725
J. F. Nuño Siles https://orcid.org/0000-0001-8304-8066
L. K. Nuttall https://orcid.org/0000-0002-8599-8791
J. Oberling https://orcid.org/0009-0001-4174-3973
E. Oelker https://orcid.org/0000-0002-3916-1595
M. Oertel https://orcid.org/0000-0002-1884-8654
M. Ohashi https://orcid.org/0000-0001-8072-0304
F. Ohme https://orcid.org/0000-0003-0493-5607
R. Oliveri https://orcid.org/0000-0002-7497-871X
K. Oohara https://orcid.org/0000-0002-7518-6677
B. O'Reilly https://orcid.org/0000-0002-3874-8335
M. Orselli https://orcid.org/0000-0003-3563-8576
R. O'Shaughnessy https://orcid.org/0000-0001-5832-8517
S. Oshino https://orcid.org/0000-0002-2794-6029
I. Ota https://orcid.org/0000-0001-5045-2484
D. J. Ottaway https://orcid.org/0000-0001-6794-1591
B. J. Owen https://orcid.org/0000-0003-3919-0780
A. E. Pace https://orcid.org/0009-0003-4044-0334
R. Pagano https://orcid.org/0000-0001-8362-0130
M. A. Page https://orcid.org/0000-0002-5298-7914
A. Pai https://orcid.org/0000-0003-3476-4589
S. Pal https://orcid.org/0000-0003-2172-8589
M. A. Palaia https://orcid.org/0009-0007-3296-8648
C. Palomba https://orcid.org/0000-0002-4450-9883
P. Palud https://orcid.org/0000-0002-5850-6325
K. C. Pan https://orcid.org/0000-0002-1473-9880
F. Pannarale https://orcid.org/0000-0002-7537-3210
F. Paoletti https://orcid.org/0000-0001-8898-1963
A. Paolone https://orcid.org/0000-0002-4839-7815
A. Papadopoulos https://orcid.org/0009-0006-1882-996X
L. Papalini https://orcid.org/0000-0002-5219-0454
G. Papigkiotis https://orcid.org/0009-0008-2205-7426
A. Parisi https://orcid.org/0000-0003-0251-8914
J. Park https://orcid.org/0000-0002-7510-0079
W. Parker https://orcid.org/0000-0002-7711-4423
D. Pascucci https://orcid.org/0000-0003-1907-0175
A. Pasqualetti https://orcid.org/0000-0003-0620-5990
R. Passaquieti https://orcid.org/0000-0003-4753-9428
O. Patane https://orcid.org/0000-0002-4850-2355
A. V. Patel https://orcid.org/0000-0001-6872-9197
L. Pathak https://orcid.org/0000-0002-9523-7945
B. Patricelli https://orcid.org/0000-0001-6709-0969
K. Paul https://orcid.org/0000-0002-8406-6503
S. Paul https://orcid.org/0000-0002-4449-1732
E. Payne https://orcid.org/0000-0003-4507-8373
A. Pele https://orcid.org/0000-0002-1873-3769
F. E. Peña Arellano https://orcid.org/0000-0002-8516-5159
S. Penn https://orcid.org/0000-0003-4956-0853
A. Perego https://orcid.org/0000-0002-0936-8237
C. Périgois https://orcid.org/0000-0002-9779-2838
G. Perna https://orcid.org/0000-0002-7364-1904
A. Perreca https://orcid.org/0000-0002-6269-2490
J. Perret https://orcid.org/0009-0006-4975-1536
S. Perriès https://orcid.org/0000-0003-2213-3579
H. P. Pfeiffer https://orcid.org/0000-0001-9288-519X
K. A. Pham https://orcid.org/0000-0002-7650-1034
K. S. Phukon https://orcid.org/0000-0003-1561-0760
L. Piccari https://orcid.org/0009-0000-0247-4339
O. J. Piccinni https://orcid.org/0000-0001-5478-3950
M. Pichot https://orcid.org/0000-0002-4439-8968
M. Piendibene https://orcid.org/0000-0003-2434-488X
F. Piergiovanni https://orcid.org/0000-0001-8063-828X
L. Pierini https://orcid.org/0000-0003-0945-2196
G. Pierra https://orcid.org/0000-0003-3970-7970
V. Pierro https://orcid.org/0000-0002-6020-5521
M. Pillas https://orcid.org/0000-0003-3224-2146
F. Pilo https://orcid.org/0000-0003-4967-7090
L. Pinard https://orcid.org/0000-0002-8842-1867
I. M. Pinto https://orcid.org/0000-0002-2679-4457
M. Pinto https://orcid.org/0009-0003-4339-9971
B. J. Piotrzkowski https://orcid.org/0000-0001-8919-0899
M. D. Pitkin https://orcid.org/0000-0003-4548-526X
A. Placidi https://orcid.org/0000-0001-8032-4416
E. Placidi https://orcid.org/0000-0002-3820-8451
M. L. Planas https://orcid.org/0000-0001-8278-7406
W. Plastino https://orcid.org/0000-0002-5737-6346
C. Plunkett https://orcid.org/0000-0002-1144-6708
R. Poggiani https://orcid.org/0000-0002-9968-2464
L. Pompili https://orcid.org/0000-0002-0710-6778
C. Posnansky https://orcid.org/0009-0009-7137-9795
R. Poulton https://orcid.org/0000-0003-2049-520X
J. Powell https://orcid.org/0000-0002-1357-4164
M. Pracchia https://orcid.org/0009-0001-8343-719X
B. K. Pradhan https://orcid.org/0000-0002-2526-1421
T. Pradier https://orcid.org/0000-0001-5501-0060
K. Prasai https://orcid.org/0000-0001-6552-097X
G. Pratten https://orcid.org/0000-0003-4984-0775
G. Principe https://orcid.org/0000-0003-0406-7387
G. A. Prodi https://orcid.org/0000-0001-5256-915X
A. Puecher https://orcid.org/0000-0003-1357-4348
J. Pullin https://orcid.org/0000-0001-8248-603X
M. Pürrer https://orcid.org/0000-0002-3329-9788
H. Qi https://orcid.org/0000-0001-6339-1537
J. Qin https://orcid.org/0000-0002-7120-9026
G. Quéméner https://orcid.org/0000-0001-6703-6655
B. Rajbhandari https://orcid.org/0000-0001-7568-1611
K. E. Ramirez https://orcid.org/0000-0003-2194-7669
F. A. Ramis Vidal https://orcid.org/0000-0001-6143-2104
M. Ramos Arevalo https://orcid.org/0009-0003-1528-8326
A. Ramos-Buades https://orcid.org/0000-0002-6874-7421
S. Ranjan https://orcid.org/0000-0001-7480-9329
P. Rapagnani https://orcid.org/0000-0002-1865-6126
A. Ray https://orcid.org/0000-0002-7322-4748
V. Raymond https://orcid.org/0000-0003-0066-0095
M. Razzano https://orcid.org/0000-0003-4825-1629
D. H. Reitze https://orcid.org/0000-0002-5756-1111
B. Revenu https://orcid.org/0000-0002-7629-4805
L. Ricca https://orcid.org/0009-0002-1638-0610
F. Ricci https://orcid.org/0000-0001-5475-4447
M. Ricci https://orcid.org/0009-0008-7421-4331
A. Ricciardone https://orcid.org/0000-0002-5688-455X
J. W. Richardson https://orcid.org/0000-0002-1472-4806
K. Riles https://orcid.org/0000-0002-6418-5812
S. Rinaldi https://orcid.org/0000-0001-5799-4155
A. Rocchi https://orcid.org/0000-0002-1382-9016
L. Rolland https://orcid.org/0000-0003-0589-9687
J. G. Rollins https://orcid.org/0000-0002-9388-2799
A. E. Romano https://orcid.org/0000-0002-0314-8698

R. Romano https://orcid.org/0000-0002-0485-6936
A. Romero https://orcid.org/0000-0003-2275-4164
S. Ronchini https://orcid.org/0000-0003-0020-687X
T. J. Roocke https://orcid.org/0000-0003-2640-9683
D. Rosińska https://orcid.org/0000-0002-3681-9304
M. P. Ross https://orcid.org/0000-0002-8955-5269
M. Rossello-Sastre https://orcid.org/0000-0002-3341-3480
S. Rowan https://orcid.org/0000-0002-0666-9907
S. K. Roy https://orcid.org/0000-0001-9295-5119
S. Roy https://orcid.org/0000-0003-2147-5411
D. Rozza https://orcid.org/0000-0002-7378-6353
E. Ruiz Morales https://orcid.org/0000-0002-0995-595X
S. Sachdev https://orcid.org/0000-0002-0525-2317
P. Saffarieh https://orcid.org/0009-0000-7504-3660
S. Safi-Harb https://orcid.org/0000-0001-6189-7665
M. R. Sah https://orcid.org/0009-0005-9881-1788
S. Saha https://orcid.org/0000-0002-3333-8070
T. Sainrat https://orcid.org/0009-0003-0169-266X
S. Sajith Menon https://orcid.org/0009-0008-4985-1320
Y. Sakai https://orcid.org/0000-0001-8810-4813
M. Sakellariadou https://orcid.org/0000-0002-2715-1517
S. Sakon https://orcid.org/0000-0002-5861-3024
O. S. Salafia https://orcid.org/0000-0003-4924-7322
F. Salces-Carcoba https://orcid.org/0000-0001-7049-4438
M. Saleem https://orcid.org/0000-0002-3836-7751
F. Salemi https://orcid.org/0000-0002-9511-3846
M. Sallé https://orcid.org/0000-0002-6620-6672
S. Salvador https://orcid.org/0000-0003-3444-7807
A. Samajdar https://orcid.org/0000-0002-0857-6018
N. Sanchis-Gual https://orcid.org/0000-0001-5375-7494
E. M. Sänger https://orcid.org/0009-0003-6642-8974
F. Santoliquido https://orcid.org/0000-0003-3752-1400
A. Sasli https://orcid.org/0000-0001-7357-0889
P. Sassi https://orcid.org/0000-0002-4920-2784
B. Sassolas https://orcid.org/0000-0002-3077-8951
B. S. Sathyaprakash https://orcid.org/0000-0003-3845-7586
O. Sauter https://orcid.org/0000-0003-2293-1554
R. L. Savage https://orcid.org/0000-0003-3317-1036
T. Sawada https://orcid.org/0000-0001-5726-7150
M. G. Schiworski https://orcid.org/0000-0001-9298-004X
P. Schmidt https://orcid.org/0000-0003-1542-1791
S. Schmidt https://orcid.org/0000-0002-8206-8089
R. Schnabel https://orcid.org/0000-0003-2896-4218
K. Schouteden https://orcid.org/0000-0002-5975-585X
E. Schwartz https://orcid.org/0000-0001-8922-7794
M. Scialpi https://orcid.org/0009-0007-6434-1460
J. Scott https://orcid.org/0000-0001-6701-6515
S. M. Scott https://orcid.org/0000-0002-9875-7700
R. M. Sedas https://orcid.org/0000-0001-8961-3855
M. Seglar-Arroyo https://orcid.org/0000-0001-8654-409X
Y. Sekiguchi https://orcid.org/0000-0002-2648-3835
A. S. Sengupta https://orcid.org/0000-0002-3212-0475
E. G. Seo https://orcid.org/0000-0002-8588-4794
J. W. Seo https://orcid.org/0000-0003-4937-0769
M. Serra https://orcid.org/0000-0002-6093-8063
U. S. Shah https://orcid.org/0000-0001-8249-7425
M. A. Shaikh https://orcid.org/0000-0003-0826-6164
L. Shao https://orcid.org/0000-0002-1334-8853
A. K. Sharma https://orcid.org/0000-0003-0067-346X
A. Sharma https://orcid.org/0009-0007-2194-8633
P. Shawhan https://orcid.org/0000-0002-8249-8070
N. S. Shcheblanov https://orcid.org/0000-0001-8696-2435
H. Shinkai https://orcid.org/0000-0003-1082-2844
D. H. Shoemaker https://orcid.org/0000-0002-4147-2560
D. M. Shoemaker https://orcid.org/0000-0002-9899-6357
H. Siegel https://orcid.org/0000-0001-5161-4617
D. Sigg https://orcid.org/0000-0003-4606-6526
L. Silenzi https://orcid.org/0000-0001-7316-3239
L. Silvestri https://orcid.org/0009-0008-5207-661X
L. P. Singer https://orcid.org/0000-0001-9898-5597
D. Singh https://orcid.org/0000-0001-9675-4584
M. K. Singh https://orcid.org/0000-0001-8081-4888
N. Singh https://orcid.org/0000-0002-1135-3456
A. M. Sintes https://orcid.org/0000-0001-9050-7515
V. Skliris https://orcid.org/0000-0003-0902-9216
B. J. J. Slagmolen https://orcid.org/0000-0002-2471-3828
J. R. Smith https://orcid.org/0000-0003-0638-9670
L. Smith https://orcid.org/0000-0002-3035-0947
R. J. E. Smith https://orcid.org/0000-0001-8516-3324
W. J. Smith https://orcid.org/0009-0003-7949-4911
K. Somiya https://orcid.org/0000-0003-2601-2264
I. Song https://orcid.org/0000-0002-4301-8281
S. Soni https://orcid.org/0000-0003-3856-8534
V. Sordini https://orcid.org/0000-0003-0885-824X
H. Sotani https://orcid.org/0000-0002-3239-2921
F. Spada https://orcid.org/0000-0001-5664-1657
V. Spagnuolo https://orcid.org/0000-0002-0098-4260
A. P. Spencer https://orcid.org/0000-0003-4418-3366
P. Spinicelli https://orcid.org/0000-0001-8078-6047
F. Stachurski https://orcid.org/0000-0002-8658-5753
D. A. Steer https://orcid.org/0000-0002-8781-1273
N. Steinle https://orcid.org/0000-0003-0658-402X
S. Steinlechner https://orcid.org/0000-0003-4710-8548
N. Stergioulas https://orcid.org/0000-0002-5490-5302
S. Sudhagar https://orcid.org/0000-0001-8578-4665
L. Suleiman https://orcid.org/0000-0003-3783-7448
J. M. Sullivan https://orcid.org/0009-0006-6238-9594
J. Sun https://orcid.org/0009-0008-8278-0077
L. Sun https://orcid.org/0000-0001-7959-892X
J. Suresh https://orcid.org/0000-0003-2389-6666
P. J. Sutton https://orcid.org/0000-0003-1614-3922
B. L. Swinkels https://orcid.org/0000-0002-3066-3601
A. Syx https://orcid.org/0009-0000-6424-6411
M. J. Szczepańczyk https://orcid.org/0000-0002-6167-6149
P. Szewczyk https://orcid.org/0000-0002-1339-9167
M. Tacca https://orcid.org/0000-0003-1353-0441
H. Tagoshi https://orcid.org/0000-0001-8530-9178
H. Takahashi https://orcid.org/0000-0003-0596-4397
R. Takahashi https://orcid.org/0000-0003-1367-5149
A. Takamori https://orcid.org/0000-0001-6032-1330
S. Takano https://orcid.org/0000-0002-1266-4555
H. Takeda https://orcid.org/0000-0001-9937-2557
N. Tamanini https://orcid.org/0000-0001-8760-5421
S. J. Tanaka https://orcid.org/0000-0002-8796-1992
S. Tanioka https://orcid.org/0000-0003-3321-1018
L. Tao https://orcid.org/0000-0003-4382-5507
E. N. Tapia San Martín https://orcid.org/0000-0002-4817-5606
A. Taruya https://orcid.org/0000-0002-4016-1955
J. D. Tasson https://orcid.org/0000-0002-4777-5087
J. G. Tau https://orcid.org/0009-0004-7428-762X
R. Tenorio https://orcid.org/0000-0002-3582-2587

A. Theodoropoulos https://orcid.org/0000-0003-4486-7135
L. M. Thomas https://orcid.org/0000-0003-3271-6436
J. E. Thompson https://orcid.org/0000-0002-0419-5517
E. Thrane https://orcid.org/0000-0002-4418-3895
S. Tibrewal https://orcid.org/0000-0003-4159-8291
J. Tissino https://orcid.org/0000-0003-2483-6710
S. Tiwari https://orcid.org/0000-0003-1611-6625
V. Tiwari https://orcid.org/0000-0002-1602-4176
A. M. Toivonen https://orcid.org/0009-0008-9546-2035
K. Toland https://orcid.org/0000-0001-9537-9698
A. E. Tolley https://orcid.org/0000-0001-9841-943X
T. Tomaru https://orcid.org/0000-0002-8927-9014
T. Tomura https://orcid.org/0000-0002-7504-8258
H. Tong https://orcid.org/0000-0002-4534-0485
A. Torres-Forné https://orcid.org/0000-0001-8709-5118
I. Tosta e Melo https://orcid.org/0000-0001-5833-4052
E. Tournefier https://orcid.org/0000-0002-5465-9607
A. Trapananti https://orcid.org/0000-0001-7763-5758
R. Travaglini https://orcid.org/0000-0002-5288-1407
F. Travasso https://orcid.org/0000-0002-4653-6156
M. C. Tringali https://orcid.org/0000-0001-5087-189X
A. Tripathee https://orcid.org/0000-0002-6976-5576
G. Troian https://orcid.org/0000-0001-6837-607X
A. Trovato https://orcid.org/0000-0002-9714-1904
T. Tsang https://orcid.org/0000-0003-3666-686X
S. Tsuchida https://orcid.org/0000-0001-8217-0764
L. Tsukada https://orcid.org/0000-0003-0596-5648
K. Turbang https://orcid.org/0000-0002-9296-8603
M. Turconi https://orcid.org/0000-0001-9999-2027
H. Ubach https://orcid.org/0000-0002-0679-9074
N. Uchikata https://orcid.org/0000-0003-0030-3653
T. Uchiyama https://orcid.org/0000-0003-2148-1694
R. P. Udall https://orcid.org/0000-0001-6877-3278
T. Uehara https://orcid.org/0000-0003-4375-098X
K. Ueno https://orcid.org/0000-0003-3227-6055
V. Undheim https://orcid.org/0000-0003-4028-0054
T. Ushiba https://orcid.org/0000-0002-5059-4033
M. Vacatello https://orcid.org/0009-0006-0934-1014
H. Vahlbruch https://orcid.org/0000-0003-2357-2338
N. Vaidya https://orcid.org/0000-0003-1843-7545
G. Vajente https://orcid.org/0000-0002-7656-6882
J. Valencia https://orcid.org/0000-0003-2648-9759
M. Valentini https://orcid.org/0000-0003-1215-4552
S. A. Vallejo-Peña https://orcid.org/0000-0002-6827-9509
V. Valsan https://orcid.org/0000-0003-0315-4091
M. van Dael https://orcid.org/0000-0002-6061-8131
E. Van den Bossche https://orcid.org/0009-0009-2070-0964
J. F. J. van den Brand https://orcid.org/0000-0003-4434-5353
M. van der Sluys https://orcid.org/0000-0003-1231-0762
J. van Dongen https://orcid.org/0000-0003-0964-2483
H. van Haevermaet https://orcid.org/0000-0003-2386-957X
J. V. van Heijningen https://orcid.org/0000-0002-8391-7513
P. Van Hove https://orcid.org/0000-0002-2431-3381
N. van Remortel https://orcid.org/0000-0003-4180-8199
A. F. Vargas https://orcid.org/0000-0001-8396-5227
V. Varma https://orcid.org/0000-0002-9994-1761
A. Vecchio https://orcid.org/0000-0002-6254-1617
J. Veitch https://orcid.org/0000-0002-6508-0713
P. J. Veitch https://orcid.org/0000-0002-2597-435X
R. C. Venterea https://orcid.org/0000-0003-3299-3804
P. Verdier https://orcid.org/0000-0003-3090-2948
D. Verkindt https://orcid.org/0000-0003-4344-7227
Y. Verma https://orcid.org/0000-0003-4147-3173
S. M. Vermeulen https://orcid.org/0000-0003-4227-8214
A. Veutro https://orcid.org/0009-0002-9160-5808
A. Viceré https://orcid.org/0000-0003-0624-6231
A. D. Viets https://orcid.org/0000-0002-4241-1428
A. Vijaykumar https://orcid.org/0000-0002-4103-0666
V. Villa-Ortega https://orcid.org/0000-0001-7983-1963
E. T. Vincent https://orcid.org/0000-0002-0442-1916
S. Vitale https://orcid.org/0000-0003-2700-0767
H. Vocca https://orcid.org/0000-0002-1200-3917
D. Voigt https://orcid.org/0000-0001-9075-6503
L. Vujeva https://orcid.org/0000-0001-7697-8361
S. P. Vyatchanin https://orcid.org/0000-0002-6823-911X
M. Wade https://orcid.org/0000-0002-5703-4469
K. J. Wagner https://orcid.org/0000-0002-7255-4251
H. Wang https://orcid.org/0000-0002-6589-2738
Y. F. Wang https://orcid.org/0000-0002-2928-2916
G. Waratkar https://orcid.org/0000-0003-3630-9440
M. Was https://orcid.org/0000-0002-1890-1128
T. Washimi https://orcid.org/0000-0001-5792-4907
N. L. Weickhardt https://orcid.org/0000-0002-3923-5806
A. J. Weinstein https://orcid.org/0000-0002-0928-6784
L. Wen https://orcid.org/0000-0001-7987-295X
K. Wette https://orcid.org/0000-0002-4394-7179
J. T. Whelan https://orcid.org/0000-0001-5710-6576
B. F. Whiting https://orcid.org/0000-0002-8501-8669
C. Whittle https://orcid.org/0000-0002-8833-7438
D. Wilken https://orcid.org/0000-0002-7290-9411
D. Williams https://orcid.org/0000-0003-3772-198X
M. J. Williams https://orcid.org/0000-0003-2198-2974
N. S. Williams https://orcid.org/0000-0002-5656-8119
J. L. Willis https://orcid.org/0000-0002-9929-0225
B. Willke https://orcid.org/0000-0003-0524-2925
M. Wils https://orcid.org/0000-0002-1544-7193
G. Woan https://orcid.org/0000-0003-0381-0394
H. T. Wong https://orcid.org/0000-0003-4145-4394
I. C. F. Wong https://orcid.org/0000-0003-2166-0027
M. Wright https://orcid.org/0000-0003-1829-7482
C. Wu https://orcid.org/0000-0003-3191-8845
D. S. Wu https://orcid.org/0000-0003-2849-3751
H. Wu https://orcid.org/0000-0003-4813-3833
Z. Wu https://orcid.org/0000-0002-0032-5257
D. M. Wysocki https://orcid.org/0000-0001-9138-4078
V. A. Xu https://orcid.org/0000-0002-3020-3293
Y. Xu https://orcid.org/0000-0001-8697-3505
N. Yadav https://orcid.org/0009-0009-5010-1065
H. Yamamoto https://orcid.org/0000-0001-6919-9570
K. Yamamoto https://orcid.org/0000-0002-3033-2845
T. S. Yamamoto https://orcid.org/0000-0002-8181-924X
T. Yamamoto https://orcid.org/0000-0002-0808-4822
R. Yamazaki https://orcid.org/0000-0002-1251-7889
K. Z. Yang https://orcid.org/0000-0001-8083-4037
Y. Yang https://orcid.org/0000-0002-3780-1413
Z. Yarbrough https://orcid.org/0000-0002-9825-1136
A. B. Yelikar https://orcid.org/0000-0002-8065-1174
J. Yokoyama https://orcid.org/0000-0001-7127-4808
H. Yuzurihara https://orcid.org/0000-0002-3710-6613
M. Zeeshan https://orcid.org/0000-0002-6494-7303

M. Zeoli https://orcid.org/0009-0007-1898-4844
M. Zevin https://orcid.org/0000-0002-0147-0835
R. Zhang https://orcid.org/0000-0001-8095-483X
C. Zhao https://orcid.org/0000-0001-5825-2401
Z.-C. Zhao https://orcid.org/0000-0001-5180-4496
Y. Zheng https://orcid.org/0000-0002-5432-1331
H. Zhong https://orcid.org/0000-0001-8324-5158
Z.-H. Zhu https://orcid.org/0000-0002-3567-6743
A. B. Zimmerman https://orcid.org/0000-0002-7453-6372
M. E. Zucker https://orcid.org/0000-0002-2544-1596
J. Zweizig https://orcid.org/0000-0002-1521-3397

## References

Aasi, J., Abbott, B. P., Abbott, R., et al. 2015, CQGra, 32, 074001
Abac, A. G., Abbott, R., Abe, H., et al. 2024a, ApJ, 973, 132
Abac, A. G., Abbott, R., Abouelfettouh, I., et al. 2024b, ApJL, 970, L34
Abac, A. G., Abouelfettouh, I., Acernese, F., et al. 2025a, ApJL, 995, L18
Abac, A. G., Abouelfettouh, I., Acernese, F., et al. 2025b, ApJL, 993, L25
Abac, A. G., Abouelfettouh, I., Acernese, F., et al. 2025c, PhRvD, 112, 102005
Abac, A. G., Abouelfettouh, I., Acernese, F., et al. 2026a, ApJL, 1004, L21
Abac, A. G., Abouelfettouh, I., Acernese, F., et al. 2026b, ApJ, 998, 70
Abac, A. G., Abouelfettouh, I., Acernese, F., et al. 2026c, arXiv:2509.07348
Abac, A. G., Abouelfettouh, I., Acernese, F., et al. 2026d, arXiv:2508.18083
Abac, A. G., Abouelfettouh, I., Acernese, F., et al. 2026e, arXiv:2509.04348
Abac, A. G., Abouelfettouh, I., Acernese, F., et al. 2026f, arXiv:2603.19019
Abac, A. G., Abouelfettouh, I., Acernese, F., et al. 2026g, arXiv:2603.19020
Abac, A. G., Abouelfettouh, I., Acernese, F., et al. 2026h, arXiv:2603.19021
Abac, A. G., Abouelfettouh, I., Acernese, F., et al. 2026i, arXiv:2512.16347
Abbott, B. P., Abbott, R., Abbott, T. D., et al. 2016a, PhRvL, 116, 061102
Abbott, B. P., Abbott, R., Abbott, T. D., et al. 2016b, PhRvL, 116, 241102
Abbott, B. P., Abbott, R., Abbott, T. D., et al. 2016c, PhRvX, 6, 041015
Abbott, B. P., Abbott, R., Abbott, T. D., et al. 2016d, PhRvX, 6, 041014
Abbott, B. P., Abbott, R., Abbott, T. D., et al. 2017a, PhRvL, 119, 161101
Abbott, B. P., Abbott, R., Abbott, T. D., et al. 2017b, PhRvD, 95, 042003
Abbott, B. P., Abbott, R., Abbott, T. D., et al. 2017c, ApJL, 848, L12
Abbott, B. P., Abbott, R., Abbott, T. D., et al. 2017d, PhRvL, 118, 221101
Abbott, B. P., Abbott, R., Abbott, T. D., et al. 2017e, ApJL, 851, L16
Abbott, B. P., Abbott, R., Abbott, T. D., et al. 2018, PhRvL, 121, 231103
Abbott, B. P., Abbott, R., Abbott, T. D., et al. 2019a, PhRvX, 9, 031040
Abbott, B. P., Abbott, R., Abbott, T. D., et al. 2019b, ApJ, 875, 161
Abbott, B. P., Abbott, R., Abbott, T. D., et al. 2019c, PhRvD, 100, 024017
Abbott, B. P., Abbott, R., Abbott, T. D., et al. 2019d, PhRvX, 9, 011001
Abbott, B. P., Abbott, R., Abbott, T. D., et al. 2019e, PhRvL, 123, 161102
Abbott, B. P., Abbott, R., Abbott, T. D., et al. 2020a, LRR, 23, 3
Abbott, B. P., Abbott, R., Abbott, T. D., et al. 2020b, CQGra, 37, 055002
Abbott, B. P., Abbott, R., Abbott, T. D., et al. 2020c, CQGra, 37, 045006
Abbott, B. P., Abbott, R., Abbott, T. D., et al. 2020d, ApJL, 892, L3
Abbott, R., Abbott, T. D., Abraham, S., et al. 2021a, ApJL, 915, L5
Abbott, R., Abbott, T. D., Abraham, S., et al. 2021b, PhRvX, 11, 021053
Abbott, R., Abbott, T. D., Abraham, S., et al. 2021c, ApJ, 921, 80
Abbott, R., Abbott, T. D., Abraham, S., et al. 2021d, PhRvL, 126, 241102
Abbott, R., Abbott, T. D., Abraham, S., et al. 2021e, PhRvD, 104, 022004
Abbott, R., Abbott, T. D., Abraham, S., et al. 2021f, PhRvD, 104, 022005
Abbott, R., Abbott, T. D., Acernese, F., et al. 2021g, PhRvD, 104, 122004
Abbott, R., Abbott, T. D., Acernese, F., et al. 2022a, PhRvL, 129, 061104
Abbott, R., Abbott, T. D., Acernese, F., et al. 2022b, ApJ, 932, 133
Abbott, R., Abbott, T. D., Acernese, F., et al. 2023a, PhRvX, 13, 041039
Abbott, R., Abbott, T. D., Acernese, F., et al. 2023b, PhRvX, 13, 011048
Abbott, R., Abbott, T. D., Acernese, F., et al. 2024, PhRvD, 109, 022001
Abbott, R., Abe, H., Acernese, F., et al. 2023c, MNRAS, 524, 5984
Acernese, F., Agathos, M., Agatsuma, K., et al. 2015, CQGra, 32, 024001
Adams, T., Buskulic, D., Germain, V., et al. 2016, CQGra, 33, 175012
Ade, P. A. R., Aghanim, N., Arnaud, M., et al. 2016, A&A, 594, A13
Ajith, P., Hannam, M., Husa, S., et al. 2011, PhRvL, 106, 241101
Akutsu, T., Ando, M., Arai, K., et al. 2019, NatAs, 3, 35
Alléné, C., Aubin, F., Bentara, I., et al. 2025, CQGra, 42, 105009
Alsing, J., Silva, H. O., & Berti, E. 2018, MNRAS, 478, 1377
Álvarez-López, S., Liyanage, A., Ding, J., Ng, R., & McIver, J. 2024, CQGra, 41, 085007
Andres, N., Assiduo, M., Aubin, F., et al. 2022, CQGra, 39, 055002
Antoniadis, J., Tauris, T. M., Ozel, F., et al. 2016, arXiv:1605.01665
Apostolatos, T. A., Cutler, C., Sussman, G. J., & Thorne, K. S. 1994, PhRvD, 49, 6274
Ashton, G., Hübner, M., Lasky, P. D., et al. 2019, ApJS, 241, 27
Ashton, G., Thiele, S., Lecoeuche, Y., McIver, J., & Nuttall, L. K. 2022, CQGra, 39, 175004
Aubin, F., Brighenti, F., Chierici, R., et al. 2021, CQGra, 38, 095004
Bailyn, C. D., Jain, R. K., Coppi, P., & Orosz, J. A. 1998, ApJ, 499, 367
Baird, E., Fairhurst, S., Hannam, M., & Murphy, P. 2013, PhRvD, 87, 024035
Barkat, Z., Rakavy, G., & Sack, N. 1967, PhRvL, 18, 379
Belczynski, K., Heger, A., Gladysz, W., et al. 2016, A&A, 594, A97
Berry, C. P. L., Mandel, I., Middleton, H., et al. 2015, ApJ, 804, 114
Biswas, B., & Rosswog, S. 2025, PhRvD, 112, 023045
Blanchet, L. 2014, LRR, 17, 2
Brandes, L., & Weise, W. 2025, PhRvD, 111, 034005
Brown, J. C. 1991, ASAJ, 89, 425
Cannon, K., Cariou, R., Chapman, A., et al. 2012, ApJ, 748, 136
Cannon, K., Caudill, S., Chan, C., et al. 2021, SoftX, 14, 100680
Chatterji, S., Blackburn, L., Martin, G., & Katsavounidis, E. 2004, CQGra, 21, S1809
Chatziioannou, K., Cornish, N., Klein, A., & Yunes, N. 2015, ApJL, 798, L17
Chatziioannou, K., Cornish, N., Wijngaarden, M., & Littenberg, T. B. 2021, PhRvD, 103, 044013
Chia, H. S., Olsen, S., Roulet, J., et al. 2022, PhRvD, 106, 024009
Christensen, N., & Meyer, R. 2022, RvMP, 94, 025001
Chu, Q., Kovalam, M., Wen, L., et al. 2022, PhRvD, 105, 024023
Colleoni, M., Ramis Vidal, F. A., García-Quirós, C., Akçay, S., & Bera, S. 2025, PhRvD, 111, 104019
Cornish, N. J., & Littenberg, T. B. 2015, CQGra, 32, 135012
Cornish, N. J., Littenberg, T. B., Bécsy, B., et al. 2021, PhRvD, 103, 044006
Cutler, C., Apostolatos, T. A., Bildsten, L., et al. 1993, PhRvL, 70, 2984
Cutler, C., & Flanagan, É. E. 1994, PhRvD, 49, 2658
Dal Canton, T., Nitz, A. H., Gadre, B., et al. 2021, ApJ, 923, 254
Dal Canton, T., Nitz, A. H., Lundgren, A. P., et al. 2014, PhRvD, 90, 082004
Damour, T. 2001, PhRvD, 64, 124013
Davies, G. S., Dent, T., Tápai, M., et al. 2020, PhRvD, 102, 022004
Del Pozzo, W., Berry, C. P. L., Ghosh, A., et al. 2018, MNRAS, 479, 601
Dietrich, T., Coughlin, M. W., Pang, P. T. H., et al. 2020, Sci, 370, 1450
Dietrich, T., Samajdar, A., Khan, S., et al. 2019, PhRvD, 100, 044003
Drago, M., Klimenko, S., Lazzaro, C., et al. 2021, SoftX, 14, 100678
El-Badry, K., Rix, H.-W., Latham, D. W., et al. 2024, OJAp, 7, 58
Essick, R., Coughlin, M. W., Zevin, M., et al. 2025, PhRvD, 112, 102001
Essick, R., Godwin, P., Hanna, C., Blackburn, L., & Katsavounidis, E. 2020, Mach. Learn. Sci. Technol., 2, 015004
Estellés, H., Colleoni, M., García-Quirós, C., et al. 2022a, PhRvD, 105, 084040
Estellés, H., Husa, S., Colleoni, M., et al. 2022b, ApJ, 924, 79
Ewing, B., Huxford, R., Singh, D., et al. 2024, PhRvD, 109, 042008
Fairhurst, S. 2009, NJPh, 11, 123006
Fairhurst, S. 2011, CQGra, 28, 105021
Fan, Y.-Z., Han, M.-Z., Jiang, J.-L., Shao, D.-S., & Tang, S.-P. 2024, PhRvD, 109, 043052
Farr, B., Berry, C. P. L., Farr, W. M., et al. 2016, ApJ, 825, 116
Farr, W. M., Sravan, N., Cantrell, A., et al. 2011, ApJ, 741, 103
Farrow, N., Zhu, X.-J., & Thrane, E. 2019, ApJ, 876, 18
Finn, L. S., & Chernoff, D. F. 1993, PhRvD, 47, 2198
Fishbach, M., Holz, D. E., & Farr, B. 2017, ApJL, 840, L24
Fisher, R. P., Hemming, G., Bizouard, M.-A., et al. 2021, SoftX, 14, 100677
Flanagan, É. É., & Hinderer, T. 2008, PhRvD, 77, 021502
Fowler, W. A., & Hoyle, F. 1964, ApJS, 9, 201
Fryer, C. L., Woosley, S. E., & Heger, A. 2001, ApJ, 550, 372
García-Bellido, J., Nuño Siles, J. F., & Ruiz Morales, E. 2021, PDU, 31, 100791
Ghonge, S., Chatziioannou, K., Clark, J. A., et al. 2020, PhRvD, 102, 064056
Glanzer, J., Banagiri, S., Coughlin, S. B., et al. 2023, CQGra, 40, 065004
Golomb, J., Legred, I., Chatziioannou, K., & Landry, P. 2025, PhRvD, 111, 023029
Grover, K., Fairhurst, S., Farr, B. F., et al. 2014, PhRvD, 89, 042004
Gupta, T., & Cornish, N. J. 2024, PhRvD, 109, 064040
Hamilton, E., London, L., Thompson, J. E., et al. 2021, PhRvD, 104, 124027
Hanna, C., Caudill, S., Messick, C., et al. 2020, PhRvD, 101, 022003
Harris, C. R., Millman, K. J., van der Walt, S. J., et al. 2020, Natur, 585, 357
Hourihane, S., Chatziioannou, K., Wijngaarden, M., et al. 2022, PhRvD, 106, 042006
Hoy, C., & Raymond, V. 2021, SoftX, 15, 100765
Huang, Y., Middleton, H., Ng, K. K. Y., Vitale, S., & Veitch, J. 2018, PhRvD, 98, 123021

Hunter, J. D. 2007, CSE, 9, 90
Huth, S., Pang, P. T. H., Tews, I., et al. 2022, Natur, 606, 276
Huxford, R., George, R., Trevor, M., Yarbrough, Z., & Godwin, P. 2024, arXiv:2412.04638
Johnson-McDaniel, N. K., Ghosh, A., Ghonge, S., et al. 2022, PhRvD, 105, 044020
Joshi, P., Tsukada, L., Hanna, C., et al. 2025, arXiv:2506.06497
Kafka, P. 1988, ESASP, 283, 121
Kalogera, V., & Baym, G. 1996, ApJL, 470, L61
Kasliwal, M. M., & Nissanke, S. 2014, ApJL, 789, L5
Kidder, L. E. 1995, PhRvD, 52, 821
Klimenko, S. 2022, arXiv:2201.01096
Klimenko, S., & Mitselmakher, G. 2004, CQGra, 21, S1819
Klimenko, S., Mohanty, S., Rakhmanov, M., & Mitselmakher, G. 2005, PhRvD, 72, 122002
Klimenko, S., Vedovato, G., Drago, M., et al. 2011, PhRvD, 83, 102001
Klimenko, S., Vedovato, G., Drago, M., et al. 2016, PhRvD, 93, 042004
Klimenko, S., Yakushin, I., Mercer, A., & Mitselmakher, G. 2008, CQGra, 25, 114029
Koloniari, A. E., Koursoumpa, E. C., Nousi, P., et al. 2025, MLS&T, 6, 015054
Kovalam, M., Kaium Patwary, M. A., Sreekumar, A. K., et al. 2022, ApJL, 927, L9
Kreidberg, L., Bailyn, C. D., Farr, W. M., & Kalogera, V. 2012, ApJ, 757, 36
Krolak, A., & Schutz, B. F. 1987, GReGr, 19, 1163
Kumar, P., & Dent, T. 2024, PhRvD, 110, 043036
Kyutoku, K., Shibata, M., & Taniguchi, K. 2021, LRR, 24, 5
Landry, P., Essick, R., & Chatziioannou, K. 2020, PhRvD, 101, 123007
Landry, P., & Read, J. S. 2021, ApJL, 921, L25
Legred, I., Chatziioannou, K., Essick, R., Han, S., & Landry, P. 2021, PhRvD, 104, 063003
LIGO Scientific Collaboration and Virgo Collaboration 2018, Data quality report user documentation https://docs.ligo.org/detchar/data-quality-report/
LIGO–-Virgo–-KAGRA Collaboration 2025b, GWTC-4.0: Parameter Estimation Data Release, v1, Zenodo, doi:10.5281/zenodo.16053484
LIGO–Virgo–KAGRA Collaboration 2018, LVK Algorithm Library—LALSuite, Free Software (GPL), doi:10.7935/GT1W-FZ16
LIGO–Virgo–KAGRA Collaboration 2025a, https://emfollow.docs.ligo.org/userguide
LIGO–Virgo–KAGRA Collaboration 2025b, GWTC-4.0: Candidate Data Release, v2, Zenodo, doi:10.5281/zenodo.17014083
LIGO–Virgo–KAGRA Collaboration 2025c, GWTC-4.0: Parameter Estimation Data Release, v2, Zenodo, doi:10.5281/zenodo.17014085
LIGO–Virgo–KAGRA Collaboration 2025d, GWTC-4.0: Glitch Modelling for Events, v1, Zenodo, doi:10.5281/zenodo.16857060
LIGO–Virgo–KAGRA Collaboration 2025e, GWTC-4.0: Data Quality Products for Transient Gravitational Wave Searches, v1, Zenodo, doi:10.5281/zenodo.16856919
LIGO–Virgo–KAGRA Collaboration 2025f, GWTC-4: O4a Search Sensitivity Estimates, v1, Zenodo, doi:10.5281/zenodo.16740117
LIGO–Virgo–KAGRA Collaboration 2025g, GWTC-4: Cumulative Search Sensitivity Estimates, v1, Zenodo, doi:10.5281/zenodo.16740128
Lim, Y., Bhattacharya, A., Holt, J. W., & Pati, D. 2021, PhRvC, 104, L032802
Littenberg, T. B., & Cornish, N. J. 2015, PhRvD, 91, 084034
Littenberg, T. B., Kanner, J. B., Cornish, N. J., & Millhouse, M. 2016, PhRvD, 94, 044050
Macleod, D. M., Areeda, J. S., Coughlin, S. B., Massinger, T. J., & Urban, A. L. 2021a, SoftX, 13, 100657
Macleod, D., Goetz, E., Davis, D., et al. 2021b, gwdetchar/gwdetchar, v2.3.1, Zenodo, doi:10.5281/zenodo.2575786
Magee, R., Fong, H., Caudill, S., et al. 2019, ApJL, 878, L17
Margalit, B., & Metzger, B. D. 2017, ApJL, 850, L19
Matas, A., Dietrich, T., Buonanno, A., et al. 2020, PhRvD, 102, 043023
Mehta, A. K., Buonanno, A., Gair, J., et al. 2022, ApJ, 924, 39
Messick, C., Blackburn, K., Brady, P., et al. 2017, PhRvD, 95, 042001
Miller, M. C., Lamb, F. K., Dittmann, A. J., et al. 2019, ApJL, 887, L24
Miller, M. C., Lamb, F. K., Dittmann, A. J., et al. 2021, ApJL, 918, L28
Mishra, T., Bhaumik, S., Gayathri, V., et al. 2025, PhRvD, 111, 023054
Mishra, T., O'Brien, B., Szczepańczyk, M., et al. 2022, PhRvD, 105, 083018
NASA 2025, GCN, gcn.nasa.gov
Nathanail, A., Most, E. R., & Rezzolla, L. 2021, ApJL, 908, L28
Ng, K. K. Y., Vitale, S., Zimmerman, A., et al. 2018, PhRvD, 98, 083007
Nissanke, S., Kasliwal, M., & Georgieva, A. 2013, ApJ, 767, 124
Nissanke, S., Sievers, J., Dalal, N., & Holz, D. 2011, ApJ, 739, 99
Nitz, A. H., Capano, C., Nielsen, A. B., et al. 2019, ApJ, 872, 195
Nitz, A. H., Capano, C. D., Kumar, S., et al. 2021, ApJ, 922, 76
Nitz, A. H., Dal Canton, T., Davis, D., & Reyes, S. 2018, PhRvD, 98, 024050
Nitz, A. H., Dent, T., Dal Canton, T., Fairhurst, S., & Brown, D. A. 2017, ApJ, 849, 118
Nitz, A. H., Dent, T., Davies, G. S., et al. 2020a, ApJ, 891, 123
Nitz, A. H., Kumar, S., Wang, Y.-F., et al. 2023, ApJ, 946, 59
Nitz, A. H., Schäfer, M., & Dal Canton, T. 2020b, ApJL, 902, L29
Nitz, A. H., & Wang, Y. F. 2021a, PhRvL, 126, 021103
Nitz, A. H., & Wang, Y. F. 2021b, ApJ, 915, 54
Nuttall, L. K. 2018, RSPTA, 376, 20170286
Olsen, S., Venumadhav, T., Mushkin, J., et al. 2022, PhRvD, 106, 043009
Ottaway, D. J., Fritschel, P., & Waldman, S. J. 2012, OExpr, 20, 8329
Özel, F., & Freire, P. 2016, ARA&A, 54, 401
Özel, F., Psaltis, D., Narayan, R., & McClintock, J. E. 2010, ApJ, 725, 1918
Paek, G. S. H., Im, M., Jeong, M., et al. 2025, ApJ, 981, 38
Pannarale, F., Rezzolla, L., Ohme, F., & Read, J. S. 2011, PhRvD, 84, 104017
Pillas, M., Antier, S., Ackley, K., et al. 2025, PhRvD, 112, 083002
Poisson, E., & Will, C. M. 1995, PhRvD, 52, 848
Pompili, L., Buonanno, A., Estellés, H., et al. 2023, PhRvD, 108, 124035
Powell, J. 2018, CQGra, 35, 155017
Pratten, G., García-Quirós, C., Colleoni, M., et al. 2021, PhRvD, 103, 104056
Pratten, G., Schmidt, P., Buscicchio, R., & Thomas, L. M. 2020, PhRvR, 2, 043096
Pürrer, M., Hannam, M., & Ohme, F. 2016, PhRvD, 93, 084042
Raaijmakers, G., Greif, S. K., Hebeler, K., et al. 2021, ApJL, 918, L29
Racine, É. 2008, PhRvD, 78, 044021
Ramos-Buades, A., Buonanno, A., Estellés, H., et al. 2023, PhRvD, 108, 124037
Ray, A., Niu, W., Sakon, S., et al. 2023, arXiv:2306.07190
Relton, P., & Raymond, V. 2021, PhRvD, 104, 084039
Rezzolla, L., Most, E. R., & Weih, L. R. 2018, ApJL, 852, L25
Rhoades, C. E., & Ruffini, R. 1974, PhRvL, 32, 324
Riley, T. E., Watts, A. L., Bogdanov, S., et al. 2019, ApJL, 887, L21
Riley, T. E., Watts, A. L., Ray, P. S., et al. 2021, ApJL, 918, L27
Robinet, F., Arnaud, N., Leroy, N., et al. 2020, SoftX, 12, 100620
Romero-Shaw, I. M., Talbot, C., Biscoveanu, S., et al. 2020, MNRAS, 499, 3295
Ruiz, M., Shapiro, S. L., & Tsokaros, A. 2018, PhRvD, 97, 021501
Rutherford, N., Mendes, M., Svensson, I., et al. 2024, ApJL, 971, L19
Sachdev, S., Caudill, S., Fong, H., et al. 2019, arXiv:1901.08580
Sachdev, S., Magee, R., Hanna, C., et al. 2020, ApJL, 905, L25
Sakon, S., Tsukada, L., Fong, H., et al. 2024, PhRvD, 109, 044066
Salemi, F., Milotti, E., Prodi, G. A., et al. 2019, PhRvD, 100, 042003
Santamaría, L., Ohme, F., Ajith, P., et al. 2010, PhRvD, 82, 064016
Schmidt, P., Ohme, F., & Hannam, M. 2015, PhRvD, 91, 024043
Schutz, B. F. 1986, Natur, 323, 310
SCiMMA 2025, SCiMMA Hopskotch, https://scimma.org/hopskotch
Singer, L. P., Chen, H.-Y., Holz, D. E., et al. 2016, ApJL, 829, L15
Singer, L. P., Price, L. R., Farr, B., et al. 2014, ApJ, 795, 105
Smith, J. R., Abbott, T., Hirose, E., et al. 2011, CQGra, 28, 235005
Soni, S., Berger, B. K., Davis, D., et al. 2025, CQGra, 42, 085016
Speagle, J. S. 2020, MNRAS, 493, 3132
Spera, M., & Mapelli, M. 2017, MNRAS, 470, 4739
Stevenson, S., Berry, C. P. L., & Mandel, I. 2017, MNRAS, 471, 2801
Stevenson, S., Sampson, M., Powell, J., et al. 2019, ApJ, 882, 121
Talbot, C., Biscoveanu, S., Zimmerman, A., et al. 2025, CQGra, 42, 235023
Talbot, C., & Thrane, E. 2017, PhRvD, 96, 023012
Thompson, J. E., Fauchon-Jones, E., Khan, S., et al. 2020, PhRvD, 101, 124059
Thompson, J. E., Hamilton, E., London, L., et al. 2024, PhRvD, 109, 063012
Thrane, E., & Talbot, C. 2019, PASA, 36, e010
Tsukada, L., Joshi, P., Adhicary, S., et al. 2023, PhRvD, 108, 043004
Usman, S. A., Nitz, A. H., Harry, I. W., et al. 2016, CQGra, 33, 215004
Vajente, G., Huang, Y., Isi, M., et al. 2020, PhRvD, 101, 042003
Varma, V., Field, S. E., Scheel, M. A., et al. 2019, PhRvR, 1, 033015
Vazsonyi, L., & Davis, D. 2023, CQGra, 40, 035008
Veitch, J., Mandel, I., Aylott, B., et al. 2012, PhRvD, 85, 104045
Veitch, J., Raymond, V., Farr, B., et al. 2015, PhRvD, 91, 042003
Venumadhav, T., Zackay, B., Roulet, J., Dai, L., & Zaldarriaga, M. 2019, PhRvD, 100, 023011
Venumadhav, T., Zackay, B., Roulet, J., Dai, L., & Zaldarriaga, M. 2020, PhRvD, 101, 083030
Viets, A. D., Wade, M., Urban, A. L., et al. 2018, CQGra, 35, 095015
Villa-Ortega, V., Dent, T., & Barroso, A. C. 2022, MNRAS, 515, 5718
Vines, J., Flanagan, É. É., & Hinderer, T. 2011, PhRvD, 83, 084051

Virgo Collaboration 2021, PythonVirgoTools, v5.1.1, https://git.ligo.org/virgo/virgoapp/PythonVirgoTools
Virtanen, P., Gommers, R., Oliphant, T. E., et al. 2020, NatMe, 17, 261
Vitale, S., Lynch, R., Raymond, V., et al. 2017a, PhRvD, 95, 064053
Vitale, S., Lynch, R., Sturani, R., & Graff, P. 2017b, CQGra, 34, 03LT01
Vitale, S., Lynch, R., Veitch, J., Raymond, V., & Sturani, R. 2014, PhRvL, 112, 251101
Waskom, M. 2021, JOSS, 6, 3021
Wette, K. 2020, SoftX, 12, 100634
Williams, D., Veitch, J., Chiofalo, M., et al. 2023, JOSS, 8, 4170
Woosley, S. E., & Heger, A. 2021, ApJL, 912, L31
Wysocki, D., O'Shaughnessy, R., Lange, J., & Fang, Y.-L. L. 2019, PhRvD, 99, 084026
Zackay, B., Dai, L., Venumadhav, T., Roulet, J., & Zaldarriaga, M. 2021, PhRvD, 104, 063030
Zackay, B., Venumadhav, T., Dai, L., Roulet, J., & Zaldarriaga, M. 2019, PhRvD, 100, 023007
Zevin, M., Bavera, S. S., Berry, C. P. L., et al. 2021, ApJ, 910, 152
Zevin, M., Berry, C. P. L., Coughlin, S., Chatziioannou, K., & Vitale, S. 2020, ApJL, 899, L17
Zweizig, J. 2006, The Data Monitor Tool Project, https://labcit.ligo.caltech.edu/~jzweizig/DMT-Project.html